\newcommand{\gaz}{\theta}

\newcommand{\vsys}{V_{\rm sys}}
\newcommand{\vobs}{V_{\rm obs}}

\newcommand{\vt}{V_\theta}

\newcommand{\vrot}{V_{\rm rot}}
\newcommand{\vc}{V_{\rm c}}
\newcommand{\vdisk}{V_{\rm disk}}
\newcommand{\vdiskstar}{V^{\rm disk}_\ast}
\newcommand{\vdiskstarmax}{V^{\rm disk}_{\ast,\rm max}}

\newcommand{\vsphere}{V_{\rm sphere}}
\newcommand{\sobs}{\sigma_{\rm obs}}
\newcommand{\slos}{\sigma_{\rm LOS}}

\newcommand{\sigr}{\sigma_R}
\newcommand{\sigp}{\sigma_{\gaz}}
\newcommand{\sigz}{\sigma_z}
\newcommand{\sinst}{\sigma_{\rm inst}}
\newcommand{\sinstT}{\sigma^{\rm inst}_{\rm T}}
\newcommand{\sinstG}{\sigma^{\rm inst}_{\rm G}}
\newcommand{\dsinst}{\delta\sigma_{\rm inst}}
\newcommand{\sbs}{\sigma_{\rm beam}}
\newcommand{\stm}{\sigma_{\rm tpl}}

\newcommand{\sddisk}{\Sigma_{\rm dyn}}
\newcommand{\sds}{\Sigma_{\rm \ast}}
\newcommand{\sdhi}{\Sigma_{\rm HI}}

\newcommand{\sda}{\Sigma_{\rm atom}}
\newcommand{\sdm}{\Sigma_{\rm mol}}

\newcommand{\msol}{\mathcal{M}_\odot}
\newcommand{\mtot}{\mathcal{M}^{\rm tot}_{\rm dyn}}
\newcommand{\mdisk}{\mathcal{M}^{\rm disk}_{\rm dyn}}
\newcommand{\mdiskstar}{\mathcal{M}^{\rm disk}_{\rm \ast}}

\newcommand{\mhalo}{\mathcal{M}^{\rm halo}_{\rm dyn}}

\newcommand{\mhi}{\mathcal{M}_{\rm HI}}
\newcommand{\matom}{\mathcal{M}_{\rm atom}}
\newcommand{\mmol}{\mathcal{M}_{\rm mol}}

\newcommand{\mbar}{\mathcal{M}_{\rm bar}}

\newcommand{\mldyndisk}{\Upsilon_{\rm dyn}^{\rm disk}}

\newcommand{\mls}{\Upsilon_\ast}
\newcommand{\mlsdisk}{\Upsilon_\ast^{\rm disk}}

\newcommand{\mlsldisk}{\Upsilon_{\ast,\lambda}^{\rm disk}}

\newcommand{\mlskdisk}{\Upsilon_{\ast,K}^{\rm disk}}

\newcommand{\fsd}{f_\ast^{\rm disk}}
\newcommand{\Fdisk}{\mathcal{F}_{\ast,{\rm max}}^{\rm disk}}
\newcommand{\Fb}{\mathcal{F}_{\rm bar}}

\documentclass[12pt,preprint]{aastex}

\slugcomment{To appear in ApJ}

\begin{document}

\title{The DiskMass Survey. II. Error Budget}

\author{Matthew A. Bershady,\altaffilmark{1}
Marc A. W. Verheijen,\altaffilmark{2} 
Kyle B. Westfall,\altaffilmark{1,2,3}
David R. Andersen,\altaffilmark{4}
Rob A. Swaters,\altaffilmark{5} and
Thomas Martinsson\altaffilmark{2} 
}

\altaffiltext{1}{University of Wisconsin, Department of Astronomy, 475
N. Charter St., Madison, WI 53706; mab@astro.wisc.edu}

\altaffiltext{2}{University of Groningen, Kapteyn Astronomical
  Institute, Landleven 12, 9747 AD Groningen, Netherlands;
  verheyen@astro.rug.nl}

\altaffiltext{3}{National Science Foundation (USA) International Research Fellow}

\altaffiltext{4}{NRC Herzberg Institute of Astrophysics, 5071 W
Saanich Road, Victoria, BC V9E 2E7}

\altaffiltext{5}{University of Maryland, Dept. of Astronomy, College
Park, MD 20742}

\smallskip
\begin{abstract}
  We present a performance analysis of the DiskMass Survey. The survey
  uses collisionless tracers in the form of disk stars to measure the
  surface-density of spiral disks, to provide an absolute calibration
  of the stellar mass-to-light ratio ($\mls$), and to yield robust
  estimates of the dark-matter halo density profile in the inner
  regions of galaxies. We find a disk inclination range of
  25-35$^\circ$ is optimal for our measurements, consistent with our
  survey design to select nearly face-on galaxies. Uncertainties in
  disk scale-heights are significant, but can be estimated from radial
  scale-lengths to 25\% now, and more precisely in the future. We
  detail the spectroscopic analysis used to derive line-of-sight
  velocity dispersions, precise at low surface-brightness, and
  accurate in the presence of composite stellar populations. Our
  methods take full advantage of large-grasp integral-field
  spectroscopy and an extensive library of observed stars. We show
  that the baryon-to-total mass fraction ($\Fb$) is not a well-defined
  observational quantity because it is coupled to the halo mass model.
  This remains true even when the disk mass is known and
  spatially-extended rotation curves are available.  In contrast, the
  fraction of the rotation speed supplied by the disk at 2.2 scale
  lengths (disk maximality) is a robust observational indicator of the
  baryonic disk contribution to the potential. We construct the
  error-budget for the key quantities: dynamical disk mass
  surface-density ($\sddisk$), disk stellar mass-to-light ratio
  ($\mlsdisk$), and disk maximality ($\Fdisk\equiv \vdiskstarmax / \vc$).
  Random and systematic errors in these quantities for individual
  galaxies will be $\sim$ 25\%, while survey precision for sample
  quartiles are reduced to 10\%, largely devoid of systematic errors
  outside of distance uncertainties.
\end{abstract}


\section{INTRODUCTION}

Paper I of this series (Bershady et al. 2010) presented the DiskMass
Survey (DMS), a study designed to break the disk-halo degeneracy (van
Albada et al. 1985), and bypass a major roadblock in testing galaxy
formation models. Without an independent measurement of the
mass-to-light ratio of the stellar disk ($\mlsdisk$), it is not
possible to determine the structural properties of dark matter halos
from rotation curve decompositions. The DMS is an effort to make a
direct, and absolute kinematic measurement of the mass surface-density
of intermediate-type spiral disks ($\sddisk$), calibrate
$\mls$, and determine the density profiles of dark matter halos
in these systems. In a nutshell, the question we aim to answer is
this: How maximal are normal spiral disks?  Specifically, we want to
know how much of the observed disk rotation within the inner 2 to 2.5
disk radial scale-lengths is in response to the mass of the disk
itself.

In this survey, the amplitude of the vertical motions of collisionless
tracers in 46 galaxies are measured via integral-field spectroscopy
(IFS) of the integrated star-light, in conjunction with a photometric
estimate of the vertical scale height of these tracers. The former is
measured as a velocity dispersion using absorption-lines in the
stellar continuum sensitive to old but luminous disk stars, typically
K giants. The latter is based on the correlation between disk
oblateness and radial scale-length. An estimate of the disk-mass
surface density then follows dimensionally from these length and
velocity scales.  Our program is inspired by the insights of van der
Kruit \& Searle (1981) and Bahcall \& Casertano (1984), and the
pioneering observations of van der Kruit \& Freeman (1984, 1986) and
Bottema (1993). Similar surveys are also underway (Herrmann \&
Ciardullo 2009) using different collisionless tracers, albeit with the
same dynamical approach.

Galaxies in the DMS were selected, as described in Paper I, based on
apparent size, inclination and visual (qualitative) morphology. Given
these constraints, selection from the UGC (Nilson 1973) resulted in a
sample with disk central surface-brightness in range $\mu_{0,R} = 20.5
\pm 1.1$ mag, comparable to Freeman's (1970) result (translated to the
R band) of $\mu_{0,R} = 20.65 \pm 0.3$ mag for what are commonly
referred to as ``normal'' spiral disks.  Galaxies in the DMS lie at
distances of 15 to 200 Mpc, with morphological types mostly between Sb
and Scd. The sample spans factors of 100 in $K$-band luminosity, 8 in
blue-to-infrared color, and 10 in disk size and central
surface-brightness. {\it A posteriori}, we find the sample almost
entirely has rotation velocities $>$ 120 km s$^{-1}$. Several studies
(e.g., Dalcanton et al. 2004; Yoachim \& Dalcanton 2006) have shown
that disk properties above and below this rotation-speed have distinct
structural properties, including different fractions of thick to thin
disk components, as well as different dust-to-star vertical
scale-heights. Faster rotators tend to have relatively smaller
thick-to-thin disk luminosity ratios, and shorter dust
scale-heights. While these and other differences within the spiral
galaxy population preclude global dynamical conclusions about disks
systems based on the DMS sample, this survey does target the heart of
the spiral population: Galaxies in the DMS are typical of Tully-Fisher
surveys, and well-sample the knee of the spiral luminosity function,
in which most disk stars are contained. As such, the DMS provides a
particularly relevant calibration of $\mls$ in massive, star-forming
systems in the nearby universe.

While computing $\sddisk$ from stellar velocity-dispersions and
scale-heights is conceptually simple, the actual measurements are
demanding and the analysis complex.  To compute disk maximality (here,
$\Fdisk$) we also need to measure the rotation speed ($\vrot$),
orthogonally projected to the vertical stellar motions ($\sigz$) used
to determine $\sddisk$. This orthogonality presents an observational
quandry, requiring us to choose modest inclinations for measuring {\it
both} $\vrot$ and $\sigz$. At such inclinations it is also a challenge
to determine the inclination angle itself. Finally, to complete our
scientific objectives, we also must determine $\mlsdisk$. This
requires further measurement and analysis to correct $\sddisk$ for the
atomic and molecular gas content of the disk, and to correct
surface-photometry for dust extinction. Disk dark-matter, if it
exists, we assume is distributed with a scale-height similar to old
disk stars.

Accurate $\mls$ values are critical for inferring the dark-halo
profiles in the wide range of galactic systems required to trace the
cosmic history of the stellar baryon fraction. Hence the calibration
of $\mls$ is of prime relevance for understanding galaxy structure and
formation. Ideally $\mls$ would be calibrated for stellar populations
spanning as wide a range of properties (age, metallicity, abundance)
as possible. While the DMS does not sample extreme populations likely
found in giant ellipticals, low-mass dwarf irregulars and spheroidals,
and low-surface-brightness disks, the DMS contains a well-defined
sample spanning a wide range of properties. This is ample for testing
the mass zero-point of stellar population synthesis models as well as
trends with color and star-formation rate. We argued in Paper I
that $\mls$ should be known to 30\% or better in order to make
substantial progress in determining, e.g., the maximality of spiral
disks.

The goals for this second paper in the DMS series are to estimate, and
present an accounting of, the error-budget of the primary derived
quantities: $\sddisk$, $\mlsdisk$ and $\Fdisk$. Specifically, we
verify we can reach the stated goal of 30\% uncertainty in $\mls$. As
a secondary goal we substantiate our assertion that low-inclination
disks are optimal for this type of measurement. To achieve these goals
we step through every major aspect of our measurement and analysis,
starting with considerations that informed our survey strategy,
namely, uncertainties in the disk stellar velocity ellipsoid (SVE; \S
2.1), disk oblateness (\S 2.2), and inclination (\S 2.3).  Within the
methodological framework established in Paper I, we detail all facets
required to arrive at an accurate and reliable estimate for $\sigz$
(\S 3). We focus here on the challenging aspects of the spectral
analysis unique and central to the survey. Distance errors are
considered briefly in \S 4. The development in these preliminary
sections allows us to establish our expected error budget for the
primary derived quantities from our survey (\S 5): $\sddisk$,
$\mlsdisk$, and $\Fdisk$. The error-budget analysis is summarized
in \S 6. In Paper III (Westfall et al. 2010, in preparation) we
present the cross-correlation technique used to derive the SVE from
pilot observations of UGC 6918. The same galaxy is used here to
illustrate central features of our analysis. All distant-dependent
quantities are scaled to H$_0$ = 73 km s$^{-1}$ Mpc$^{-1}$.
Logarithmic errors are specified as $\Delta \ln
X \equiv \epsilon(X)/X$, where $\epsilon(X)$ is the error in quantity
$X$.

\section{SURVEY DESIGN OPTIMIZATION}

We argued in Paper I that a nearly face-on approach to measuring both
the disk mass surface-density and the total mass is optimal because of
the quadratic dependence of $\sddisk$ on $\sigz$, compared to
the linear dependence on the disk scale-height. We develop this
argument by comparing (\S 2.1 and \S 2.2) the constraints available on
the shape of the SVE compared to those on the disk oblateness. We
quantify in \S 2.2 the contribution of disk-oblateness uncertainties
to our error budget.  Because of the uncertainties in the SVE there is
a detailed balancing that can be done between the amount of projection
of the vertical versus tangential motions of the stars into the
observers line-of-sight.  We tie together the uncertainties in the SVE
and inclination in \S 2.3 to arrive at an optimized disk inclination
range for the DMS.

\subsection{Disk Stellar Velocity Ellipsoid}

The SVE is described by its radial, tangential and vertical
components: $\sigr$, $\sigp$, and $\sigz$. Here we do not
entertain the effects of a tilted ellipsoid. What is relevant for the
DMS, in order to deproject $\sigz$ from the observed line-of-sight
velocity dispersion, is the shape of the SVE. This shape is
conveniently parameterized by the axial ratios $\alpha \equiv
\sigz/\sigr$ and $\beta \equiv
\sigp/\sigr$. Expectations from the Solar Neighborhood are
that $\sigr>\sigp>\sigz$, specifically with
$0.5<\alpha<0.6$ and $0.6<\beta<0.7$ for the thin disk, depending on
what tracers are used (Binney \& Merrifield, 1998). However, little is
actually known about these values in external galaxies, and no
measurement exists to indicate if there is a radial dependence to
these ratios. Radial dependencies are likely, based on
dynamical arguments, including the simple observation that galaxies
tend to be dynamically hotter in their interior.

Extant knowledge of $\alpha$ and $\beta$ for external spiral galaxies
(summarized in Shapiro et al. 2003) consists of integrals over major-
and minor-axis kinematic data within the inner 1-3 radial
scale-lengths of 6 moderately inclined galaxies, requiring assumptions
regarding the form and validity of the epicycle approximation and
asymmetric drift equation; and 40 edge-on galaxies, requiring further
dynamical assumptions and scaling arguments (van der Kruit \& de Grijs
1999). All measured values of $\alpha$ and $\beta$ for external
galaxies are global quantities.

These external-galaxy studies have used the epicycle approximation to
measure $\beta$, namely: $\beta = \onehalf [ \partial( \ln \vt) /
  \partial( \ln R) + 1]$, where $\vt$ is the tangential speed of
the stars (see Westfall 2009). Data presented in Shapiro et al. (2003)
imply $0.6<\beta<0.8$.  Since $\beta$ depends on the derivative of the
tangential speed, in general we do not expect $\beta$ to be constant
with radius. To obtain global values for $\beta$ these same studies
have parameterized the radial dependence of the tangential speed as a
power-law. A parameterization serves to minimize errors associated
with the derivative in the epicycle approximation. Adopting a more
realistic functional form for the tangential speed (e.g., the
``universal rotation curve'' of Courteau, 1997) would be
preferable. With a suitable model for the tangential speed, the
uncertainty in the value of $\beta$ is not of concern because the very
stellar measurements needed to determine the line-of-sight velocity
dispersion ($\slos$) can be used to estimate $\beta$.

To derive $\sigz$ from $\slos$ then is largely a matter of
determining $\alpha$, and herein lies the problem.  From existing data
there is some hint of a trend in the ellipsoid ratio $\alpha$ to
larger values (0.8) for types earlier than Sb, but this is based on a
few points with substantial errors. It is conceivable this trend is
due to increasing bulge contamination in earlier types. For later
types, the mean ratio falls in the range $0.5<\alpha<0.7$, with less
indication of trend, but a 50\% spread (1$\sigma$), i.e., $0.3 <
\alpha < 0.9$. For an edge-on approach to measuring disk mass, this
translates into a 100\% systematic error in $\sddisk$ from the
velocity component alone. In short, the SVE in disk galaxies is not
well-known.

This brief discussion concludes that the experimental design to
measure disk mass via the stellar $\slos$ must either determine
the SVE shape very well, or choose an inclination range where
$\sigz$ is favorably projected and uncertainties in the SVE lead to
little error in the correction. It turns out that a nearly face-on
inclination answers both desiderata. Future papers in this series will
show the dependence on inclination for the SVE determination (see also
Westfall 2009). Here we provide estimates for typical $\alpha$ and
$\beta$ values and uncertainties in our survey in \S 3.5.4.

\subsection{Disk Oblateness}

We show it is possible to estimate $z_0$, the vertical scale-height
used to determine $\sddisk$ (equation 1 of Paper I), reliably
from $h_R$, the exponential radial disk scale-length.  Several
independent, photometric surveys exist of edge-on galaxies, linking
scale-height and scale-length to galaxy type, rotation speed, and
other readily observable global properties. In contrast to the
situation for the SVE, a relatively clear picture emerges.

\subsubsection{Disk vertical structure}

To place disk oblateness properly into the context of $\sddisk$
error budget, it is necessary to clarify the definition of
scale-height in terms of the vertical structure of a disk.
Anticipating the generalization in \S 5 we begin by noting the
formulation for $\sddisk$ in equation (1) of Paper I assumes a
locally isothermal disk with a vertical mass-density distribution
function of ${\rm sech}^2(z/z_0)$. Other density distributions are also
appropriate, including ${\rm sech}(z/z_{\rm sech})$ and $\exp(-z/h_z)$ (van
der Kruit 1988).  A more general expression of $\sddisk$ can be
written to include the three vertical distribution functions:
$\sddisk = \sigz^2 / \pi k {\rm G} h_z$, where $k = 3/2$, 1.7051, 2
for exponential, sech, or sech$^2$ vertical mass
distributions. Since all three functions behave as exponential
distributions at large scale-heights, we can relate $z_0 = \sqrt{2} \,
z_{\rm sech} = 2 h_z$ in our specific functional formulation. We use these
equalities throughout the discussion here, and define oblateness as $q
\equiv h_R / h_z \equiv 2 h_R / z_0 \equiv \sqrt{2} h_R / z_{\rm sech}$.

Unfortunately, the issue of what is the actual vertical mass-density
distribution of disks remains outstanding. In the DMS, we parameterize
this ignorance in the possible range of $k$. An isothermal
distribution is conceptually preferable given a simple picture of a
single, relaxed population of disk stars. However, the non-isothermal
density distributions, which have discontinuous potentials at the
mid-plane, appear to be a suitable approximation to a combination of a
thin and thick stellar disk plus a very thin mid-plane distribution of
atomic and molecular gas and very young stars. This is a model
qualitatively consistent with our picture of the Milky Way.  Indeed,
recent studies of resolved stellar populations in nearby,
edge-on-galaxies also show such three-component stellar systems, with
scale-heights increasing with population age (Seth et al. 2005).
Observations in the near-infrared, capable of penetrating the disk
mid-plane dust layer, indicate that an exponential vertical
distribution is likely the best functional form (Wainscoat et
al. 1989, Aoki et al. 1991, de Grijs \& van der Kruit 1997), although
Aoki et al. point out that the steepness of the density distribution
inferred from the $K$-band light near the mid-plane that they observed
in NGC 891 could be due to an excess of red super-giants with low
M/L. Yet even if the light near the galaxy mid-plane is dominated by
massive stars with low $\Upsilon$, the additional gas components still
make an exponential vertical {\it mass}-density distribution a
plausible approximation.

Luckily, the uncertainties in the vertical mass-density distribution
can be decoupled from uncertainties in estimating a characteristic
scale-height, $h_z$, for the purposes of measuring
$\sddisk$. This is true so long as the light-weighting of the
kinematic signal in face-on galaxies is the same as what defines the
photometric vertical profile in edge-on samples. This is a reasonable
assumption when coupling the near-infrared light distribution in
edge-on systems to gravity-insensitive kinematic signal dominated by
cool stars in face-on samples. (The lines in our spectroscopic regions
of \ion{Mg}{1b} and \ion{Ca}{2} meet these desiderata.) In this
situation the relative contribution of different disk components
(e.g., thin and thick) is statistically self-consistent in photometric
estimates of scale-heights (observed in edge-on samples) and in
kinematic estimates of $\slos$ (observed in face-on galaxies);
$k$ parameterizes the dynamical variation in the disk vertical
mass-density to the estimate of $\sddisk$. Accordingly, we focus
now on what we know about the correlation of the vertical to the
radial scale-lengths of the disk light distribution, and then explore
how the correlation might be biased by wavelength or the presence of
multiple disk components.

\subsubsection{A fiducial relationship}

Figure 1 illustrates our compilation of four studies (Kregel et
al. 2002; Pohlen et al. 2000; Schwarzkopf \& Dettmar 2000; and
Xilouris et al. 1997, 1999) of the vertical to radial disk
scale-length ($h_z$ and $h_R$, respectively) based on photometry of
edge-on galaxies. All absolute values have been rescaled consistently
to our choice of H$_0$. For contrast and application to the DMS, the
two panels break out the spirals into intermediate-types (top) and
early and late types (bottom). The majority of DMS galaxies in the
Phase-B sample (Paper I), for which there are stellar spectroscopic
observations, have morphological types consistent with those in the
top panel.

For the Kregel. et al. (2002) sample, we have adopted their $I$-band
results as they recommend, included 3 more-nearby galaxies from their
later work (NGC 891, 5170, and 5529; Kregel et al. 2004), but excluded
ESO555-G36 because of contamination due to a bright fore-ground star
(de Grijs 1998).  

Of the seven edge-on galaxies we include in Figure 1 from Xilouris et
al. (1997, 1999), two overlap with the Kregel et al. sample; however,
these galaxies have been observed independently and subjected to
different modeling techniques. The Xilouris et al. observations
include in $B,V,I,J,$ and $K$ bands. Since we are interested
primarily in the distribution of old, luminous stars also used as
dynamical tracers, and to be consistent with the Kregel et al sample we
therefore focus on the red and near-infrared bands (including
$I$). Plotted are the $I$-band values, after adjusting their
distance-estimates based on recession velocities corrected for Virgo
in-fall, consistent with the procedures in Kregel et al. (2002).

Accordingly, for the Schwarzkopf \& Dettmar (2000) sample, for which
they have observed in $r,R,H,K$ bands, we have restricted their sample
to those 15 galaxies with near-infrared $H$- or $K$-band measurements.
Similarly, for the Pohlen et al. (2000) sample, for which they have
observed in $g,r/R,i$ bands, we have restricted their sample to only
those 5 including $i$-band measurements. We find the scatter increases
substantially if we include bluer measurements. These last two studies
are particularly interesting because unlike the two previous studies
which use only an exponential vertical distribution, they allow the
vertical distribution to be characterized by either exponential, sech,
or sech$^2$ functions. Their tabulated values represent the
corresponding effective exponential scale-heights regardless of the
fitted functional form. By limiting their sample to just the
near-infrared measurements, the trends of radial to vertical
scale-lengths appear identical for all functional forms.

While the scatter in Figure 1 may at first glance look large, there
are two significant degrees of coherence. First, disk oblateness for
intermediate-types Sb-Scd correlates with the scale of the disk
($h_R$; top panel). Second, there is an offset in the relation with
galaxy types earlier than Sb or later than Scd (top versus bottom
panels). The correlations are in the sense that later-type galaxies
(less bulge-dominated systems) have thinner disks, and larger disks
(at a given bulge-dominance) are thinner. This makes some
astrophysical sense in that whatever produces a bulge or pseudo-bulge
either represents a merging or disk instability process that would heat
the disk. The fact that disk thickness does not scale linearly with
disk scale-length must reflect a more complicated interplay between
angular momentum and the mechanisms responsible for disk heating.
While previous studies have noted the type-dependence, the correlation
with scale appears as, if not more, fundamental, and is likely
associated with the general scale-dependence of galaxy properties
(e.g., van den Bergh 1960). Key here for minimizing scatter is this at
least {\it bivariate} correlation.

Our fit to the Sb-Scd galaxies typical of the DMS for the Kregel et
al. subset (shown in Figure 1) statistically matches both the slope,
zero-point and scatter of the other three samples; these four samples
are statistically indistinguishable in this regard.  The Kregel et al.
sample is a good match to our own both in the distribution of physical
size, surface-brightness, rotation velocity, and morphological type
(compare Table 1 of Kregel et al. [2002] with our Table 3 of Paper I.)
Hence we adopt this fit as the operational relation for the DMS at
this time:
\begin{equation}
\log (q_{R}) \ \equiv \ \log \ (h_R/h_z) \ = \ 0.367 \ \log \ (h_R/{\rm kpc})  \ + \ 0.708 \ \pm \ 0.095.
\end{equation}
This is consistent with $h_z \propto h_R^{2/3}$.  We conclude that the
effective oblateness of the disk can be reasonably estimated to about
25\% (1$\sigma$ systematic error for any one galaxy) for face-on
galaxies typical of the DMS sample, simply via measurement of $h_R$ in
the $I$-band. We do not distinguish between sub-types for Sb-Scd since
there is presently no solid statistical basis (e.g., K-S test) to do
so. For error-budgeting purposes we adopt the logarithmic derivatives
$\Delta \ln q_R = 0.25$ and $\Delta \ln h_R = 0.03$, the latter
following the analysis of MacArthur et al. (2003). We believe $\Delta
\ln q_R = 0.25$ is likely an over-estimate for reasons given in \S 2.2.3 
and \S 2.2.5. For the 3 galaxies in the DMS earlier than Sb and the 2
galaxies later than Sd, we use the data points in the bottom panel of
Figure 1 to estimate offsets from the above fiducial, assuming the
same slope.  This adds 15\% additional uncertainty in $q_R$ (a total
of 29\% instead of 25\% uncertainty in the oblateness correction), but
only for $\sim$10\% of the DMS sample.  We expect a more comprehensive
compilation of the literature or future analysis will add to, and
improve, the calibration of $q_R$ for all types.

\subsubsection{Wavelength dependence}

While we have calibrated $q_{R}$ in the $I$-band, we also explored the
wavelength dependence of this relationship. This is important because
we know disks in external galaxies have radial color gradients
presumably due to changes in mean age and metallicity with radius;
expectations from the solar neighborhood are that vertical color
gradients should exist as well due to increasing scale-heights with
stellar population age. {\it A priori}, it is unclear if the vertical
and radial gradients scale such that disk oblateness appears constant
with wavelength. If not, in order to avoid systematics in deriving
$\sddisk$ the oblateness--radial scale-length relation must be
calibrated at the same wavelength used to measure radial scale-length,
and at a wavelength appropriate to measure the scale-height of the
kinematic tracers.

Inspection of the sizes ($h_R$) and oblateness ($q$) as a function of
band-pass in the Xilouris et al. sample shows clearly that size
increases while oblateness increases significantly
at shorter wavelengths. Larger radial scale-lengths in the blue are
not unexpected: galaxy disks have color gradients in the sense that
they become bluer with radius, e.g., see de Jong (1996a) in the
face-on context. In fact, an inspection of the radial scale-lengths
tabulated by de Jong (1996b) in $B$ and $K$ bands shows just this
effect. Likewise a decrease of $q$ in blue light is consistent with
expectations that star-formation is concentrated toward the mid-plane,
despite the impact of extinction and scattering on the apparent light
distribution. While the Xilouris et al. data shows an effect from $I$
to $B$ bands, there is also some hint of an effect between $I$ and $V$
as well, but there is insufficient data to probe if differences exist
between $I$ and redder bands.

Yoachim \& Dalcanton (2006; hereafter YD) find a similar result in a
larger sample of 34 edge-on galaxies observed in $B,R,K$
bands. However, they also see a decrease in $h_z$ in the $K$-band,
which they tentatively interpret as a combination of extinction,
observational depth, and stellar-population effects. This data set is
rewarding to work with because, like the work of Kregel et al., the
authors have published their uncertainties. We have plotted their
measurements in Figure 2 (top panels) for all of their
band-passes. The histogram to the right shows the residuals about the
relationship given by equation (1).  From these figures we draw two
conclusions.

First, our calibration with independent data provides an excellent
description of the data. The mean residuals in the $R$ and $K$ bands
are under 4\%, and the scatter is somewhat lower (0.07 dex, or 18\%).
If we assume their sample is similar to what we have compiled from the
above studies, this indicates that our zero-point for $q_R$ is likely
accurate to better than 7\% in the red and near-infrared, and that our
estimate of $\Delta \ln q_R$ can be reduced. Because we cannot verify
this assumption, we retain $\Delta \ln q_R = 0.25$.

Second, the mean residual in the $B$-band differ by at most 0.05 dex
(12\%), and while the $R$ and $K$ band mean residuals differ formally
by 5\% they are indistinguishable on the basis of a K-S test. This
result requires no assumptions because it is a differential
comparison. In other words, our finding is that while oblateness and
scale-length change with wavelength, to first order this simply shifts
galaxies along the regression given by equation (1). This implies that
$q_R$ can be readily estimated via any accurate scale-length
measurement from the blue to near-infrared wavelengths.  Since we are
interested in the scale-height of the old stars, this still requires
scale-length measurements in the red or near-infrared.

\subsubsection{Thick disk component}

The YD data set can also be used to explore what impact a thick-disk
component has on our relation for $q_R$.  This is important because
the presence of a significant thick disk, if hidden due to lack of
projection in a face-on system, would systematically increase
$\slos$ relative to a system dominated by a thin disk.  Unless
the effective oblateness, $q_R$, accounts for this thick component,
the result would be an overestimate of $\sddisk$.

Ever since the early work of van der Kruit \& Searle (1981), there
have been on-going searches for thick disks around external spiral
galaxies. For example, with the advent of deep CCD photometry,
Morrison et al. (1994) found NGC 5097 was absent a Milky Way-like
thick disk, with at most a 2\% contribution (by light) from a thicker
component. Fry et al. (1999) found no evidence for a thick disk in NGC
4244 also down to very faint light levels. However, even the earlier
photographic work of van der Kruit \& Searle (1981) sometimes showed
small departures from a single exponential vertical light distribution
at very low light levels. The work of YD convincingly shows the need
for more than a single component in a wide range of late-type spirals,
at least qualitatively consistent with the work of Seth et
al. (2005). The impact of the YD two-component disk fits in the $R$
band are shown in the bottom panel of Figure 2. Compared to the
one-component fits, the oblateness of the two-component thin plus
thick disk increases 36\% for the thin disk and decreases 50\% for the
thick disk. If the kinematic signal in face-on systems could be
uniquely identified with thin or thick components, then these data
could be used to recalibrate equation (1). The scatter in the $q_R$
relation for the thin-disk component is comparable to that for the
Kregel et al. sample (25\%), while for the thick-disk component the
scatter is only slightly larger (32\%).

For the DMS, however, we expect both thin and thick disks to
contribute to our kinematic signal. For example, YD find the ratio of
thick to thin scale-heights is roughly a factor of 2.4. This is much
larger than the value of 1.1 Seth et al. (2005) found for the ratio of
scale-heights for red giant-branch (RGB) to asymptotic giant-branch
(AGB) stars in similar galaxies. Hence YD's findings are likely not
coupled to population-age effects, but some other mechanism. In other
words, it is plausible to assume that RGB stars (our expected primary
kinematic tracer) are well mixed in both thin and thick components as
parameterized by YD. In this case, while superficially the changes in
disk oblateness going from one to two-disk components appear
alarmingly large, we show the single-disk fits are a suitable
characterization of the vertical light-distribution for mass-modeling.

A rough assessment of the importance of the thick disk to the
integrated light can be gleaned from YD's Figure 7: The observed
excess light departing from a single-disk fit becomes appreciable
above $\sim 3 h_z$. The excess accounts for no more than about 5\% of
the total light enclosed within their measuring window in scale-height
(their fitting region is $R/h_R < 4$ and $z/h_z = 6$ to 8), but the
excess is at large heights where the dynamical impact in terms of the
effective scale-height is more significant. This percentage appears
fairly independent of rotation velocity. On the other hand, YD show
that the luminosity ratio of thick to thin disks decreases
substantially at higher rotation speeds, while the ratio of vertical
scale-heights is constant. (This requires the thick disk radial
scale-length to increase at lower circular velocities relative to the
thin disk, which is what they see.)  For $\vrot>100$ km s$^{-1}$, the
contribution of the thick-disk to the overall disk surface-brightness
is below 10\% in the $R$ band.  This is consistent with previous
studies finding little thick-disk contributions or evidence for
disk-flaring within 4 radial scale-lengths (equivalent to larger
thick-disk scale-lengths). For example, NGC 5097 has a rotation speed
of 220 km s$^{-1}$ (Casertano 1983), while NGC 4244 has a rotation
speed of 100 km s$^{-1}$ (Olling 1996). For the DMS, 92\% of the
sample has $\vrot > 120$ km s$^{-1}$. This is in a regime where YD
estimate the thick-disk component is below 10\% of the total disk
light contribution and thick and thin disk radial scale-lengths are
equivalent. (Only 4\% of our sample would be expected to have $>20$\%
thick-disk contributions based on their calibration.) Therefore we
anticipate the single-disk fits provide an fairly accurate
characterization of disk oblateness for the DMS.

We refine this initial assessment by calculating the first moment of
the vertical light profile,\footnote{This is the moment at a specific
  radial and azimuthal location ($R,\theta$) in the disk, i.e., {\it
    not} the projected edge-on light distribution. In this case,
  $z_1(R,\theta) = \int_0^\infty I(R,z,\theta) \ z \ dz \ /
  \ \int_0^\infty I(R,z,\theta) \ dz$, where $I(R,z,\theta) =
  \sum_{1,...,j} \ I_j(R,z,\theta)$, and $I_j(R,z,\theta) \ = \ I_{0,j}
  \ e^{-R/h_{R,j}} \ {\rm sech}^2(z/z_{0,j})$ is the light density for
  a single isothermal component, $j$.}  $z_1$, for the one- and two-disk
model parameters in YD's Tables 3 and 4. This moment is a
non-parametric proxy for $h_z$, sensitive to the shape and extent of
the light distribution. For single-component sech$^2$ vertical light
distribution $z_1/h_z = 2 \ln 2$.  Because YD provide the median
parameters over a set of different fitting schemes, where the median
is taken for each parameter individually, these values do no make
self-consistent sets. For example, if one integrates the thin and
thick disk profiles specified in their Table 4, their tabulated ratio
of thick to thin disk luminosity is not recovered. We proceed by
adopting the median scale-lengths and scale-heights, but renormalize
the central surface-brightness of the thin and thick disks to
simultaneously yield (i) the tabulated thick-to-thin luminosity ratio
and (ii) the same vertical surface-brightness profile in the mean for
single and two-disk models in the region for $z<3h_z$ and
$R/h_R<4$. The latter essentially reproduces their Figure 7. We also
implement their correction for internal extinction for the 2-disk
model, which in their scheme is a correction to the thin-disk
luminosity only, but still require condition (ii) to be met.

The difference between the face-on, radial surface-brightness profiles
for the one- and two-disk models, renormalized as outlined above, are
shown in the top panel of Figure 3.  Without extinction, the
differences are of order 5\%, as anticipated, increasing at larger
radii. The effect of extinction is to decrease the contribution from
the thin-disk component; with our renormalization this brings the one-
and two-disk face-on surface-brightness profiles into closer agreement.
The ratio of vertical first moments in the bottom panels of Figure 3
show significant scatter consistent with measurement errors, and a
trend to relatively smaller $z_1$ for the one-disk models at slower
rotation speeds. In the regime of interest to the DMS between
$1<R/h_R<3$ for the fast-rotating disks ($\vrot > 120$ km s$^{-1}$), the
$z_1$ ratio is consistent with unity; with no extinction the weighted
mean ratio is 0.94, and with extinction this increases to 1.06. In the
context of the concerns framed at the beginning of this subsection, it
is clear that the presence of a thick-disk component has an
insignificant impact on the effective disk scale-height
relevant for disk-mass measurement.

In Figure 4 we show there is a correlation between offsets from the
fiducial oblateness relation of the thick and thin components, as well
as between the one- and two-disk models. For the later, the two-disk offsets
are computed as the luminosity weighted mean offset of the thick and
thin components. In other words, when the thick disk tends to be more
or less oblate than what would be inferred by its radial scale-length,
so too does the thin disk component; together these offsets are in
lock-step with the oblateness variation of the one-disk model. Whether
due to astrophysics or the fitting process, both models represent the
same departures from the fiducial oblateness relation. This result
combined with the insensitivity of $z_1$ to one- versus two-component
disk fits leads us to conclude that the single-disk fits provide a
statistically accurate characterization of disk oblateness for the
DMS.

\subsubsection{Additional correlations and concerns}

It is reasonable to expect that the estimation of $h_R/h_z$ will
improve in the near future.  Measurement precision will inrease with
deeper images, sample sizes will increase with the extensive imaging
surveys now in hand, and sample homogeneity will increase with better
classification.  In this context, it is relevant to recall that the
single-disk fits of YD yield 30\% less scatter than our calibrators.

It is also likely that the estimation of $h_R/h_z$ will improve
because of additional correlations between disk oblateness and other
observables. For example, Dalcanton et al. (2004), Kregel et
al. (2005) and YD show there is a correlation between disk rotation
speed and $h_R/h_z$. Kregel et al.'s sample also shows a possible
correlation of $h_R/h_z$ with \ion{H}{1} mass. Size, rotation-speed,
and mass are all proxies for galaxy scale (in the van den Bergh sense),
so these correlations may not be independent. Indeed, we do not see
any correlations in the residuals from $q_R$ with these other
quantities. However, a principal-component analysis may further reduce
the scatter in Figure 1. The significant residuals from the
correlation of $h_R/h_z$ with $h_R$ (in excess of observational error)
indicates a further reduction in scatter of $q_R$ may be possible.

One other relation that is being used in the literature to estimate
$h_z$ (Herrmann \& Ciardullo 2009) is an apparent correlation of disk
oblateness with surface-brightness (Bizyaev \& Mitronova 2002, 2009).
Taken at face value, the formula provided by Bizyaev \& Mitronova
(2002) produces estimates of $h_z$ based on measurements of $h_R$ and
the disk central surface-brightness (in the $K$ band) with 22\%
precision for their sample. Applying equation (1) to their sample we
find $h_z$ is estimated only to 30 to 40\% precision, depending on
what subset of their data is used. The subset with larger scatter is
their preferred subset of {\it larger} galaxies, for which they
believe their measurements are more robust. Somewhat worrisome is the
fact that this same sample has an offset of about 20\% in the mean
from equation (1) as well. Furthermore, the formulae relating
$h_R/h_z$ to $K$-band central disk surface-brightness from their two
studies (2002 versus 2009) do not give consistent results, with
$h_R/h_z$ differing by over 60\% near the Freeman value (assuming
typical $B-K$ colors for disks). The sample and photometric data are
similar between these two studies by the same authors, but there is
insufficient information to determine why their derived relation has
changed. More puzzling is the fact that we do not find a convincing
correlation in the data sets we have analyzed here (see for example
Kregel et al. 2005). For these reasons equation (1) remains our
estimator for disk oblateness, and we would caution against using
surface-brightness as a proxy for disk oblateness at this time.

\subsubsection{Contributions to the error budget}

The uncertainties in estimating $h_z$ from applying the calibration of
$h_R/h_z$ in equation (1) have both systematic and random
components. Random errors in $h_z$ arise from the propagation of random
errors in measuring $h_R$ in an individual galaxy. There are also
likely to be stochastic, astrophysical variations in disk oblateness
from galaxy to galaxy of a given size and type. These variations
contribute, no doubt, to some of the scatter about the mean relation
seen in Figure 1 even for the restricted subset of Sb-Scd systems.
However, while such variation leads to systematic errors in estimating
$h_z$ for individual galaxies, the effect is random for the sample as
a whole.  Hence errors in $h_z$ that arise from real variation in disk
oblateness will be substantially reduced in a statistical sense for
the survey as a whole, and will be {\em random} errors. The
propagation of these errors for $\sddisk$, $\mlsdisk$ and
$\mhalo$ are discussed in \S 5, where we also consider the impact of
a thick disk.

\subsection{Disk Inclination}

How close to a face-on orientation is best? Low inclinations minimize
(a) line-of-sight contamination to $\sigz$ from $\sigr$ and
$\sigp$, (b) beam-smearing from velocity-field shear, and (c)
extinction effects on the derived surface-brightness, color and
velocity dispersion of the disk. However, in order to obtain the full
mass-budget of the galaxy, we must measure not only $\sddisk$
but the total mass ($\mtot$) via measurements of the projected
circular rotation and disk inclination ($i$).

Nominally, we would prefer galaxies which are minimally inclined to
derive kinematic inclinations from their velocity fields, with
sufficient accuracy such that errors on the derived total masses from
rotation curves are comparable to those of the disk mass
surface-density from $\sigz$. However, the detailed trade-off
depends on the specific science goal. Within our survey there are two
related ones, namely (i) measurement of $\sddisk$ and
calibration of $\mlsdisk$, where low inclination is preferred
($<30^\circ$) to minimize line-of-sight contamination to $\sigz$;
and (ii) measurement of disk-to-total mass ratios (from rotation-curve
decompositions), which favors modest inclination (25$^\circ$ to
45$^\circ$) to balance disk- and total-mass errors. The
$\sddisk$ error-budget depends on inclination primarily through
the correction to $\slos$ for the projected $\sigr$ and
$\sigp$ components of the SVE. The degradation in precision of
measuring $\sigz$ is shallower than a simple 1/$\cos i$ function
because the SVE shape can be measured directly from the data, and this
measurement improves with increasing inclination (up to moderate
inclinations). Westfall (2009) addresses the optimum inclination for
SVE decomposition. The following discussion encapsulates the arguments
used to arrive at these quantitative inclination ranges, and how we
optimized our selection.

Our base-line approach has been to rely on kinematic estimates of
inclination (Paper I), which we find to be superior at low
inclination to photometric estimates based on apparent disk
ellipticity (Andersen \& Bershady 2003). Indeed, one of the
motivations of Andersen's (2001) survey was to establish the efficacy
of using H$\alpha$ velocity fields measured with coarsely-sampled IFUs
to constrain disk inclination in preparation for the current DMS.
Because of the high S/N and spectral resolution of the H$\alpha$ data,
we are able to determine kinematic inclinations with a precision
better than 2 degrees at the relatively low inclination of 28 degrees
(see Appendix A).

As noted in Paper I, inclination can also be estimated by inverting
the Tully-Fisher relation (TF; Tully \& Fisher 1977), with the
advantage that random errors in so-called inverse--Tully-Fisher (iTF)
inclinations do not blow up at low inclination, as do both photometric
ellipticity and kinematic isovelocity methods. For nearly face-on
galaxies the iTF-inclination random errors in percentage terms are
simply proportional to the quadrature sum of the projected velocity
and luminosity errors (the latter including distance uncertainties),
while the systematic errors are proportional to the scatter in the TF
relationship. Because of the potential for large systematic errors
(i.e., a priori: What is the TF scatter for the sample or source in
hand, and how accurate are the non-Hubble flow corrections to the
observed systemic velocity?), our preference is to work in a regime of
inclination where kinematic inclinations can be measured with small
enough errors to reduce the overall error budget (random and
systematic), and where we can directly verify a galaxy's location on
the Tully-Fisher relation. While our survey sample was selected
accordingly, post-facto, high-precision kinematic inclinations are not
obtainable for all survey galaxies. Consequently we have also utlized
additional inclination constraints from iTF in some cases (Andersen et
al. in preparation).

Figure 5 illustrates the trade-offs with inclination between
logarthmic errors in total-mass and disk mass surface-density,
accounting for errors in inclination only (formulae are given in the
Appendices as noted).  Left-hand panels show errors assuming
inclination is measured from fitting the H$\alpha$ velocity fields
with a model of an inclined, rotating disk (Appendix A.1). In the top
panel, data points represent a preliminary analysis of 70 galaxies in
the DMS, yielding quantitative expectations for errors introduced by
inclination uncertainties (Andersen 2001). Black points and curves
(decreasing left to right) represent total-mass errors ($\Delta \ln
\mtot$, Appendix B.3), with dashed lines enclosing the range of
galaxies with good velocity fields and regular kinematics, and the
solid curve representing the mid-point of this distribution. The
dotted curve contains the most deviant points with the largest
kinematic inclination errors. Open points and dark-gray curves in the
top panel represent the disk mass surface-density errors ($\Delta \ln
\sddisk$, Appendix B.1) assuming the SVE is known to 10\% -- an
optimistic scenario -- and is flattened with $\alpha =
0.7$. Line-types have the same meaning as for total mass; the
dark-shaded area shows the range of $\Delta \ln \sddisk$ for
$0.4<\alpha<1.0$ and the mid-point of the inclination error
distribution.  The light-gray shaded area represents systematic errors
due to deprojection of $\slos$, discussed in \S 3.5.4.  The
bottom panel repeats this calculation for $\Delta \ln \sddisk$ assuming the
SVE uncertainty is 50\% (\S 2.1.1).  Total mass errors are the same in
the top and bottom panels.

The right panels of Figure 5 repeat the logarthmic errors in
total-mass and disk surface-density, but adopt inclination and
inclination errors from inverting the Tully-Fisher relation (Appendix
A.2). Logarithmic errors in total mass (black horizontal lines) are
given for two assumptions of the observed TF scatter (0.1 and 0.3 mag,
labeled) and a TF-slope of $-7$, suitable in red bands.  Black,
dotted, horizontal lines give the fractional error for 0.3 mag scatter
and TF slopes of $-5$ and $-9$. This range of slopes and scatter cover
the band-dependent results in the literature. Best results are for the
$K$-band TF found by Verheijen (2001) with a -9 slope and a scatter of
order 0.2 mag. Logarithmic errors in disk mass surface-density are
indicated by gray, shaded regions for 10\% (top) and 50\% (bottom)
uncertainty in the SVE for a range of ellipsoid ratios
$0.4<\alpha<1.0$, a TF slope of -7, and 0.3 mag scatter. The solid,
bisecting line assumes $\alpha=0.7$. Dashed curves show the full range
of TF scatter and slopes for $\alpha=1$, while the dotted curve does
the same for $\alpha=0.4$. For both $\Delta \ln \mtot$ and $\Delta \ln
\sddisk$, the precision is more sensitive to the scatter than the slope
of the TF relation.

Disk-mass errors are rather flat for $i>15^\circ$ for 10\% ellipsoid
errors. This would argue for going to inclinations larger than
40$^\circ$ to reduce total-mass errors.  However, working against this
arguement is the fact that systematic errors in $\sddisk$
continue to increase rapidly with inclination. Hence we conclude that
if we can determine the SVE to 10\%, we should select
galaxies with $30^\circ<i<40^\circ$ to equalize total- and disk-mass
errors each at about 13\%. A more reasonable estimate is that our
SVE errors will be closer to 20-30\% (Westfall 2009). This would
indicate lower inclinations of 25-35$^\circ$ to match total- and
disk-mass errors at about 1.5$\times$ higher levels (20\%).

Note, however, that errors in Figure 5 are lower limits. Although
additional photometric and kinematic errors are unlikely to dominate
the total-mass error budget for inclination-contributed errors above
$\sim$10\%, we know already that there is a 25\% error contribution to
disk-mass errors from estimating the vertical scale-height $z_0$ alone
(\S 2.2). Using iTF inclinations looks very powerful for reducing
errors in both disk-mass and total mass to levels well below other
contributed errors (10\%).

For iTF inclinations to be effective (i) very low inclination galaxies
are needed to reduce disk-mass errors, and (ii) a clean identification
of what kind of galaxies lie on a low-dispersion TF relation is
needed to reduce total-mass errors.  The latter implies the {\it
  shape} of the outer rotation-curve needs to be well-defined, which
means \ion{H}{1} velocity map is required, and that occurring warps
are well understood (see discussion in Verheijen 2001). With better
SVE estimates, a wider range of inclinations can be used below a given
error threshold in disk-mass. For 10\% errors in both mass quantities,
inclinations below 15$^\circ$ and 25$^\circ$ are needed for SVE errors
of 50\% and 10\%, respectively. Targets cannot be so face-on that their
projected velocity gradients are unobservable.  For $V_{\rm flat} = 250$
km s$^{-1}$, inclinations have to be above 11$^\circ$ to keep the
projected velocity a factor of 2 above the turbulent motions ($\sim$
25 km s$^{-1}$).  However, since we are not observing a line-width but
a velocity field, the averaging of many spatial position allows for a
velocity-centroid to be determined in the flat part of the rotation
curve well below the actual dispersion due to, e.g., turbulent
motions, as illustrated in Figure 16 of Paper I. Whatever the true
inclination and circular speed, clearly the galaxy in that figure has
a very regular velocity field and a flat asymptotic rotation curve
from which an iTF-based inclination can be {\it precisely}
derived. Despite this promise, in the absence of (ii), i.e., known
{\it accuracy}, using iTF remains a compelling yet potentially fatal
path if used in isolation.

To summarize, an optimal compromise for our survey goals can be
reached by targeting galaxies with kinematically determined
inclinations between 25$^\circ$ and 35$^\circ$.  We do, however, take
advantage of the iTF method to reduce errors in some situations
(Andersen et al. 2010, in preparation).

\section{THE BROADENING FUNCTION}

Since $\sigz$ is at the heart of our measure of $\sddisk$,
its estimation is arguably where we can have the greatest impact on
minimizing errors. In Paper I we outlined two complementary approaches
to deriving $\sigz$, with the intent of applying both in an
iterative fashion to (i) optimize the determination of the SVE (using
$\slos$ on individual fibers), and (ii) minimize template
mismatch (using stacks of fibers in radial bins).  We focus here on
the latter because this method ultimately allows us to probe
$\sigz$ at the largest radii. However, most of what is discussed is
generic to both, or a coupled approach to determining $\sigz$.

Following the order of analysis, we proceed with determining the
impact on the error budget given our method of fiber stacking (\S
3.1), deriving $\slos$ from cross-correlation methods (\S 3.2),
and estimating random errors on $\slos$ due to spectral errors
(\S 3.3) and template mismatch (\S 3.4).  Corrections to
$\slos$ that enable us to arrive at a reliable $\sigz$
estimate are described in (\S 3.5). These corrections compensate for
effects of beam smearing, instrumental resolution, projection, and
internal extinction. As a tertiary topic we consider the uncertainties
in the spatial registration (including effects due to seeing
variations) in \S 3.6. While positional registration of the IFU
pointings is actually relevant to the initial fiber stacking, the
uncertainties are negligible. Nevertheless, the discussion is included
for completeness.

\subsection{Fiber Averaging: Ring ``Stacking''}

Two-dimensional spectroscopic data can be binned spatially in a
variety of ways to maximize S/N. Even for an analysis of the ellipsoid
ratio, for example, we can divide the fibers into quadrants along
major and minor axes, and average spectra in bins of radius. For the
purpose of estimating $\sddisk$, with an estimate of the SVE
shape already in hand, we take advantage of the near face-on geometry
by combining fibers in azimuthal rings to measure a single
$\slos$ at a given radius. The detailed assignment of fibers to
a radial bin depends on a galaxy's inclination and position angle
(PA), which we determine kinematically. Nominally the binning is done
by radius, but we have also found it useful to bin instead by
surface-brightness as this can be useful in isolating arm and
inter-arm regions.

In any binning scheme, spectra are registered to take out the effects
of projected rotation before co-addition. This is achieved by an
iterative cross-correlation of each fiber spectrum against a
suitably-broadened template. (The cross-correlation technique is
described in detail in Paper III, and summarized here in the next
section.) We work progressively starting with bins at the highest S/N,
e.g., as a function of radius.  On the first pass, we adopt a template
between G9~III and K1~III, broadened to the dispersion of the previous
bin. After determining the best template and broadening function for
the stack (below), we adopt these as input for the registration, and
iterate to convergence. The advantage of fiber averaging is that we
are only required to determe a velocity centroid for each fiber in the
average, instead of both centroid and width.

The spectral stacking algorithm has these specific steps: (1)
Determine the best-fitting velocity offset for each fiber within a
radial ring for a fixed velocity dispersion and provided template
spectrum; (2) Shift all ring spectra to rest-frame velocity and
combine them using weights defined by $(S/N)^2$; (3) Using the
combined ring spectrum and the same template, determine the
best-fitting velocity dispersion after fixing the velocity offset to
be in the rest frame; and (4) Repeat steps 1-3, restarting with the
updated velocity dispersion. Convergence is reached when the
difference in subsequent determinations of the velocity offsets is
below a specified threshold; here we set a threshold of 1 km s$^{-1}$
except for rings with substantially low-S/N components.  Noordermeer
et al. (2008) have presented an alternate algorithm that alters the
individual velocity offsets in order to minimize the velocity {\it
  dispersion} of the stacked spectrum. This is in contrast to relying
on the best estimates of the velocity {\it centroid} for each fiber as
we have done here. While the algorithms should be roughly
equivalent at high S/N, we prefer our approach for low S/N
applications because velocity centroids are more precise.  We discuss
the impact of errors in the velocity registration on the derived
broadening function in \S 3.3.

One of the ways to aid in the convergence and make the stacking
process more robust at lower S/N is to provide a prior estimate of
the velocity offsets. We have compared priors including (a) the measured
velocities of gas determined from the same spectra containing the
stellar absorption (e.g., [\ion{O}{3}]$\lambda$5007 in the
\ion{Mg}{1b}-region); (b) model velocities of the gas, based on a
projected rotation-curve fit; (c) or the same, except for the stellar
data in the inner region, extrapolated to larger radii.  In all cases,
we find that the resulting stacked spectra are identical over the full
range of S/N and radial bins. This implies that even though the
stellar and gas velocities systematically differ (i.e., due to
asymmetric drift), the iterative, cross-correlation process is robust
to tune up the spectral registration. Indeed, without such iterations,
the stellar spectra broaden systematically at radii where
asymmetric-drift is large or the S/N is low. This is significant
because it means we may apply such priors to aid in the convergence of
the velocity-registration of, e.g., stellar spectra in the
\ion{Ca}{2}-triplet region stellar spectra where emission lines are
observed.

Figure 4 of Paper I shows examples of azimuthal rings for SparsePak
and PPak observations of UGC 6918. The table inset gives the number of
fibers in each radial bin as a function of radius, scaled to the
radial scale-length. Figures 6 and 7 of this paper contrast the
``before- and after-registration'' spectra and cross-correlations for
the \ion{Mg}{1b} region in one radial bin (ring 4) of UGC 6918. This
ring is between 2.6$<$ R/h$_R < 3.4$, at a mean $V$-band
surface-brightness of 21.6 mag arcsec$^{-2}$. The data represent a
cumulation of 2.25 hours of exposure, achieving a mean S/N of 21
pixel$^{-1}$ in the coadded spectrum at roughly 0.5 mag below the
Freeman disk central surface-brightness (Freeman 1970; UGC 6918 has a
high surface-brightness disk).

As expected, the averaged spectrum and its associated
cross-correlation have substantially lower noise than their individual
elements. Particularly impressive is that after velocity registration
not only is the correlation tightened (lines narrowed) but the noise
is further reduced. This ``noise'' is due to the unregistered
superposition of absorption lines. Indeed, the registration reveals
many, weaker lines in the spectrum, and produces a cross-correlation
that better matches that of the broadened template even outside of the
correlation-peak where the broadening-function is determined. This
comparison demonstrates the power of cross-correlation to pull out
signal from the multi-fiber data distributed over a range of projected
velocities. For comparison, if we were to limit ourselves to only
those fibers that could be fit individually, the S/N would drop by
$\sim$20\% because 40-50\% of the fibers could not be fit at the
limiting surface brightness of our data.

A final advantage of fiber averaging is the ability to mask out
sky-lines.  While these foregrounds are nominally subtracted from the
data, there are often residuals due to imperfect match of the
spectrograph aberrations between object and sky spectra, detector
under-sampling, or simply the enhanced root-mean-square (RMS) due to
the large number of counts in the sky line.  All of these `features'
lower the S/N in the specific wavelength region of the line; however
the feature shifts in the galaxy rest-frame in a spatially-dependent
manner because of the galaxy internal motions. Therefore by masking
out the sky-line regions from the stack, it is possible to recover a
continuous, line-free spectrum, optimized for S/N. This is
particularly important in the \ion{Ca}{2}-triplet region where sky
lines are strong and prevalent. Figure 8 illustrates the masking
method. 

In this example, we have chosen only to mask out the strongest lines,
typically those with peak flux more than twice the level of the sky
continuum. Mask widths are 0.3 nm (roughly three times the
instrumental full-width at half-maximum, FWHM), except for
closely-spaced lines where the mask-width is reduced to leave
inter-line gaps. Mask-widths are never less than 3 pixels.  These mask
widths were found empirically to maximize the resultant S/N and, based
on simulations, to minimize systematics in the derived
broadening. (The sky-lines are unresolved, and the internal velocity
shifts of the galaxy are always larger than the instrumental
resolution.) The only region which does not appear to benefit from
masking is in the molecular-band region from 860 to 870 nm.
Masking requires $\geq 105$ km s$^{-1}$ velocity spread in the
\ion{Ca}{2}-triplet region between fibers in a ring, achieved except
in the inner-most ring for 80\% of our \ion{Mg}{1b} sample and all but
one galaxy observed in the \ion{Ca}{2}-triplet region. The velocity
spread is set simply by the largest mask-wdith.

\subsection{Deriving the Broadening Function $\slos$}

We apply a cross-correlation technique, rather than direct-fitting of
the spectral data in wavelength space, to determine the broadening
function and random errors on this broadening. While the two methods
are equivalent in principle, in practice the information is projected
in different ways (Simkin 1974). For example, in the direct-fitting
approach, an assessment of template (mis)match is grossly evident in
the detailed {\it depth} of various lines. In contrast, assessment of
the broadening (mis)match is more readily obtained via inspection of
the cross-correlation function since all of the signal for line
profile {\it shape} is consolidated. Because the broadening is the
primary signal of interest, we prefer the cross-correlation method,
particularly because information on template mismatch (relevant for
systematic errors) is still available in the cross-correlation using
information outside of the peak (\S 3.4). 

We have developed a new cross-correlation method, optimizing several
technical attributes relevant to the accuracy and precision of the
broadening measurement in our program. The analysis code is very
general in the sense that it allows for Gauss-Hermite series
decomposition of the cross-correlation peak (van der Marel \& Frank
1993) and input of any spectral template. For clarity, we focus in
this paper on results for simple Gaussian broadening and templates
based on measurements of single, Galactic stars.  A complete
discussion of the method is presented in Westfall (2009) and Paper
III. We summarize here the salient features germane to the
error-budget.

First, we construct a differential formulation based on convolution
(rather than deconvolution; see Franx \& Illingworth 1988) to derive
the broadening in a way that treats the templates and galaxy data in
an identical, and symmetric fashion. To describe this we adopt the
nomenclature in Paper III where $\circ$ denotes cross-correlation,
$\otimes$ denotes convolution, $G$ is the galaxy spectrum, $T$ is the
template spectrum, and $B$ is the broadening function. We compare the
cross-correlation of a broadened and redshifted template with an
un-broadend, un-shifted template (BTXC $\equiv X_T \equiv (T \otimes
B) \circ T$) to the cross-correlation of the galaxy spectrum with the
un-broadened, un-shifted template (XC $\equiv X \equiv G \circ T$).
We compare, in a $\chi^2$ sense, only the core of the correlation
peaks, finding a region of 1.7 times the FWHM of the XC peak to be
optimal in terms of precision. Our approach contrasts with earlier
applications (see Rix \& White 1992 and Statler 1995 for reviews)
which compare the XC with the un-broadend, un-shifted template
auto-correlation (AC $\equiv A \equiv T \circ T$), or even deconvolve
the XC peak directly.  While mathematically equivalent, our approach
starts with broadening the template, as is done in nature to the ideal
template of the galaxy spectrum.  Moreover, our approach should be
immune to systematic effects due to ``detector-censored'' data, i.e.,
where the observed band-pass is finite. What this means in practice is
that we do not have information from our templates about flux
contributions due to broadening from outside of observed spectral
window. We find this is particularly important when strong features of
interest (e.g., \ion{Mg}{1b} or \ion{Ca}{2}) are near the edge of the
detected band-pass in either the template or galaxy spectrum. Tests
indicate the impact of detector-censoring can lead to systematics in
$\slos$ of order 10\% (Westfall 2009 and Paper III). By using
convolution rather than deconvolution, we avoid filtering problems
associated with Fourier transforms of noisy data.

Second, the fitting procedure allows for the masking of source
emission-lines and sky-lines, handled symmetrically for template and
galaxy spectra. Emission-line masking is critical in the \ion{Mg}{1b}
region not only for the [\ion{O}{3}]$\lambda\lambda$4959,5007 doublet
but also for the weaker [\ion{N}{1}]$\lambda\lambda$5198,5200 doublet
(Figure 6).  Sky-line masking is critical in the \ion{Ca}{2}-triplet
region (Figure 8). Masking also enable us to isolate spectral regions
of interest, e.g., the \ion{Mg}{1b}-triplet versus weaker Fe features in
this region, or \ion{Ca}{2}-triplet versus Paschen-series lines in the
near-infrared, as we discuss below. The inability to mask spectral
features with previous versions of cross-correlation software has
often been touted as a primary advantage of directing-fitting methods
(e.g., Rix \& White 1992).  However, in our cross-correlation
formulation masking is neither conceptually difficult or
computationally challenging.

A relevant detail for Fourier-transform cross-correlation techniques
and masking concerns tapering (apodization) of the mask edges to avoid
high-frequency ripples. When the number of masks is large relative to
the number of spectral channels, the tapering function appreciably
diminishes the cross-correlation signal. Because of the identical and
differential way we treat the template and galaxy correlation
functions, we expect that a ripple should not have an impact on the
derived broadening parameters. Simulations bear out the expectation
that changing the tapering function does not alter the accuracy of the
recovered parameters, but precision is improved by eliminating
tapering, or at most by applying a 2-pixel cosine taper. (Each pixel
is between 7 and 12 km s$^{-1}$.)

Third, the fitting is iterative in two significant ways: (1) in the
optimization of the broadening, velocity-shifting of the template and
the mask placement in the two (template and galaxy) reference frames;
and (2) in fitting a low-order spectral continuum to the residuals
between the broadened, shifted template and the galaxy spectrum in
wavelength space, and then removing this residual from the observed
galaxy spectrum (mathematically equivalent to adding the residual to
the broadened template, but computationally simpler). The latter
accounts for any low-frequency spectral mismatch between the template
and galaxy due to, e.g., in order of likely significance: stellar mix,
illumination correction, color-terms in flux calibration, reddening,
or nebular continuum emission. This continuum correction is important
because while the low order spectral shape does not contribute
directly to the width or shape of the cross-correlation peak, it does
impact the cross-correlation function at lower frequencies, which in
turn alters the amplitude of the correlation peak. This mismatch has
impact on the differential comparison of the BTXC and XC even for the
high spectral-frequency component (e.g., consider the effect of
baseline variations near the correlation-peak core), which
fundamentally is the only quantity of interest for $\slos$. The
mismatch also alters the goodness-of-fit assessment of the template
(\S 3.4).

Finally, the fitting process includes evaluation of the error spectra,
and the computation of a full covariance matrix for determining errors
on the fitted parameters. These error estimates have been tested
against simulations, and found to be accurate and robust. Comparison
of the simulations with real data is described below.

\subsection{Random and Systematic Errors on $\vsys$ and $\slos$ due to Spectral Noise}

We focus here and in the next subsection on simulations matched to the
SparsePak observations of two galaxies from our pilot program, taken
in both \ion{Mg}{1b} and \ion{Ca}{2}-triplet regions: UGC 11356, a
well-studied giant elliptical galaxy; and UGC 6918, a high
surface-brightness spiral galaxy in our sample. The former was
observed for the purpose of comparing our measurements to those in the
literature. It should also contain a relatively simple stellar
population yielding similar kinematics in the two spectral regions --
albeit with $\slos$ substantially larger than for our survey
sources. UGC 6918 has $\slos$ values typical of our survey
sources, but potentially illustrates a composite stellar system where
systematic differences arise in the derived $\slos$ between the
two wavelength regions. These difference might occur due to variations
in the dynamics of disk stellar populations correlating with age, and
hence color.

We have carried out a set of Monte Carlo simulations to determine the
accuracy and reliability of extracting the centroid velocity
($\vsys$) and line-wdith ($\slos$) using our
cross-correlation technique. These simulations use our stellar
templates, observed at high S/N. These spectra are velocity shifted,
broadened and noise-aberrated to span a range of S/N, $\slos$,
and $\vsys$ encompassed by our survey data.  We extended the range
of simulated $\slos$ to higher values typical of the cores of
giant elliptical galaxies.  Independent, and more detailed simulations
in the \ion{Mg}{1b} region applicable to UGC 6918 are given in Paper
III.

As part of our analysis, we divide the \ion{Mg}{1b}-region into two
spectral subregions, and considered these in addition to the full
\ion{Mg}{1b} and \ion{Ca}{2}-triplet spectral regions. These two
subregions contain, respectively, the \ion{Mg}{1b} triplet, and
everything but the \ion{Mg}{1b} triplet, as indicated in the K1~III
spectrum in Figure 9.  As this figure shows, the latter subregion is
dominated by signal from many weak lines
of \ion{Fe}{1}, \ion{Ti}{1}, \ion{Cr}{1}, \ion{Fe}{2}, \ion{Ti}{2},
and TiO, in decreasing importance, prevalent in the cool stars
expected to dominate the detailed line-signature in the integrated
light of galaxies. This division was motivated by the results of
Barth, Ho, \& Sargent (2002) indicating the \ion{Mg}{1b}-triplet was
problematic for $\slos$ measurements -- plausibly due to abundance
variations between stellar templates and integrated galaxy
spectra. Inspection of Figure 9 shows that our \ion{Mg}{1b} subregion
still contains narrow, weaker lines of \ion{Fe}{1} and \ion{Fe}{2}, as
well as the TiO band-head. In particular, the bluer two lines of
the \ion{Mg}{1b} triplet are significantly contaminated, as seen in
the real galaxy spectra of Figure 6. The bluest line of the triplet
coincides nearly with the TiO molecular band-head which is strong in
stars cooler than M0 (see Figure 15 in Paper I). In terms of random
errors on $\slos$, we find from our simulations that the two
subregions have comparable S/N in a cross-correlation sense, or about
$1/\sqrt{2}$ that of the full \ion{Mg}{1b} region. Otherwise both
subregions yield similar systematic trends with S/N. For this reason,
we do not distinguish these subregions further in discussion of S/N.

While we have studied simulations using a large range of stellar
templates, we illustrate results using K1~III and M3~III templates
here for clarity. These two stars, respectively, appear to be the
best, or close to the best single-star templates in the \ion{Mg}{1b}
and \ion{Ca}{2}-triplet regions. This holds for all of the galaxies in
our sample analyzed to date, as well as the elliptical UGC 11356, as
we demonstrate below. 
%
%
The results of these simulations are shown in Figure 10a (for
the K1~III template) and Figure 10b (for the M3~III template). Measurements
from galaxy observations are also shown for comparison.  These include
all individual fibers in the \ion{Mg}{1b} region in UGC 6918 for which
cross-correlation yielded successful measurements; the inner 14 fibers
for UGC 11356 in both \ion{Mg}{1b} and \ion{Ca}{2}-triplet regions;
the same set of fibers in UGC 6918 for the \ion{Ca}{2}-triplet region;
and the 5 rings defined in Figure 4 of Paper I for UGC 6918 in both
spectral regions.

The accuracy of the derived velocity centroid and broadening based on
the simulations is superb, and well below the random errors for S/N
$>$ 3 pix$^{-1}$ (refer to panels in first and third columns of
Figures 10a and 10b).  This result is independent of broadening and
centroid velocity. Below this S/N level there is a hint that
systematics begin to become significant, with positive velocity and
velocity-dispersion offsets for smaller line-widths, and negative
offsets at larger line-widths.  This is clearly demonstrated by the
more detailed simulations presented in Paper III. We conclude that at
S/N $>$ 3 pix$^{-1}$ systematic errors in the derived velocities and
widths are negligible. We defer discussion of the accuracy of
$\slos$ derived from the galaxy observations until after
consideration of the effects of template mismatch.

In terms of precision, to first order we find that $\vsys$ and
$\slos$ errors scale inversely with S/N and $\slos$
as expected (wider profiles yield less precise measures at a given S/N;
refer to panels in second and fourth columns of Figures 10a and 10b).
At very large $\slos$ there is some indication that
the dependence on S/N is somewhat stronger, at least for
$\vsys$. There is very little dependence of these results on the
simulation template. However, the simulations were fit with the
correct template, so the effects of template-mismatch are absent in
these results.

In contrast, the templates used to derive centroid velocities and
broadenings for the galaxy observations may be mismatched. Indeed,
comparison of measurements of simulations to those of real galaxy
spectra shows the latter have errors twice as large in the
\ion{Mg}{1b} region, yet similar errors in the \ion{Ca}{2}-triplet
region. Interestingly, at a given S/N the errors derived from the
simulations are 2 to 2.5 times larger in the \ion{Ca}{2}-triplet
region than in the \ion{Mg}{1b} region. Assuming no template mismatch,
we conclude that the \ion{Mg}{1b} region in principle yields more
precise kinematic measurements than the \ion{Ca}{2}-triplet region at
a given spectral continuum S/N. However, in practice the two regions
yield comparable precision.  It is plausible that the additional
contribution of {\it random} error to the kinematic measurements in
the \ion{Mg}{1b} region is due to template mismatch. This conjecture
has some basis in the fact that errors in $\vsys$ and
$\slos$ in the galaxy spectral data in the \ion{Mg}{1b} region
increase substantially when they are fit with a template (M3~III) that
is clearly not a good representation of the spectrum in that region.
In contrast, the errors derived in the \ion{Ca}{2}-triplet region
appear relatively immune to the template applied.

On the basis of these simulations we conclude that the typical random
error in our survey for $\Delta \ln \slos$ is 3\%, given a typical
spectral continuum S/N in the \ion{Mg}{1b} region of 40 for an
azimuthally-averaged spectrum (see Paper I and Figure 4 therein).
Since the measurement of the broadening function is not yet corrected
for other effects (\S 3.5), for the purpose of book-keeping we refer
to this quantity as $\sobs$, and hence $\Delta \ln \sobs =
0.03$. 

One final consideration concerns the impact of velocity centroid erros
on the registration precision of the azimuthal averaging. In general
centroid errors will systematically broaden the stacked spectrum. Most
of our data has spectral continuum S/N $ > 2$ in the individual fiber
spectra. Since the errors on $\vsys$ remain well below
$\slos$ (i.e., $\epsilon(\vsys)/\slos<0.3$, Figures 10a
and 10b) in this S/N regime, azimuthal averaging introduces less than a
few percent increase to the derived broadening. This is negligible for
our purposes in this paper. However we do note that in our outer-most
radial rings, where the S/N for individual fibers is 1 to 2, our
simulations indicate there could be as much as a 20\% increase in the
measured $\slos$ due to registration errors.

\subsection{Errors on $\slos$ due to Template Mismatch}

Perusal of the literature reveals that single stars typically
have been used as templates for cross-correlation analysis to study the
dynamics of disk and spheroidal stellar systems. One critical question
for our analysis is whether $\slos$ is sensitive to the
specific choice of template. Substantial discussion of the issue of
template-mismatch can be found in Rix \& White (1992), Statler (1995),
and references therein. Late-G or early-K giants are
usually adopted, with the (often unstated) assumption that these stars
dominate the kinematic signal in the integrated light of early and
intermediate-type galaxies. This is certainly reasonable given the
luminosity of red-giant and horizontal branch stars, and their
apparent dominance of the integrated light of old stellar populations.
For later-type disks (especially near their outskirts, or in the cores
of vigorously star-forming systems), the relative youth of their
stellar populations may alter the picture, both due to the prevalence
of luminous, hot, young stars on, or near the tip of the Main Sequence
and cool, intermediate-age giants (e.g., the AGB).  The concern
regarding hot stars may be tempered by virtue of their decreasing
line-strengths from metals. Nonetheless, the question remains whether
there are substantial systematics in $\slos$ from
template-mismatch. We define the template-mismatch error to be $\Delta \ln
\stm \equiv \Delta\stm/\slos$, where
$\Delta\stm$ is the half-width of the full range of
$\slos$ for all viable templates.

To motivate the importance of answering this question we illustrate in
Figure 11 the measured $\slos$ for fibers in the core of UGC
11356 and UGC 6918 using a range of template stars from F0 to M5, all
luminosity-class II-III (giants). The specific stars and their spectra
are illustrated in Figure 15 of Paper I.  Values are means over the
individual fiber measurements, with errors given as the standard
deviation of these measurements. The errors are within a factor of two
from the mean estimated errors from the cross-correlation analysis,
indicating little intrinsic variance between the regions sampled by
the individual fibers. The range of template spectral types was chosen
on the basis of direct visual inspection of the template and galaxy
spectra.  In the \ion{Mg}{1b} region, types earlier than F0 have
insufficient line-strengths in both \ion{Mg}{1b} and
\ion{Ca}{2}-triplet regions. While types later than M0 have a strong
molecular band-head in the \ion{Mg}{1b} region which is not observed,
types as late as M5 are acceptable in the \ion{Ca}{2}-triplet region.
We extended our template range accordingly. We have limited templates
here to luminosity class II-III stars, based on astrophysical
prejudice for what stellar types with strong lines are most likely to
dominate the integrated light of galaxies.  Furthermore, we choose
mostly solar metallicity stars given the reasonable assumption that
the integrated light of disks is dominated by Population I
stars. Nevertheless, we include one sub-solar metallicity star (HR
4695), at intermediate spectral type, to probe the validity of the
latter assumption.

The left-hand column of Figure 11 demonstrates that template mismatch
is very significant in spirals {\it and} ellipticals, and in both
spectral regions. While trends of $\slos$ with spectral type
are different for the two spectral regions, they are qualitatively
similar for both galaxies. Compared to the full \ion{Mg}{1b} region,
we find the systematic trends in $\slos$ with template are
twice as large in the subregion isolating the \ion{Mg}{1b}-triplet,
while the sub-region excluding the \ion{Mg}{1b} triplet has a smaller
range. The two \ion{Mg}{1b} sub-regions also have different
qualitative trends.  These differences are largely due to the
appearance of the TiO band-head in cool stars later than mid-K,
located near the bluest of the \ion{Mg}{1b} triplet lines. Given the
increased random errors (\S 3.3) in $\slos$ by limiting the
spectral range to either of these subregions, we do not considered
them further here. However, we note the added information by definning
such subregions can be exploited to further optimize template-matching
in the highest S/N regimes.

One method for limiting the impact of template mismatch on
$\slos$ is to restrict the template spectral range with a
notional argument, e.g., based on colors or stellar population
synthesis (SPS) models. For example, were we to restrict the templates
to F8-M2 or G8-K4 ranges, we would obtain $\Delta \ln \stm =
0.12$ and 0.07 respectively.  Indeed, more recent studies (e.g.,
Falc\'{o}n-Barroso et al. 2006) use SPS models to directly fit the
spectral continuum (effectively color) and line-strength.  In
principle this option is open to us, but until recently the stellar
libraries have had insufficient spectral resolution for our purposes.
The one exception is PEGASE-HR (Le Borgne et al. 2004), based on the
ELODIE library observed at a resolution of $R =
42,000$. Unfortunately, the models degrade the resolution to $R =
10,000$ for a Gaussian instrumental profile; this is too low for many
of our \ion{Mg}{1b} observations, which often have non-Gaussian
instrumental profiles. Further, the library does not extend far enough
to the red to reach the \ion{Ca}{2}-triplet. Nonetheless, improved
models like these are highly desirable in the future.

Even with suitable high-resolution SPS models, properly modified for
our instrumental broadening, there remains the issue of degeneracy --
in a photometric sense -- between equally suitable models with a wide
range of model parameters (e.g., age, metallicity, star-formation
history). The problem here is that it has not yet been demonstrated
that this photometric degeneracy has an equivalent kinematic
degeneracy. Specifically, the amplitude of $\Delta \ln \stm$ has
{\it not} been quantified using the direct-fitting SPS approach in any
study presented in the literature.

For the above reasons we proceed here with a simple analysis based on
a set of single template stars. These are observed with the same
instrument and same instrumental configuration (often observed on the
same night) as used for our target galaxies. We define a set of
indices that allow us to minimize the impact of template-mismatch on
$\slos$, quantify $\Delta \ln \stm$ in this context, and
conclude with a brief discussion of how this approach can be further
improved.

\subsubsection{Template mismatch indices}

The function $\chi^2_\nu(XC)$ which is minimized to determine the
optimum broadening, $\slos$, is the error-normalized RMS
between the XC and BTXC, taken in the usual $\chi^2$ sense, but
measured {\it only} within a small fitting window of the
cross-correlation (1.7 times the FWHM of the cross-correlation
peak). We find that $\chi^2_\nu(XC)$ is highly insensitive to changes
in the template, even though the derived $\slos$ varies
substantially. This appears a worrisome fact for the cross-correlation
approach, but since the Fourier transform does not throw out intrinsic
information, sensitivity to template mismatch must be present
somewhere in the cross-correlation function outside the fitting
window. Inspecting simulations, we concluded that (i) the relative
heights of the XC and BTXC give information on the match of the {\it
  average} line-depth (equivalent width) in the template versus galaxy
spectra; while (ii) the ``RMS'' amplitude and asymmetry of the
cross-correlation outside the fitting window give information on the
match of the {\it relative} line-depths between the two. Based on
this, we developed two indicators based on the cross-correlation
function, and a third based on the direct spectrum to compare
direct-fitting vs cross-correlation approaches. All three of these
indices are illustrated in Figure 11 for the optimum broadening for
each template:

\begin{enumerate}

\item XC-rms is the RMS between the XC and BTXC.  It is like
  $\chi^2_\nu(XC)$, except it's not error-normalized, i.e., it is not
  $\chi^2$, and is computed over the full correlation range. This
  range is nominally the same for all templates unless the fitted
  velocities are substantially different. However, XC-rms is
  normalized by the amplitude of the cross-correlation peak to take
  into account the trend of stronger correlation peaks (and hence
  asymmetry with later spectral types).

\item A$_{\rm N,c}$ is the RMS asymmetry (A) of the XC when mirrored about
  its fitted velocity centroid. It is an index of the
  ``lopsidedness,'' or lack of mirror-symmetry of the
  cross-correlation, both at small and large lags.  Like XC-rms, it is
  also normalized (N) by the amplitude of the cross-correlation peak.
  It is further corrected (c) for the asymmetry of the
  similarly-normalized cross-correlation of the broadened template
  with the un-broadened template (BTXC). This accounts for the
  non-intuitive (but mathematically correct) result that in the
  presence of detector censoring (i.e., any finite spectral window)
  the cross correlation of a template with its broadened counter-part
  has some intrinsic non-zero amount of asymmetry. The correction is
  small ($<10$\%).  With the exception of the correction, A$_{\rm N,c}$ is
  equivalent to the inverse of the term {\it ``R''} defined by Nelson
  \& Whittle (1995).

\item $\chi^2_\nu (\lambda)$ is the error-weighted RMS between the
  observed galaxy spectrum and the best-fitting broadened template,
  based on the cross-correlation and continuum fitting techniques, as
  described above. This index is independent of the fitting process
  that determines the velocity broadening, but is otherwise equivalent
  to what is used in direct-fitting methods.

\end{enumerate}

Both XC-rms and A$_{\rm N,c}$ have larger dynamic range given their
relative scatter than $\chi^2_\nu (\lambda)$. The
insensitivity of $\chi^2_\nu (\lambda)$ to variations in template
raises the possibility that template-mismatch errors in direct-fitting
methods may be substantial. The indices XC-rms and A$_{\rm N,c}$ exhibit
similar template resolution for both galaxies and for a given spectral
region. Consequently, application of these indices (described below)
to a large number of galaxy spectra yield quantitatively comparable
results for $\Delta \ln \stm$. Because of the greater simplicity
and intuitive nature of the XC-rms definition we adopt it in
preference to A$_{\rm N,c}$.

\subsubsection{Index application and performance}

In practice, the above indices can be used to minimize $\Delta \ln
\stm$ by identifying the template with the minimum index value,
defining a confidence interval based on the errors in that index, and
then averaging $\slos$ for all templates with index values
within this confidence interval.  Variance in $\slos$ can also
be determined for this same template subset, and $\Delta \ln \stm$
quantified in a well-defined manner. For the data in Figure 11, error
bars on these indices are based on the measured variance between
groups of fibers, and are likely over-estimates, for reasons given
below. Nonetheless they are suitable for demonstrating the outlined
technique to minimize the impact of template-mismatch. We adopt a
``1$\sigma$'' confidence interval in the sense that a template index
must be no greater than the quadrature sum of the minimum index value
and the 1$\sigma$-errors on both the minimum and template index
values.

As anticipated in the preceding discussion, XC-rms and A$_{\rm N,c}$ are
substantially superior to $\chi^2_\nu (\lambda)$ in terms of $\Delta \ln
\stm$. A feature which appears to be problematic for
$\chi^2_\nu (\lambda)$ is the selection of templates at disparate
temperatures in each spectral region. This not only increases $\Delta \ln
\stm$, but without some deeper understanding of what causes
these selection discontinuities in spectral type using $\chi^2_\nu
(\lambda)$, it is hard to understand how to move forward to improve
the situation; one is tempted to abandon $\chi^2_\nu (\lambda)$ and
its associated direct-fitting approach as we have done here. One clue
for future efforts may be that the low-metallicity star HR 4695 shows
unusually low $\chi^2_\nu (\lambda)$. It is highly unlikely that
massive and luminous galaxies have integrated light with sub-solar
metallicity. Indeed, many galaxies in our sample have high values of
XC-rms and A$_{\rm N,c}$ for this same star, i.e., it provides a
worse-fitting template, consistent with astrophysical
expectations. Template-mismatch sensitivity to metallicity is
definitely a desirable feature of any analysis attempting to
understand the dynamics and stellar populations of galaxies.

Because we have a large, comprehensive template library, we can safely
assume there exists a linear combination of templates that, once
broadened, accurately represent any observed galaxy absorption-line
spectrum. Here, we will make the further, simplifying, assumption that
at least one of the templates alone is a suitable match to the
observed galaxy spectrum. By this we mean that there exists one
template which produces minimal systematic error in the derived
broadening due to template-mismatch. In future papers in this series
we substantiate that these are both good assumptions. In using an
index-minimization approach, where the indices themselves are subject
to random errors, the impact of template-mismatch on $\slos$
uncertainties -- what we have defined as $\Delta \ln \stm$ -- is a
random process. In other words, we would get a different value for
$\slos$ for the same set of templates for repeat measurement of
a galaxy spectrum. Hence for our application $\Delta \ln \stm$ is a
{\em random} error.

Using the individual fiber measurements for UGC 11356 and UGC 6918, we
illustrate in Figure 12 the relative amplitudes of random errors due to
shot noise in the spectra ($\Delta \ln \slos$) and random errors
due to template mismatch ($\Delta \ln \stm$).  As we found earlier,
$\Delta \ln \slos$ is inversely proportional to S/N. By applying
the XC-rms index, we dramatically lower the errors associated with
template mismatch, essentially eliminating it for \ion{Ca}{2}-triplet
region measurements, and making these errors comparable to $\Delta \ln
\slos$ in the \ion{Mg}{1b} region.  The correlation between the
two sets of errors is related to the fact that errors on
$\slos$ and XC-rms for a single spectrum and template correlate
in the same way with S/N.

As verification of our template-mismatch minimization method, we show
in the right panel of Figure 12 the difference between individual fiber
measurements made in the \ion{Mg}{1b} and \ion{Ca}{2}-triplet regions
at similar physical locations for our two example galaxies. The
velocity dispersion measurements are statistically identical as we
would expect for the elliptical and also the spiral -- if
age-dynamical variations in disks are small. Figure 11 shows that this
result need not necessarily be the case without some proper
identification of suitable template.  By extension, we may postulate
this result as an initial confirmation that age-dynamical variations
for spirals are in fact small, as observed in their integrated
optical--near-infrared light.  It is also worth noting that our
measurement of the absolute value of the velocity dispersion in UGC
11356 agrees within the errors with previous results in both the
\ion{Mg}{1b} region (Bender et al. 1994, Fisher 19997, Gerhard et
al. 1998) and \ion{Ca}{2}-triplet region (Nelson \& Whittle
1995). This is significant since our best-fitting template in the
\ion{Mg}{1b} region is similar to what was used in previous
studies. In the \ion{Ca}{2}-triplet region our best-fitting template
is considerably latter than those observed by Nelson \& Whittle
(1995), but we find little change in the derived broadening when going
to these earlier stellar types.

To apply our method in general we need to calibrate XC-rms errors as a
function of S/N. Based on simulations, we illustrate in Figure 13 how
the logarithmic errors on XC-rms depend on S/N. Errors for the other
indices were computed in the same way. These errors are only due to
random errors added to the simulated spectra, i.e., there is no
template mismatch between the simulated BTXC and XC. We find
logarithmic errors are only weakly dependent on line-width and S/N,
which is convenient for the application we have described.

To determine if these simulations are realistic we took groups of
fibers at similar radii and S/N (hence surface-brightness) in UGC 11356
and UGC 6918 separately, as illustrated in Figure 11. We computed the
variance in the indices, and plotted them accordingly in Figure 13.
We interpret these as upper limits to the index errors since real
variations between the spectra may exist, e.g., stellar population
non-uniformity. These ``measurements'' of $\Delta \ln$ XC-rms are roughly
twice as large as the values estimated from simulations.  We adopt $\Delta
\ln$ XC-rms $ = 0.06\pm0.02$ and $0.12\pm0.04$ based on simulations
and measurements, respectively, over a S/N range typical of observed,
azimuthally-averaged spectra.

Results of applying this calibration of XC-rms errors to
azimuthally-averaged spectra of 7 representative galaxies\footnote{UGC
  463, 4555, 5180, 6128, 6869, 6918, and 10443.}  from our survey are
shown in Figure 14.  The ring spectra have S/N from 7 to 200 per
pixel.  The median random error on $\slos$ due to shot noise is
3\%, with 75\% of the sample having errors $<$5\%; random errors
follow the trend and zero-point with S/N as seen for the individual
fibers.  The median value is a restatement of our result in the
previous section for $\Delta \ln \slos$.  The median random errors
due to template mismatch are 4\%, with 75\% of the sample having
errors $<$5\% assuming $\Delta \ln$ XC-rms is what is observed, i.e.,
about 0.12, or a factor of two larger than the simulation value. We
adopt the median value, i.e., $\Delta \ln \stm = 0.04$.

Were we to adopt a single template for all galaxies and all radial
bins, e.g., a K1~III star for the \ion{Mg}{1b} region, our derived
$\slos$ would be low, on average, by about 2\%, with a standard
deviation of 8\%. There is a hint of a trend with radius such that a
K1~III star is either too early or too late at the inner-most and
outermost radii, respectively. Adopting a K1~III star is not a bad
choice, but with the indices described here, we are able to largely
eliminate systematic error due to template mismatch.

\subsubsection{Future improvements}

Our analysis can be improved in the following ways: (a) by including a
wider range of metallicity for stars already in our existing spectral
library; (b) by allowing for multiple templates in each radial bin,
either single-star templates or multi-component SPS model templates
based on our library; (c) by enabling each of these spectral
components to have their own separate kinematic broadening (e.g., de
Bruyne et al. 2004); and lastly (d) by treating the
\ion{Ca}{2}-triplet and \ion{Mg}{1b} regions simultaneously, in a
self-consistent way such that the stellar population synthesis is
consistent with the observed, de-reddened colors and spectra, while
yielding an identical velocity dispersion for a given template
component in both spectral regions (see \S 4.3.2 and Figure 7 of
Paper-1). These are refinements for future work.

\subsection{Corrections to $\slos$}

\subsubsection{Instrumental resolution}

The instrumental dispersion, $\sinst$, is due to a combination of the
finite fiber aperture and spectrograph optical
aberrations.\footnote{The instrumental resolution defined as the FWHM,
  or 2.35 $\sinst$.} We measure $\sinst$ via Gaussian line-fitting to
the line-lamp spectra. For any given observational configuration we
typically identify 50 or more lines over the full spectral range which
have adequate S/N and are un-blended. The widths of these lines are
determined to high precision (a few \%), and provide an exquisite map
of the instrumental broadening as a function of spatial and spectral
position (the former is equivalent to fiber \#).  We find a mean
broadening due to instrumental resolution of $\sinst = 10.8 \pm 1.5
\pm 2.2$ and $13.2 \pm 1.9 \pm 2.4$ km s$^{-1}$ in the \ion{Mg}{1b}
and \ion{Ca}{2}-triplet regions respectively for SparsePak, and
$\sinst = 18.8 \pm 1.7 \pm 1.4$ km s$^{-1}$ in the \ion{Mg}{1b} region
for PPak, where the two sets of ``errors'' for each value are the
characteristic range of broadening in the spatial and spectral
dimensions. Subtle changes in spectrograph camera focus on different
observing runs produced different patterns of instrumental resolution
with wavelength and spatial position; these variations are within the
quoted range.

Instrumental-broadening corrections for the ionized-gas line-widths
are determined directly based on the widths from line-lamp spectra
closest to the observed wavelength of the ionized-gas line in the
galaxy spectrum. Since our measurement of line-width from the stellar
spectroscopic data are achieved via cross-correlation against stellar
templates observed with the same or similar instrumental
configuration, {\it to first order} the instrumental broadening is
taken into account. This accounting is imperfect due to redshift, the
fact that stellar templates were not observed in every fiber, and
because of focus variations between observing runs.

The first two of these effects produce very modest differences between
the instrumental broadening of template and galaxy spectra: Galaxy
redshifts are low ($z<0.042$; corresponding to shifts typically 25\%
of the observed wavelength range), the templates were observed with
many fibers spanning the slit, and the instrumental broadening varies
little over the spectral and spatial range of the data. Consequently,
the mismatch in $\sinst$ is $<$ 1 km s$^{-1}$ and varies from fiber to
fiber. However, run-to-run changes in spectrograph focus have led to
systematic differences between the template and galaxy instrumental
resolutions, $\sinstT$ and $\sinstG$ respectively. We define
$\dsinst^2 \equiv (\sinstG)^2 - (\sinstT)^2$, and apply a correction
of the form $\slos^2 = \sobs^2 - \dsinst^2$. Note that $\dsinst^2$ can
be positive or negative.

For SparsePak, we estimate $\dsinst^2 \sim 0.0$, and so we do not
apply a correction for instrumental mismatch. The remaining mismatch
due to redshift and fiber sampling effects are less than a quarter of
the range of observed instrumental broadening with wavelength and
fiber, or $<$ 0.5 km s$^{-1}$, from which we arrive at
$\Delta \ln \sinst = 0.04$ for SparsePak. 

For PPak, $\dsinst^2$ is often not zero. A typical value of
$\sqrt{|\dsinst^2|}$ is 6 km s$^{-1}$. With the effort of measuring
line-lamp spectra line-widths, we are able to determine this quantity
to better than 0.4 km s$^{-1}$.  The uncertainty includes the
remaining mismatch due to redshift and fiber sampling effects. Hence
$\Delta \ln \sinst = 0.02$ for PPak.

What is relevant for the error budget of $\sddisk$ is the quantity
$(\sinst/\sobs)^2
\, \Delta \ln \sinst$ (see \S 3.5.6 and Appendix B). In this context,
the uncertainties introduced by $\sinst$ are the same for both
instruments. We adopt an equivalent, median value of $\sinst = 15$ km
s$^{-1}$ and $\Delta \ln \sinst = 0.03$ for calculations below.
Overall, the effects are systematic for a single fiber for a given
galaxy, but random when averaged over a stack of fibers. Similarly,
the effects are random when averaging over observing runs for a single
fiber and galaxy.

\subsubsection{Beam smearing}

Beam-smearing arises from the projected intensity, velocity and
velocity-dispersion gradients across the SparsePak and PPak fiber
faces, or ``beams,'' suitably broadened for seeing. Features of
fiber-optical spectroscopic data that differ from aperture-synthesis
measurements at radio wavelengths and from direct-imaging spectroscopy
include the discrete nature of the beams and the azimuthal scrambling
properties of fibers. Compared to imaging-spectroscopy, the scrambling
property of fibers ensures that the observed line-width is independent
of the direction of the velocity gradient.

Begeman (1989) describes the general convolution process that relates
an observed moment of the velocity distribution function (VDF) to the
intrinsic distribution function, assuming the VDF is Gaussian at any
given spatial position. This study arrives at a Taylor-series
approximation to the convolution integral relating the observed and
intrinsic velocity fields. While this can be generalized for
non-Gaussian VDFs and higher-order moments, the validity of a
Taylor-series approximation must be evaluated in each specific
application. We follow the general convolution scheme of Begeman
(1989) but depart from this formalism in several ways.

First, we employ an iterative scheme which starts with a smooth
(polynomial model) characterization of the observed velocity and
velocity dispersion fields. (In future analysis we extend this scheme
to include higher-order moments, e.g., in the context of a
Gauss-Hermite expansion of the VDF.) We adopt this characterization as
the initial estimate of the intrinsic distribution, and as such the
scheme is well-posed. We then synthetically ``observe'' it with the
appropriate fiber foot-prints convolved with an estimate for the
seeing and instrumental resolution. Initial correction factors are
then estimated to be the ratio of the synthetic observations to the
estimated intrinsic distributions. The data is corrected, fiber by
fiber or ring by ring, based on this initial beam-smearing estimate,
re-characterized, and then re-observed. The process is repeated until
the corrections converge. The correction scheme requires no
approximation.

Second, while our beam-smearing corrections are multiplicative, and
whereas Begeman (1989) defined a linear (additive) correction, for our
error analysis here we define an equivalent quadrature beam-smearing
correction, $\sbs^2$. This quantity is equal to the quadrature
difference between the observed velocity dispersion ($\sobs$ corrected
for instrumental broadening) and the corrected velocity dispersions
($\slos$). We make this definition to parallel the corrections for
instrumental resolution and to provide intuition in terms of a
Gaussian convolution approach (Westfall 2009).

A description of the trends in $\sbs^2$ with radius and azimuth serve
to illustrate the amplitude of, and variation in, this correction. As
a percentage effect, beam-smearing increases rapidly with radius and
then declines after the roll-over from the steep inner rise of the
rotation curve; the peak and roll-over occur typically within one disk
scale-length. The corrections are typically under 4\% for PPak data
and under 7\% for SparsePak data. In an absolute sense, we find that
beam-smearing is strongly dependent on the amplitude and shape of the
rotation curve in the centers of our pilot-survey sample
galaxies. With SparsePak, beam-smearing produced by velocity gradients
alone are $<5$ km s$^{-1}$ in the center and drops to $\sim$2 km
s$^{-1}$ on the minor axis at several scale-lengths, and below $<$1 km
s$^{-1}$ on the major axis. While these numbers are representative of
the survey as a whole, this is a conservative upper limit because the
majority of observations were taken with PPak which has smaller
fibers, and because it does not account for velocity dispersion
gradients.

In general, beam-smearing corrections are determined by a complicated
interplay between velocity and velocity dispersion gradients across
the fiber faces. For example, at small radii where the rotation
velocity is small but the velocity dispersion is large, beam-smearing
corrections $\sbs^2$ can even be negative. As another example, the
crowding of isovels tends to make beam-smearing proportionally largest
along the kinematic minor axis. However, the changing projection of
the SVE dampens this effect for small values of $\alpha$ because the
instrinsic azimuthal variation in $\slos$ is maximized.

Uncertainties in beam-smearing corrections arise from three sources of
uncertainty: the shape of the SVE, seeing, and the flux distribution.
The SVE shape, however, modulates the azimuthal dependence of the
beam-smearing amplitude, not the mean value at a given radius.
Consequently there is no impact of the {\it a priori} unknown SVE
shape on the mean radial trend of $\slos$, estimated either from
stacked rings or individual fibers. In contrast, when measuring the
SVE shape itself, the azimuthal modulation in beam-smearing is crucial
to determine; we will explore this in later papers in this series. The
seeing dependence of $\sbs^2$ is quite modest for SparsePak because of
its large fibers and the good delivered image quality of the WIYN
Telescope. For PPak observations the seeing values are well known.

Finally, uncertainties in the beam-smearing corrections also arise if
the validity of our assumptions of the flux distribution across a
fiber ``beam'' are suspect. While the stellar distribution in space
and velocity is certainly smoother than that of the gas, the fibers
subtend large physical scales. Considering the distances to our
targets, the distributions of physical beam sizes have median, upper
and lower-quartile values of $1.45^{+0.65}_{-0.5}$ kpc for SparsePak
and $0.83\pm0.3$ kpc for PPak, respectively. Beam sizes are under 2.9
and 1.7 kpc, respectively, for SparsePak and PPak for 90\% of the
sample. For the galaxies in the sample, there are typically between
2-3 and 4-5 fibers per radial disk scale-length for SparsePak and PPak
respectively. Nonetheless, we find no significant difference between
$\sbs^2$ calculated assuming a uniform flux distribution and
calculated assuming an exponential gradient in the flux distribution
equivalent to the broad-band surface-brightness profile.

As an estimate for our error budget we take half the beam-smearing
correction to be the uncertainty in the correction, noting that (a)
this will tend to be an over-estimate; and (b) the correction and
therefore the uncertainty are small. Beam-smearing corrections at 2.2
h$_R$ are typically 4\% for SparsePak and 1\% for PPak. We adopt the
larger value, namely $\sqrt{|\sbs^2|}/\sobs = 0.04$, and
$\Delta \ln \sbs = 0.5$ for the calculations below. Errors in the
beam-smearing correction are systematic per fiber, but random over the
average of a stack of fibers, therefore representing a random error in
terms of budgeting.

\subsubsection{Line-of-sight integration}

There is also a broadening of $\slos$ due to the line-of-sight
integration through a differentially rotating disk of finite
thickness. This occurs even for an infinitely small beam. While the
cause is different than spatial beam-smearing, the effect is
similar. For low-inclination disks, we might expect this effect to be
negligible. The typical radial range, $\delta R$, of the line-of-sight
through a disk in the DMS can be estimated based on the typical disk
inclination (i = 28$^\circ$) and disk radial scale length ($h_R = 3.6$
kpc, or 12.6 arcsec) of the Phase-B sample (Paper I). From this the
disk oblateness can be estimated ($q_R \equiv h_R/h_z \sim 8$; \S
2.2). The radial range can be approximated as $\delta R \ \sim \ 2 \, h_z
\ \tan i \ = \ 2 \, q_R \, h_R \ \tan i $. Hence $\delta R$ is
typically 1.6 arcsec -- substantially smaller than either the
SparsePak or PPak fiber ``beams.'' Since beam-smearing is a small
effect in the DMS, we consider the effect of line-of-sight integration
to be negligible.

\subsubsection{Line-of-sight projection}

The corrected velocity dispersion, $\slos$, is a projection,
along the line-of-sight, of the SVE:
\begin{equation}
\slos^2 = (\sigr^2 \sin^2\theta + \sigp^2
\cos^2\theta)\sin^2i + \sigz^2 \cos^2i,
\end{equation}
where $i$ is the galaxy
inclination relative to the observer, and $\theta$ is the azimuthal
angle from the kinematic major axis in the plane of the galaxy. With
definitions for the principal SVE ratios $\alpha$ and $\beta$ (\S2.1),
we further define
\begin{equation}
\gamma \equiv  1 + \frac{\tan^2i}{\alpha^2} \ 
( \sin^2 \theta + \beta^2 \cos^2\theta )
\end{equation}
to write
\begin{equation}
\sigz^2 = \frac{\slos^2}{\gamma \ \cos^2 i}.
\end{equation}

To apply the correction to the co-added spectra in each radial bin,
operationally we take an azimuthal average of $\slos$ and not
$\slos^2$.  Hence equation (4) requires solving an elliptic integral
of the second kind for $\sqrt{\gamma}$. However, at low inclination we
adopt an excellent approximation\footnote{The exact relation is given
  by $\sigz \, = \, \bar{\sigma}_{\rm LOS} \, / \, \frac{\cos
    i}{2\pi} \int_0^{2\pi} \!\! \gamma^{1/2} \, d\theta.$ However,
  $\left(\int_0^{2\pi} \!\! \sqrt{\gamma} \,d \theta \right)^2 \,
  \approx \, \int_0^{2\pi} \!\! \gamma \, d\theta$ is an approximation
  good to $<0.2$\% for $i < 45^\circ$ and $\alpha = \beta = 0.7$, and
  $<0.5$\% in the same inclination range for plausible values of $0.3
  < \alpha < 1$ and $0.7 < \beta < 1$.} such that $\sigz \ \approx \
\bar{\sigma}_{\rm LOS} / \sqrt{\bar{\gamma}} \cos i$,
where $\bar{\sigma}_{\rm LOS}$ is the azimuthally-averaged
$\slos$ and
\begin{equation} \bar{\gamma} \ \equiv \
  \frac{1}{2\pi}\int_0^{2\pi} \!\! \gamma \, d\theta \ = \ 1 +
  \frac{\tan^2 i}{2\alpha^2} ( 1 + \beta^2 ).
\end{equation}

Anticipating the development in \S 5 for $\sddisk$, we estimate the
uncertainties in $\sigz^2$.  From the discussion in \S 2.1, we adopt
expected typical values and uncertainties of $\alpha = 0.6 \pm0.15$
(25\%), $\beta = 0.7 \pm 0.04$ (5\%), and $i=30 \pm 3.5^\circ$ (12\%).
We find $\bar{\gamma} \cos^2 i = 1.32 \pm 0.29$ (22\%) and $\sigz^2 =
(0.76 \pm 0.17)
\bar{\sigma}_{\rm LOS}^2$ (see Appendix B).
Uncertainties in
$\bar{\sigma}_{\rm LOS}$ (\S 3.3, 3.4), inclination and $\bar{\gamma}$ all
enter into the error budget discussion in \S 5. Here, however we see
that despite the uncertainties in $\alpha$ and $\beta$, in general the
correction is small because the inclination is small and $\gamma$ is
near unity. This means that uncertainties introduced by this
correction are modest, which is a specific reason why we have chosen
the nearly face-on regime.  The variation in $\sqrt{\bar{\gamma}}-1 =
\slos/\sigz \cos i - 1$ is illustrated in Figure 5 as a
function of inclination and SVE shape ($0.4<\alpha<1.0$,
$\alpha=\beta$); systematic errors in disk mass due to deprojection
should scale with this function.

\subsubsection{Extinction}

With the exception of the isothermal case, where $\sigz$ is
independent of scale-height, dust extinction in the disk can modulate
the observed value of $\slos$. Considering the face-on case, an
exponential vertical mass distribution is the most extreme situation
in the sense that the range of $\sigz$ from mid-plane to infinite
height above the disk changes by a factor of $\sqrt{2}$, i.e., about
41\%, increasing with height (see van der Kruit 1988, whose general
development in the absence of extinction we elaborate on here). The
effect of a dust layer, then, will be to censor the mid-plane region
and raise the observed $\sigz$.  The impact will depend on the
total optical depth and the relative scale--height of dust and stars.
For example, an infinitely thin dust layer at the mid-plane will have
no effect on $\sigz$, regardless of optical depth.  In the absence
of dust, or in the infinitely-thin dust-layer case, the integrated
$\sigz$ is $\sqrt{3/2}$ times the mid-plane value for an
exponential vertical mass-density distribution of constant $\Upsilon$.
In general, the observed $\sigz$ will increase in either of two
situations: (1) for a finite-thickness dust-layer as the optical depth
increases at a given dust-layer thickness relative to the stellar
distribution; or (2) as the thickness of the dust layer increases at a
fixed total optical depth through the disk.

A reasonable assumption is that the dust layer is comparable to the
gas layer, and much smaller than the stellar scale-height for the old
stars we are using as dynamical tracers. This is consistent with the
observations of edge-on systems by Dalcanton et al. (2004) for
fast-rotating disks like those in the DMS. Even in slow-rotating
disks, which have thicker dust distributions, Seth et al. (2005) find
the RGB population have scale-heights 1.5 to 3 times larger than the
dust. While Howk \& Savage (1999) have shown, via unsharp
image-masking, that there are pronounced extra-planar dust structures
in fast-rotating disks seen edge-on, there is good reason to believe
these structures are associated with the energetics of star-formation
in spiral arms (e.g., Kamphuis et al. 2007). The use of
unsharp-masking, however, removes the median extinction level from the
image so that the visual impression over-emphasizes the relative
strength of the extra-planar versus the near-planar extinction.
Inspection of the un-filtered images is qualitatively consistent with
the multi-band radiative-transfer analysis (e.g., Xilouris et
al. 1998): The overall extinction at large scale-heights is
substantially lower than that near the mid-plane.

More debatable is the total, face-on optical depth. Work by Domingue,
Keel \& White (2000), Boissier et al. (2004), Holwerda et al. (2005),
and Tamura et al. (2009) all point to the clumpy nature of dust
extinction in disks seen at low inclination. The different methods
used in these studies each have their own substantial uncertainties,
but there appears a consensus that dust is heavily concentrated toward
spiral arms, with patchy effective extinction in the range $0.3 < {\rm
  A}_B^{\rm i} < 2.5$ mag; the inter-arm regions are relatively
thin (A$_B^{\rm i} < 0.1$ mag) and become thinner with increasing
radius. Keel \& White (2001) estimate that for the disks in their
sample of spirals ``half the dust mass [is] contained in only 20\% of
the projected area and 95\%-98\% of the dust mass contained in half
the area.''

An essential problem with all of these observational studies is their
inability to access the full 3D geometry of the stellar and dust
distribution. This geometry, as shown by Calzetti et al. (1994), is
necessary for modeling accurately the extinction as it pertains to
different layers of the disk.  Radiative-transfer modeling using
clumpy stellar and dust distributions (e.g., Matthews \& Wood, 2001,
for the low surface-brightness galaxy UGC 7321) are the path forward,
but such models have not yet been fully developed for a wide range of
galaxy types.  However, some superb, multi-wavelength modeling using
smooth dust distributions have been carried out by Xilouris et
al. (1997, 1999). Their models have been applied to edge-on,
intermediate-type spirals, otherwise typical of DMS galaxies. They
find that the face-on optical depth is less than one, implying
galaxies seen face-on would be nearly transparent. The extinction
values are comparable to those found suitable for moderate-inclination
galaxies using a simpler, but physically similar model (Verheijen
2001).

To reconcile these results with the observational studies of clumpy
dust distributions, in a follow-up study Misiriotis et al. (2000) have
shown that from an edge-on perspective putting dust smoothly in an
exponential disk is indistinguishable from placing the dust in a
logarithmic spiral pattern of similar scale-height. Further, Popescu
et al. (2000) and Misiriotis et al. (2001) examined whether the smooth
dust-models of Xilouris et al. were consistent not only with the
apparent extinction of star-light, but also the far-infrared (FIR)
emission from the dust itself. Based on an FIR excess, they found
evidence for an additional dust component, plausibly from a very thin
layer associated with mid-plane star-formation.

The coherent picture emerging from all of the above discussion is one
where the patchy regions of high extinction in spiral arms are
associated with this nearly mid-plane dust responsible for the FIR
excess, whereas the remainder of the disk is characterized better by
the log-spiral dust-model of Misiriotis et al. (2000). Because regions
of high extinction near the mid-plane do not effect $\slos$, we
therefore adopt the smooth dust-distribution models of Xilouris et
al. as a suitable mean for the arm-interarm disk extinction modulating
$\slos$. In \S 3.7 we test our hypothetical picture by
exploring if azimuthal variations in $\slos$ might be due to
clumpy dust distributions associated with spiral structure.


Here, then, we characterize the dust to be in a smooth exponential
distribution in radius and height. We adopt the full range of values
from the work of Xilouris et al. (1997, 1998, 1999) such that the dust
has a scale-height of roughly $0.6\pm0.3$ that of the stars, an
exponential radial scale-length of roughly $1.4\pm0.2$ that of the
stars, and a face-on optical depths of $\tau = 0.35\pm0.15$ and
$0.55\pm0.20$ in the $I$ and $V$ bands, respectively, i.e., matching
the \ion{Ca}{2}-triplet and \ion{Mg}{1b} regions. We have computed the
impact on the observed $\sigz$ in the worst-case scenario where
$\tau=0.9$ (at the galaxy center), and the dust-to-star scale-height
ratio is 0.9: This is the worst case scenario in the sense that both
optical depth and dust-to-star thickness are one standard deviation
off their nominal values such that the impact on $\sigz$ is
maximized. The change in $\sigz$ is only 1.3\% higher from the
dustless case. Since our measurements are actually made at around 1
dust scale-length, with typically smaller optical depths, the impact
of extinction on the observed $\sigz$ is completely negligible. For
this reason, we ignore extinction effects on $\slos$. The impact of
extinction on the photometric aspect of $\mlsdisk$ is discussed below.

\subsubsection{Summary of corrections and uncertainties}

The final expression for the corrected vertical velocity dispersion of
a stacked ring of fibers is given by
\begin{equation}
\sigz^2 =  \bar{\sigma}_{\rm LOS}^2  \ / \ \bar{\gamma} \ \cos^2 i,
\end{equation}
where
\begin{equation}
  \bar{\sigma}_{\rm LOS}^2 = \bar{\sigma}_{\rm obs}^2 - \delta\bar{\sigma}_{\rm inst}^2 - 
  \bar{\sigma}_{\rm beam}^2;
\end{equation}
$\bar{\sigma}_{\rm obs}$ is the observed velocity dispersion from an averaged set
of velocity-registered fibers in a ring; $\delta\bar{\sigma}_{\rm inst}$ is the
average correction for instrumental broadening not naturally taken out
by the cross-correlation process (\S3.5.1); $\bar{\sigma}_{\rm beam}$ is the
average correction for beam smearing due to the finite size of our
sampling aperture (\S 3.5.2); and $\bar{\gamma}$ is the correction for
line-of-sight projection (\S3.5.4). We have ignored corrections for
line-of-sight integration (\S 3.5.3) and extinction (\S 3.5.5), and
likewise the uncertainties in the error budget for $\sigz$ and
quantities dependent upon $\sigz$.

The accounted uncertainties in $\sigz$ from measurements and
corrections are listed in Table 1 in order: random measurement errors
in $\slos$ from cross-correlation ($\Delta \ln \sobs$,
\S 3.3); random errors from template mismatch ($\Delta \ln
\stm$, \S 3.4); instrumental broadening mismatch between
template and galaxy spectra ($\Delta \ln \sinst$, \S 3.5.1);
uncertainties in the beam-smearing correction ($\Delta \ln
\sbs$, \S 3.5.2); and errors in deprojection due to
uncertainties in $\alpha$, $\beta$, and $i$ (\S 3.5.4).

\subsection{Other Considerations}

\subsubsection{Spatial registration}

IFU centering on a target galaxy can vary over a period of many
consecutive hours of observation due to guiding errors or flexure.
Such variations are small ($<$1 arcsec), and negligible over the
course of a single exposure of 30-60 minutes. Larger centering
variations can occur between observations of the same target taken on
different nights or runs due to errors in offsets during target
acquisition. We estimate the repeatability of target acquisition with
the WIYN 3.5m telescope\footnote{The WIYN Observatory, a joint
  facility of the University of Wisconsin-Madison, Indiana University,
  Yale University, and the National Optical Astronomy Observatories.}
at about 1 arcsec (RMS) in either RA or DEC. Offsets of 20-30\% of a
SparsePak fiber diameter are therefore typical of a set of
observations of a single galaxy. For PPak observations, the position
of a guide star on the guide CCD of the spectrograph is repeatable to
0.2 arcsec, i.e., the pixel size of the guide CCD, or 8\% of the PPak
fiber diameter.

Precise and accurate spatial registration of the IFU spectroscopy must
be made both internally to the different spectroscopic pointings as
well as externally to the imaging photometry. A robust determination
of the relative spatial registration of the spectral data is critical
for kinematic measurements.  Uncorrected drifts, offsets, and
mis-alignment of the pointing lead to an effective beam-smearing when
fiber spectra are coadded, and to a mismatch when modeling the data or
combining data across configurations (e.g., H$\alpha$ with
\ion{Mg}{1b} and \ion{Ca}{2}-triplet region data). Registration of the
IFU spectroscopic data to the imaging photometry is required to match
the kinematic measurements of enclosed mass and mass surface-density
to enclosed luminosity and surface-brightness. Since the IFUs sparsely
sample the spatial dimensions, it is all the more important that both
the kinematic and photometric ``footprints'' are aligned in order to
minimize random errors in matching mass to light. To prevent a
substantial contribution to our error budget, we want the spatial
mis-registration to be smaller than the seeing.  We set a
conservative upper-limit for spatial mis-registration such that it
increases the effective smearing due to seeing by no more than $\sim$20\%
in a quadrature sense.  This results in a mis-registration requirement
for SparsePak of $\leq 0.5$ arcsec for the median seeing conditions
(0.8 arcsec FWHM; \S 3.6.2), or about 10\% of the SparsePak fiber
diameter. For PPak, with median seeing of 1.4 arcsec, the
mis-registration requirement is $\leq$ 0.8 arcsec, or 30\% of the
fiber diameter.

We use two independent methods to determine the relative spatial
offsets of our spectroscopic data from SparsePak. Both take full
advantage of the IFS spatial coverage. The first method uses the
velocity-field fitting technique described in Paper I which
simultaneously solves for the barycenter, position angle, inclination,
rotation curve, and the relative offsets of the individual
spectroscopic pointings for a given source (Andersen 2001; Andersen \&
Bershady in preparation). The registration is strictly relative, but
with the assumption that the galaxy barycenter is co-spatial with the
optical center of the galaxy, this method also provides an absolute
registration. As we note below, this assumption can be checked by
comparing the kinematic center to the center derived from a comparison
of the spectral continuum distribution to broad-band images.

Application of this kinematic technique to the SparsePak data in the
H$\alpha$ and \ion{Mg}{1b} regions for the 7 galaxies noted in \S
3.4.2 (Westfall 2009) yields median centering errors of 0.2 arcsec in
the H$\alpha$ region and 0.5 arcsec in the \ion{Mg}{1b}
region. (Exposure times were 15 minutes in H$\alpha$ and 45 minutes in
\ion{Mg}{1b}, typical for our survey.)  The amplitude of these random
errors scale directly with the product of the number of lines fit in
each region (we simultaneously fit [\ion{N}{2}] and [\ion{S}{2}] lines
in addition to H$\alpha$, whereas in the \ion{Mg}{1b}-region we only
fit the [\ion{O}{3}]$\lambda$5007 line) and the total number of fibers
per pointing for which line-emission is detected.\footnote{This
  product is typically 245 for an H$\alpha$ region pointing and 40 for
  the \ion{Mg}{1b} region.}  There is some indication that the
registration using [\ion{O}{3}]$\lambda$5007 is somewhat (up to 15\%)
noisier than expected from this scaling, consistent with a velocity
field that qualitatively looks more chaotic than that measured in the
H$\alpha$ region. Systematic errors can also be estimated by examining
differences in the registration when fitting a range of rotation curve
models, parameterized with two to three variables that control the
steepness of the rise, peak velocity, and asymptotic behavior at large
radius.  For a set of models yielding comparable minimum $\chi^2$
values, we find that these systematic errors are 2 to 2.5 times
smaller than the random errors. There is a hint that systematic errors
become proportionately smaller as random errors become larger,
suggesting that we are overestimating our systematic errors. This
sample of 7 galaxies, six of which range from $24^\circ < i <
38^\circ$, are typical of our Phase B sample in terms of quality of
the velocity-field fit, inclination, and velocity-field regularity.
We conclude that relative centering errors of 5-10\% (random) and
2-4\% (systematic) of a fiber diameter are representative of this
kinematic method, provided good S/N data and a smooth velocity field.

The second registration method aims at matching the spectral continuum
in the IFS data-cubes, in a least-squares sense, with photometry from
broad-band images in a similar wavelength region. One challenge is to
define a meaningful $\chi^2$ statistic, requiring a good definition of
detector gain and error propagation when deriving the extracted
spectral continuum signal. Seeing variations between image and spectra
can also be problematic (\S 3.6.2). In some early data, there were
also uncertainties in the IFU position-angle, which complicated the
$\chi^2$ mapping. However, by comparing the kinematic position angle
from the velocity-field fits to the photometric position angles
ameliorates even these uncertainties.

Two approaches have been taken to this continuum-fitting method. The
first method fits the spectral continuum data directly to aperture
photometry of the broad-band images using the same relative
two-dimensional foot-print pattern as the IFU, and apertures that
match the fiber diameter. Using this approach, Swaters et al. (2003)
were able to register the spectral continuum from SparsePak H$\alpha$
observations of a low surface-brightness galaxy (DDO 39) to an
$R$-band image with a centroid precision of 0.5 arcsec or better. This
source is substantially irregular and has considerable foreground
contamination from Galactic stars. Application of this method to
regular galaxies of normal surface-brightness at higher Galactic
latitude should be more precise. The second method matches the
spectral continuum to a one-dimensional light-profile. Bershady et
al. (2005) report continuum-fitting in the \ion{Ca}{2}-triplet region
using SparsePak to the $I$-band surface-brightness profile of UGC 6918
with a centroid precision of 0.25 arcsec or better. This precision is
comparable to the velocity-field fitting method, and allows us to
place the IFS observations on an astrometric scale without making any
assumptions about how light traces mass.

In a separate paper (Andersen\& Bershady, in preparation) we have used
DensePak H$\alpha$ data of a similar sample of 39 nearly face-on
galaxies from Andersen et al. (2006) to directly compare centers
derived with the kinematic and photometric methods described here. As
noted in Paper I, 14 of these galaxies are in the DMS. The
distribution of offset differences (RMS) based on this comparison is
characterized by a mode of 0.35 arcsec, but a mean of 0.7 arcsec due
to a tail that extends to 2 arcsec and contains 10\% of the
sample. This tail is likely due to the failure of one of the centering
methods. For example, the distribution extremum is UGC 4614, a galaxy
not in the DMS but with a H$\alpha$ velocity field showing strong
twisting associated with spiral arms. In this case the kinematic
method is likely more suspect because it is based on a fit to an
axisymmetric model. In general, galaxies with large asymmetries in
their rotation curves ($>$10\% in velocity) show systematically larger
offset differences.  By using both methods we are able to identify
sources with discrepant centering and then, by inspection, determine
which method is likely most problematic, thereby minimizing centering
errors.  The above mentioned mode and mean of the offset-difference
distribution are equivalent, respectively, to centroiding
uncertainties of 0.25 and 0.5 arcsec in both methods, consistent with
our earlier estimates. Hence the registration precision we are able to
achieve with either kinematic- or continuum-fitting methods is good
enough that it does not enter the error budget. The centering
uncertainties for PPak are expected to be significantly smaller given
its 331 fibers, each with smaller diameter than SparsePak.

\subsubsection{Seeing variations}

Although the impact of seeing-mismatch between IFU and broad-band data
on their spatial registration appears to be small for the large-fiber
IFUs used in the DMS, the commensurate impact on mismatching ``mass to
light'' is a different issue. With changes in seeing, the light
sampled by a single fiber may not represent the same effective
physical region as that of an identical photometric aperture, even if
it is correctly placed on a broad-band image. The issue is comparable
to deriving colors for sources in imaging data with variable
seeing. For measuring colors, the recourse is either to determine the
seeing in each image and degrade all images to the worst-case seeing
(if the source profile is not known a priori), or if the profile is
known (e.g., from stars), to carry out profile fitting on each source
in each image, and take a comparable fraction of the total light from
the derived profile fit to each image. Unfortunately, we do not know
the intrinsic light profile sampled by each fiber, and for SparsePak
we do not have estimates of the seeing conditions during the IFU
observations. For PPak observations, the CCD image of each guide star
exposure is stored and at our disposal. From each series of guide star
images, we have reconstructed the effective seeing during each of the
60 minute PPak exposures (Martinsson et al., in preparation).  The
distribution of PPak seeing conditions (FWHM), suitably averaged over
the full exposure times for each galaxy, has a mean and standard
deviation of $1.4 \pm 0.4$ arcsec; 80\% of the data was taken in
conditions between 0.9 and 2.0 arcsec; and all of the data was taken
in conditions between 0.8 and 2.1 arcsec.

There are several factors which mitigate the impact of seeing
mismatch. First, SparsePak fiber diameters are 6 times larger than the
FWHM of the point-spread-function (PSF) in median conditions at WIYN
(0.8 arcsec), so that 70\% of the light collecting area of a fiber is
more than the FWHM away from the fiber edge. This diminishes the
impact of any unresolved sources in or near the edges of a fiber on
the encircled energy as the PSF varies. Second, by employing the
spectral-continuum registration method using an image with
sufficiently good seeing, we can degrade the image quality until the
best match is found between spectral and imaging data in a $\chi^2$
sense (e.g., Andersen et al. 2008). Finally in the case of fiber
averaging, either in spatially-adjacent regions or in an azimuthal
ring, the impact of seeing variations will be averaged out. These
mitigating factors, while only qualitatively described here, are
sufficient to ignore seeing mismatch in further discussion of our
error budget.

\subsection{Assessment of Uncorrected Systematics}

The accuracy of any measurement ultimately depends on the completeness
of the assessment of, and correction for, systematic errors. The
presence of overlooked or mis-estimated systematic effects abound in
astronomy because of the complexity of astrophysical systems and the
lack of direct laboratory verification. The troubled history of the
extragalactic distance-scale measurement is but one notorious example.
Since galaxies are complex systems, and the measurement of disk-mass
involves a delicate confluence of photometric and kinematic scales, it
would not be surprising to find that in ten years time our knowledge
has increased sufficiently to revise the correction estimates
presented in this paper. Nonetheless, at this time we do have
sufficient information to surmise that any such revisions will likely
be modest.  Of the two primary components of the disk-mass
measurement, $h_z$ and $\sigz$, we we have explored potential
systematics associated with the vertical scale height extensively in
\S 2.2. Here we present a simple inquiry on the likely amplitude of
systematics in the stellar velocity dispersion.

A primary concern we have is that our approach relies on the
assumption that to a high degree disks are axisymmetric and their
mass-density distribution is smooth. Yet we know non-axisymmetric
structure in the form of spiral arms, lopsidedness or elongation
(bi-symmetry) are also salient features of disk systems.  Specific
concerns include: patchy extinction of unknown scale-height (raising
$\slos$ due to mid-plane censoring of an exponential vertical
mass-distribution); population gradients (modulating $\slos$
due to template mismatch and $\sigma$-$h_z$ trends with age); and
velocity structure (enhancing $\slos$ due to beam smearing
from, e.g., streaming motions in spiral arms).  To check for these
effects we have picked UGC 6918 as a test case because it has the
highest surface-brightness disk in our sample (and hence is most
likely to have significant extinction at least in its inner regions);
it has among the most asymmetric rotation curves in our sample
(Andersen et al.  in preparation); and it contains a weak AGN --
activity perhaps resulting from the perturbed kinematics and also
plausibly modulating stellar-population gradients.

In Figure 15 we illustrate residuals in the spectral continuum,
stellar velocity and stellar velocity dispersion from a smooth,
axisymmetric model. Kinematic measurements have been made for
individual fibers using a single (K1~III) stellar template so that we
may maximize our sensitivity to both radial and azimuthal
variations. We have constructed the residuals from an axusymmetric
model in the simplest and least parametric way possible.  The model
consists of independent Legendre polynomial fits of low order to the
radial distributions of each of these quantities. For purposes of
illustration the velocity data is deprojected for azimuthal location
but not inclination, and we exclude from the fit point that are more
than 60$^\circ$ from the kinematic major axis.  However, the velocity
residuals are calculated from the derived rotation curve projected in
azimuth.  Similarly, the model fit to the radial trend of the stellar
velocity dispersion is modulated in azimuth, when calculating
residuals, by an SVE with $\alpha=0.6$ (expected to be typical of
late-type disks) and $\beta$ set by the epicycle approximation and the
model rotation curve.

By construction the residuals in radius have no structure, although
there is scatter at smaller radii substantially in excess of the small
estimated measurement errors. The scatter is below 12\% rms in
surface-brightness and $\slos$, and roughly 3 km s$^{-1}$ rms in
velocity. In azimuth, however there is clearly correlated structure,
most evident in the surface-brightness and $\slos$ residuals. The
amplitude of this structure is below 15\%.  One half of the galaxy
also appears to be rotating 10 km $^{-1}$ faster than the other (in
projection).

The bottom panels of Figure 15 explore whether variations in
$\slos$ are correlated with variations in surface-brightness
(as might be expected if there is patchy extinction associated with
spiral structure) or deviations from smooth rotation. There is no
evident correlation between any of these quantities.

From this analysis we conclude that in the worst possible case (a high
surface-brightness, kinematically asymmetric disk) systematics are
below 15\%, and even these amplitude are only apparent at the innermost
radii where extinction is likely greatest.  Our expectations are that
for the remainder of our survey sample, such systematics will be
smaller still, and hence negligible to our overall error budget.

\section{DISTANCE ERRORS}

The flow-corrected distances, $D$, given in Table 2 of Paper I have
$7\pm0.5$\% formal errors, which are the quadrature sum of the
estimated uncertainty in $H_0$ (NED\footnote{NASA/IPAC Extragalactic
Database: {\rm http://nedwww.ipac.caltech.edu/}} quotes $73\pm5$, or
$\Delta \ln D_{\rm H0} = 0.07$; cf.  Mould et al. (2000), who estimate
9\%) and the heliocentric velocity measurements. The latter are
insignificant. Yet flow corrections are large, and their precision and
accuracy should be somewhat suspect at least to the extent that there
are no errors associated with these corrections attributed to the
distance-estimates. Uncertainties arise both from the flow-model as
well as peculiar velocities on top of any bulk flow. For example, for
both Phase-A and Phase-B samples, the median flow-correction $\Delta
V_{\rm flow} \equiv V_{\rm hel}-73D$ is $\sim -300$ km s$^{-1}$. Flow
corrections range from -811 km s$^{-1}$ to +1301 km s$^{-1}$; 90\% of
the sample is contained within $ -620 < \Delta V_{\rm flow} < +230$ km
s$^{-1}$. These values are fairly independent of heliocentric
velocity, which means the corrections become very large at small
recession velocities. The largest corrections are in the sense of
lowering distances and related distance-dependent quantities of
luminosity and size, but raising $\sddisk$ and $\Upsilon$. However the
median correction is to increase distances, and hence the predominate
effect of flow-corrections is to raise luminosity and size estimates,
but diminish $\sddisk$ and $\Upsilon$.

Taking the quantity $\Delta \ln D_{\rm flow} \equiv |0.5[1-V_{\rm
  hel}/(73 \times D)]|$ as a conservative distance-error statistic
  associated with flow-corrections {\it and} peculiar velocities, we
  find 90\% of the Phase-A and Phase-B samples have distances errors
  $<$ 11\% and $<$ 8\% respectively; 75\% of the samples have
  distances errors $<$ 6\% and $<$ 4\%; and 50\% of the samples have
  distances errors $<$ 3\% and $<$ 2\%. Similarly, $\Delta \ln D_{\rm
  flow}$ is less than 10\% for $V_{\rm hel}>2100$ km s$^{-1}$ and less
  than 5\% for $V_{\rm hel}>5000$ km s$^{-1}$ for the Phase-A sample,
  and $\Delta \ln D_{\rm flow}$ is less than 10\% for $V_{\rm
  hel}>1450$ km s$^{-1}$ and less than 5\% for $V_{\rm hel}>3350$ km
  s$^{-1}$ for the Phase-B sample.  Adopting the 90th percentile
  errors for $\Delta \ln D_{\rm flow}$ for the Phase-B sample in
  quadrature with uncertainties in $H_0$, we arrive at a total
  systematic error of 11\% for distance.  This is applicable to
  individual galaxies.  Since sources cover a range of distances and
  directions in the sky, velocity-flow corrections to their distances
  introduce random errors in the ensemble measurements. For the survey
  as a whole, the systematics are diminished to the uncertainty in
  H$_0$ only, or 7\%.

In future papers we discuss some individual cases, e.g., UGC 6918,
where other indicators such as super-novae can be used to
better-calibrate distances. Of course, TF can be used too, but
since its use inherently biases the luminosity estimate on a
case-by-case basis (although not necessarily in the mean), wherever
possible we prefer to avoid this approach. This is necessarily the
case when using iTF to infer inclinations.

\section{SURVEY ERROR BUDGET}

The formulae for the dynamical disk mass, $\sddisk$, the disk
stellar mass-to-light ratio, $\mlsdisk$, the disk fraction,
$\Fdisk$, and the baryon fraction, $\Fb$ are
presented in terms of our observables. These formulae are used to
derive our error budget for each quantity, and to identify which
portions of this budget lead to systematic or random uncertainties for
individual galaxies and for the survey as a whole.  For illustration
purposes, we compute the errors on the disk-averaged values
$\langle\sddisk\rangle$ and $\langle\mlsdisk\rangle$. We make
the canonical distinction between ``random'' and ``systematic'' error
components to mean that if the observation were repeated under
identical conditions, random errors lead to a variance in the derived
value. Systematic errors in individual galaxy measurements which
scatter symmetrically around an accurate value of the mean for the
sample ensemble, are categorized as random uncertainties for the
survey.

\subsection{Disk Dynamical Mass, $\sddisk$}

As a nominal model we will adopt an exponential vertical distribution
function ($k=3/2$) of constant scale-height and oblateness as parameterized in
\S 2.2 by equation (1).  Scaled to our nominal model, physical units
and characteristic values expected for the DMS, the radial expression
for dynamical mass surface-density becomes
\begin{equation}
\sddisk(R) \ \left[\msol \ {\rm pc}^{-2}\right] = 100.0
\ \left(\frac{k}{3/2}\right)^{-1} 
\left(\frac{h_z}{444 \ \rm{pc}}\right)^{-1} 
\left(\frac{\sigz(R)}{30 \ \rm{km \ s}^{-1}}\right)^2.
\end{equation}
Substitute $h_R$ for $h_z$, $\sddisk$ can be written as
an explicit function of our observables:
\begin{eqnarray}
\lefteqn{\sddisk(R) \ \left[\msol \ {\rm pc}^{-2}\right] \ = } \hspace{0.25in} \nonumber \\ 
& & 100
\left(\frac{k}{3/2}\right)^{-1} 
\left(\frac{q}{q_R}\right)
\left[ 
\left(\frac{D}{60.1 \ \rm{Mpc}}\right)
\left(\frac{h_R}{12.^{\prime\prime}6}\right)
\right]^{-0.63}
\left(\frac{\sigz(R)}{30 \ \rm{km \ s}^{-1}}\right)^2,
\end{eqnarray}
where $q$ is the true axis ratio (oblateness) $h_R/h_z$, and
$q_R$ is  the estimate given in equation 1.

Equation (9) becomes our operating formula for estimating errors.  The
first two quantities ($k$ and $q$) are unity for the nominal model,
and have (known) uncertainties ($\Delta \ln k = 0.14$ and $\Delta \ln
q = \Delta \ln q_R = 0.25$).  They lead to systematic errors for
individual galaxies, but random errors over the survey as a whole. For
our calculation here, we assume that galaxies do not deviate
systematically from our adopted oblateness relation and our assumed
form for the vertical density distribution.  The last three
parenthetical quantities in equation (9) are our observables. In
order: Distance uncertainties, discussed in \S 4 above, are systematic
for individual galaxies, and contain systematic (H$_0$) and random
(flow) errors for the survey as a whole. Random errors in the measured
recession velocities are negligible. Measurement uncertainties in
$h_R$ are random. We note the reduced distance- and size-dependence
due to the correlation of oblateness with galaxy scale. Errors in
$\sigz$ are random, including the observed line-of-sight dispersion,
beam-smearing and instrumental corrections, inclination,
template-mismatch, and the SVE shape. We assume in our model that
estimates for the SVE shape and its uncertainty are made on a
galaxy-by-galaxy basis.  Survey systematic errors for $\sddisk$, then,
include only distance errors from H$_0$.

Several caveats are worth mentioning regarding the impact on our
$\sddisk$ estimate due to the presence of (i) a radial gradient
in the halo potential, (ii) a thick disk, and (iii) a thin gas-layer.
First, the impact of the halo potential on $\sigz$ should be small
because the halo density gradient is locally negligible compared to
the radial density gradient of the disk, particularly in the radial
region of interest in our survey. Both (ii) and (iii) alter the
effective $k$ value that should be used with the associated $h_z$ and
$\sigz$, and also the effective value of $h_z$. Since the effect of
a thick disk is to make $h_z$ larger and $k$ smaller (a two-component
monotonic profile is always cuspier than a one-component monotonic
profile), the effects of a thick disk on our estimates of
$\sddisk$ tend to cancel. We also demonstrated in \S 2.2 that the
impact of a thick disk on the effective value of $h_z$ is smaller than
the plausible range of $k$.  Hence we include the estimate of $\Delta
\ln k$ alone as a parenthetical entry in Table 1 to indicate an upper
limit to the potential systematics associated with variations in the
effective vertical density-distribution.

Appendix B.1 has the formulation of the random and systematic errors
for $\langle\sddisk\rangle$ under the plausible model that the
observables are independent. Table 1 collects the terms in three
categories: (i) measurement errors on $\sobs$ (\S 3.3) and $h_R$ (\S
2.2); (ii) uncertainties from correcting for systematic effects,
including instrumental resolution (\S 3.5.1), beam smearing (\S
3.5.2), line-of-sight projection (\S 3.5.4), template-mismatch (\S
3.4), and oblateness (\S 2.2); and (iii) uncorrected systematic
effects, which include the vertical distribution function and distance
(\S 4). We have excluded the effects on $\slos$ from line-of-sight
integration (\S 3.5.3) and dust extinction (\S 3.5.5) as negligible
contributions to the error budget.  A typical survey value of $\sigz =
30$ km s$^{-1}$ and $i=30^\circ$ are adopted to set the error
scale. Results depend weakly on the choice of $\sigz$ because the
dominant errors (deprojection and template-mismatch) scale.  However,
there is a significant dependence to the error in $\sddisk$ on
inclination, illustrated in Figure 16.

Typical errors for individual galaxies in our survey will be 28\%
(ran) and 26\% (sys) for $\sddisk$. The dominant error terms for
$\sddisk$ are (i) the conversion of scale-length to scale-height
(25\% systematic error); and (ii) the deprojection of $\slos$
into $\sigz$ (27\% random error).  Contributions to the latter come
nearly equally from uncertainties in the ellipsoid ratio, $\alpha$,
and the inclination. Dividing the sample into quartiles (e.g., 10
galaxies binned by color, luminosity, or surface-brightness) will
yield errors of 12\% (ran) and 4\% (sys) for $\sddisk$.

\subsection{Disk Stellar Mass, $\sds$, and Mass-to-Light Ratio, $\mlsdisk$}

The total dynamical disk M/L can be written as $\mldyndisk =
\sddisk/ I$, where the surface-intensity $I$ has units of
$L_\odot$ pc$^{-2}$, and it is implicit that $I$ is associated with
the disk. This is an empirical definition based on the {\it apparent}
luminosity distribution. It is useful for mass-decompositions when
information on other constituents of the disk (gas and dust) is
unavailable. $\mldyndisk$ has the same dependence on distance,
$\sigz$ and $h_R$ as does $\sddisk$, with an added dependence
on the surface-photometry. The latter in general is not a limiting
factor in terms of random error.

To obtain $\mlsdisk$, however, requires an estimate of the stellar
mass surface-density, $\sds$, and the extinction-corrected flux
surface-intensity, $I^c$: $\mlsdisk = \sds / I^c$.  We express the
flux surface-intensity in wavelength-specific form where $I_\lambda =
I^c_\lambda \exp(-\tau_\lambda)$, and the extinction in magnitudes,
${\rm A}_\lambda = 1.086 \tau_\lambda = -2.5 \log
(I_\lambda/I^c_\lambda) = \mu_\lambda - \mu^c_\lambda$; $\mu$ is
equivalent to $I$, except expressed in mag arcsec$^{-2}$.  The
extinction correction is the sum of the Galactic foreground extinction
$\rm A_\lambda^g$ and effective internal extinction $\rm A_\lambda^i$
within the galaxy, given the complexities of the dust geometry,
scattering, and (at long wavelengths) emissivity. Internal extinction
is modest in the outer parts of late-type disks (\S 3.5.5), and for
our reddest near-infrared bands ($K$ and 3.6 $\mu$m Spitzer bands)
extinction corrections are small, as are the uncertainties in these
corrections. There is also a $\sim$3\% zero-point uncertainty inherent
to any magnitude system (e.g., the uncertainties in the $AB_{95}$
calibration; Fukugita et al. 1996).

To arrive at $\sds$ corrections must be made for the gas mass
surface-density ($\Sigma_{\rm gas}$), comprised of atomic ($\sda$) and
molecular ($\sdm$) components, and for non-baryonic disk dark matter
($\Sigma_{\rm dark}$). We ignore ionized gas and dust contributions to
$\sddisk$, assuming they are negligible. While we mention $\Sigma_{\rm
dark}$ here for completeness, we drop it from subsequent
consideration; henceforth any non-baryonic disk component is subsumed
in the stellar component, and we write: $\sds = \sddisk -
\sda - \sdm$.

Measurements of the neutral hydrogen gas mass surface-density
($\sdhi$) is straightforward and part of our program. We scale
$\sdhi$ by a factor of 1.4 to arrive at the total atomic
gas-density. This scaling takes nucleosynthesis products into account
(a factor of 1.32 corresponds to a 24\% primordial helium
mass-fraction). Estimating $\sdm$ is more indirect and
uncertain, although Spitzer images at 8, 24, and 70 $\mu$m are at our
disposal; flux-ratios from these bands imply dust of various grain
sizes, and indirectly molecular gas phase.

First, we adopt $\eta \equiv \mmol / \matom$ as the
parameterization the total mass ratio of molecular to atomic gas, and
we assume the atomic and molecular phases have comparable
abundances. Early studies by Young \& Knezek (1989) based on
measurements of CO(1-0) line-emission indicated $\eta$ was of order
unity, but varied strongly with Hubble type between 1.8 to 0.3 going
from Sb to Scd. They found the range for a given type comparable to
the mean. More recent work by Casoli et al. (1998) shows $\eta < 1$
for all types decreasing from roughly 0.35 to 0.1 between Sb and Sd,
although this trend is reduced if only the most dynamically massive
systems are considered. This result is consistent with the most recent
literature-compilation studies, e.g., Obreschkow \& Rawlings (2009).
We adopt a values of $\eta = 0.25\pm0.1$ ($\Delta \ln \eta = 0.4$) as
typical for galaxies in our survey.  Since the atomic mass density
contributes roughly $11 \pm 6$\% of the total dynamical mass within
the luminous portion of a typical intermediate-type disk (e.g.,
Verheijen 1997; Hoekstra et al. 2001), the molecular component is of
order $3 \pm 2$\%.  Uncertainty in the mass-contribution of the
molecular component can be further reduced by accounting for the
observed correlations of $\eta$ with far-infrared luminosity and
dynamical mass, as noted by Casoli et al. (1998). More recent work by
Leroy et al. (2008) indicates even more precise estimates of the
distribution of molecular gas mass can be made via 24 $\mu$m flux
maps, which we have for our entire sample.

Second, to arrive at $\sdm$ we adopt a second parameterization, namely
$\xi \equiv \sdm / ( \sds + \sdm ),$ and make the reasonable
approximation that while the atomic and molecular total masses are
correlated, the mass surface-density of the molecular component is
better correlated with the stellar mass surface-density (Regan et
al. 2001). This approximation allows us to substitute masses for mass
surface-densities in the definition of $\xi$. Taking the sum of atomic
and molecular gas along with stars as the total dynamical disk mass,
$\mdisk$, we can then rewrite $\xi = \eta / (\mdisk/1.4 \mhi - 1)$,
which is a constant estimated for each galaxy. From this it follows
$\sds = (1-\xi)(\sddisk - 1.4 \sdhi)$. Finally, the expression for the
stellar M/L at a given wavelength and radius in the disk,
\begin{equation}
\mlsldisk(R) \ = \ \frac{(1-\xi)\ [ \ \sddisk(R) - 1.4
  \ \sdhi(R) \ ] }
{{\rm d}\!\exp[ -0.4 \ (\mu_\lambda(R)
 \ - \ {\rm A}_\lambda(R) \ - \ M_{\odot,\lambda} - 21.57)]} \ ,
\end{equation}
becomes a function of our observables and parameterization of the
molecular mass component. In this equation, $\mlsldisk$ has solar units
when the $\Sigma$'s are expressed in $\msol \ {\rm
pc}^{-2}$. The luminosity scale is set by $M_{\odot,\lambda} - 21.57
\equiv \mu_{\odot,\lambda}$, where $M_{\odot,\lambda}$ is the Solar
absolute magnitude in the relevant band and magnitude system.
Appendix B.2 contains the formulation of the random and systematic
errors for $\sds$ and $\mlsldisk$.

To set the error-scale, a number of values need to be defined.  In
terms of masses, first we assign a 3\% random measurement error to
$\sdhi$ in a given radial ring (2.5\% flux-calibration error, 2\%
random error given S/N$\sim$15 per beam and typically 10 beams per
ring). Second, we assign typical values of $N_{\rm disk} = 5$ and
$N_{\rm atom} = 10$ as the number of radial bins in the stellar $\slos$
and \ion{H}{1} maps, respectively, to relate errors in total disk and
\ion{H}{1} mass to their respective surface-density errors. Third, we
estimate a characteristic ratio of the stellar to \ion{H}{1} disks
masses, $\mdiskstar/\mhi$, to be $7.2\pm 3.8$ for a maximum-disk, or
$3.5\pm 1.9$ for a so-called ``Bottema-disk'' (Bottema
1997),\footnote{A so-called Bottema-disk has $\Fdisk = 0.63$, whereas
  a maximum disk has $0.85 < \Fdisk < 0.90$.}  based on the 13 Sb-Scd
galaxies from the study by Hoekstra et al. (2001); we adopt the mean
Bottema-disk value here. From these values and given $\eta$ above, it
follows that $\xi = 0.09$. The uncertainty in $\xi$ is large (40\%
random error and 53\% systematic error at $i=30^\circ$) due to a
combination of the uncertainty in the molecular-gas mass and the
measurement errors in the atomic and total disk masses.  However, what
is relevant to the error in $\sds$ and $\mlsdisk$ is the quantity
$[\xi/(1-\xi)] \ \Delta \ln \eta$, which is only about 4\%.  Finally,
we adopt the mean value $\sds/\sdhi = 9.0 \pm 5.4$ for a Bottema-disk,
which follows from $\mdiskstar/\mhi$ above, and characteristic radii
$R_{\rm HI}/R_* \sim 1.6\pm0.3$ based on Verheijen \& Sancisi
(2001). The variances quoted here for $\mdiskstar/\mhi$ and $R_{\rm HI}/R_*$
are astrophysical and do not enter into our error estimates because
they are measurable quantities in our survey.  For estimating
characteristic errors of individual galaxies or ensemble averages we
adopt the mean values for $M_{*}/\mhi$ and $\sds/\sdhi$. Note that
$\sds$ and $\sdhi$ have very different radial dependencies.

For photometric errors we adopt $\Delta \mu_\lambda = 0.05$ mag
arcsec$^{-2}$ as a conservative upper-limit to both photon shot-noise
and calibration uncertainty. Adopting the model in Verheijen (2001),
we find A$_\lambda^{\rm i}$ = 0.45, 0.15, 0.02 mag in the $B,I,K$ bands,
respectively. As an upper limit, we take $\Delta {\rm A}_\lambda =
0.4 {\rm A}_\lambda^{\rm i}$ based on the variance in the Xilouris et
al. studies, and assume that this uncertainty dominates over
uncertainties in correction for foreground extinction (we selected
galaxies with A$_B^{\rm g}<0.25$ mag; Paper I). Treated in this way,
correction for extinction introduces a random error assuming
extinction variations across a disk. In practice, measurement of the
broad-band spectral energy distributions (SEDs), the Balmer decrement,
or the H$\alpha$ to 24$\mu$m flux ratio will allow us to estimate the
extinction on a spatially resolved scale, thereby reducing this
uncertainty. Possible exceptions are regions of high-extinction in
spiral arms, which we will be able to identify via Spitzer 8, 24, and
70 $\mu$ maps.

Appendix B.2 has the formulation of the random and systematic errors
for $\sds$ and $\mlsdisk$ under the plausible model that the
observables are independent. Table 2 collects the terms including (i)
measurement errors on $\mu_\lambda$ and $\sdhi$, and (ii)
uncertainties from correcting for systematic effects, which include
extinction and the molecular mass-fraction. The inclination dependence
of errors in $\mlskdisk$ are illustrated in Figure 15. We have
excluded spatial registration errors (\S 3.6.1) and seeing variations
(\S 3.6.2) as negligible contributions to the error budget. The error
on $\sddisk$ is brought forward from Table 1. There are no additional
uncorrected systematic effects. With the exception of the molecular
gas-mass fraction, parameterized by $\eta$, and systematic effects in
$\sddisk$, the remaining uncertainties introduce random errors on
individual $\mlsdisk$ measurements in any one galaxy. For example, an
error in $\mu_\lambda^c$ is a random error for $\mlsdisk$ because the
$\mu_\lambda$ measurement-error is random, and $\Delta {\rm
A}_\lambda$ is treatable as a random error (extinction variations from
galaxy to galaxy and within a galaxy are stochastic but estimable).
Ensemble estimates of $\mlsdisk$ for the survey as a whole suffer only
from systematic errors as described for $\sddisk$. Typical errors in
$\mlsdisk$ for individual galaxies in our survey will be 32\% (ran)
and 30\% (sys), weakly dependent on band-pass. Dividing the sample
into quartiles will lower $\mlsdisk$ errors to 14\% (ran) and 6\%
(sys).

\subsection{Disk fraction, $\Fdisk$}

The disk mass fraction within some radius, $R$, is defined as
\begin{equation}
\left. \Fdisk(R) \equiv \frac{\vdiskstar}{\vc} \right|_R
\end{equation}
where $\vdiskstar$ and $\vc$ are the circular rotation speeds
associated with the disk stars and entire potential, respectively.
Operationally, we substitute $\vdiskstar = \sqrt{\fsd \mdiskstar \ {\rm
    G} / R}$, and adopt the disk gas tangential speed for $\vc$;
$\vc$, $\vdiskstar$ and $\mdiskstar$ are all functions of $R$. The
factor $\fsd$ accounts for the non-sphericity of the mass-distribution
in the potential symmetry-plane.  In general, $f_i$ for the $i^{th}$
mass-component can be a function of radius, but for purposes here we
assume the disk oblateness, and hence $\fsd$, is constant with radius.

$\mathcal{F}_{\rm *,disk}$ takes on the value $\Fdisk$ at $R = R_{\rm
  *,disk}^{\rm max}$ where the rotation curve from the stellar
disk component reaches a maximum, (i.e., $\vdiskstar =
\vdiskstarmax$). Using the numerical integration code of Casertano
(1983), we find that $R_{\rm *,disk}^{\rm max} \sim (2.2+1/q) \ h_R$
for an oblate, constant $\Upsilon$ disk with $q>4$. Based on equation
(1), 90\% of our sample with stellar kinematic measurements has $0.09
< 1/q_R < 0.18$. We continue to assume that $q = q_R$ is a good
statistical approximation.


We arrive at an estimate for $\fsd$ using our knowledge of the likely
range of disk oblateness.  Binney \& Tremaine (1987) illustrate that
an infinitely thin ($h_z/h_R=0$) exponential disk has a 15\% higher
peak velocity than its spherical counterpart
($\vdisk/\vsphere\sim1.15$). Using numerical integration, we find
$\vdisk/\vsphere \sim 1.16 - 0.3/q$ for $q>4$, illustrated in Figure
17. The mean disk oblateness of our sample is 0.14.  With a 25\%
uncertainty in disk oblateness for any individual galaxy, $\fsd =
1.242 \pm 0.023$. This is less than a 1\% contribution to the error
budget of $\fsd$.

Appendix B.3 contains the error formulation for $\Fdisk$ in terms of
the observational uncertainties associated with $\vc$ and $\mdiskstar$
(which, as argued in the previous section is proportional to
$\Delta \ln \sddisk$), and $\fsd$. In estimating errors on $\vc$ we
make the reasonable assumptions that velocity measurements of the gas
are made; asymmetric drift is negligible such that $\vc = \vrot$ is an
excellent approximation; $\vrot = (\vobs - \vsys) / \sin i$ along the
major axis, namely corrections for line-of-sight integration are
negligible in nearly face-on disks; and errors on $\vsys$ are
negligible. The terms are collected in Table 3. The inclination
dependence of errors in $\Fdisk$ are shown in Figure 16.  Even though
$\fsd$ is fundamentally related to $h_z/h_R$, it enters independently
in $\Fdisk$ in addition to the impact of $h_z/h_R$ directly on
$\sddisk$; uncertainty in $\fsd$ introduces systematic error for and
individual galaxy. We exclude errors on $R_{\rm *,disk}^{\rm max}$
because both the uncertainty in $h_R$ and range of $R_{\rm
*,disk}^{\rm max}/h_R$ are small, and $\Fdisk$ is relatively
insensitive to radius at $R = R_{\rm *,disk}^{\rm max}$ due both to
the flatness of $\vc$ and $\vdiskstar$ near this radius. Finally,
because there is just a single $\Fdisk$ measure per galaxy, we only
consider sample errors on this quantity. In this context, all error
terms are random with the exception of systematic errors in $H_0$
entering through dependence on $\sddisk$.  Typical $\Fdisk$ errors for
individual galaxies will be 19\% (ran) and 15\% (sys), reducing to 8\%
(ran) and 3\% (sys) for averages over 10 galaxies.

\subsection{Discussion: Baryon fraction, $\Fb$}

The baryon fraction is given by the ratio of the baryon mass to the
total dynamical mass within some radius: $\Fb \equiv \mbar^{\rm tot} /
\mtot$.  The total baryon mass, $\mbar^{\rm tot}$, is straightforward
to define given our estimate of the stellar disk mass, the atomic and
molecular mass, and reasonable estimates for the bulge mass. The
latter can be parameterized by the bulge-to-disk ratio, e.g., in the
$K$-band, with a correction for systematically different mass-to-light
ratios as determined by the disk and bulge colors and SPS models
calibrated directly by the survey.  For galaxies in our survey the
dominant term in $\mbar^{\rm tot}$ is the stellar disk mass, so in
essence the uncertainties in $\mbar^{\rm tot}$ are driven by
uncertainties in $\mlsldisk$.

However, the total mass requires reasonable estimates of the total
halo mass, which are difficult to make because of its extended nature
and the radial limits of our kinematic data. While \ion{H}{1}
measurements dramatically improve halo mass estimates, they are still
insufficient to constrain the total halo mass with any
confidence. This is poignantly illustrated by the analysis in
Verheijen (1997) of the Ursa Major galaxy sample.  Even ignoring an
isothermal halo (which has infinite mass), and considering only a
single, mathematical form for the halo density profile which yields a
finite and well-defined total mass\footnote{Here we consider a
  Hernquist (1990) profile.}  the estimated fraction of luminous to
dark matter ($\Fb$) varies significantly for individual galaxies
depending on how the dark-matter profile is constrained. Were we to
consider variants of the density profile form, the uncertainty on the
total mass for a single galaxy would increase further still.

To bring this point home more clearly, we have examined the
logarithmic derivative of changes in the estimated halo-mass
($\mhalo$) with changes in the baryon mass for a given galaxy. We
use the Ursa Major sample and fitting done by Verheijen (1997),
parameterizing the baryon mass with the $K$-band mass-to-light
ratio. Fits to the halo mass and scale were constrained by the
observed \ion{H}{1} rotation curve, the observed \ion{H}{1} mass
distribution, and a stellar mass distribution based on the $K$-band
luminosity profile and a choice for $\Upsilon_{*,K}$ (constant with
radius).  Three choices of $\Upsilon_{*,K}$ were used, corresponding
to the so-called ``maximum disk,'' ``Bottema disk,'' and an
intermediate value. Figure 17 shows the distribution of $\Delta \ln
\mhalo / \Delta \ln \Upsilon_{*,K}$ versus $\Delta \ln
\Upsilon_{*,K}$ for both high- and low-surface-brightness subsets of
the Ursa Major sample. While there is no clear differences between
these subsets, a trend of larger scatter with smaller $\Delta \ln
\Upsilon_{*,K}$ is evident.  It is reasonable to surmise that there
are uncertainties associated with the logarithmic derivative directly
related to $\Delta \ln \Upsilon_{*,K}$, i.e., the (log) size of the
interval in $\Upsilon_{*,K}$ used to measure the derivative. The
dotted lines in the plot show this error model, which looks plausible.
On this basis we calculate the mean and error in the mean for the
appropriately weighted distribution to find:
\begin{equation}
\Delta \ln \mhalo / \Delta \ln \Upsilon_{*,K} \ = \ 1.32 \pm 0.32,
\end{equation}
where $\mhalo$ is the total (dark) halo mass. These values are
consistent with the derivative being equal to unity. In other words,
$\Upsilon_{*,K}$ and total halo mass ($\mhalo$) are close to 1:1
covariant. What is happening physically with the model is that as
$\Upsilon_{*,K}$ increases the halo scale-radius is forced to
increase. The mathematical nature of the model is such that $\mhalo$
also increases with increasing scale-size to produce a `best fit.'
This is not unlike the conspiracy first noted by Bosma (1981) between
the mass of the \ion{H}{1} disk and that required to make the observed
rotation curve flat (explored more recently by, e.g., Hoekstra et
al. 2001, and references therein). Here, however, the conspiracy is
between the stellar and halo masses, and purely a result of the
fitting degeneracy.

This result emphasizes that whatever errors there are in
$\Upsilon_{*,K}$, $\Fb$ will remain constant in a statistical
sense. The nominal value of $\Fb$, however, will be set by the nature
of the rotation curve shape, the halo model, and the way it is fit.
All of these differences will lead to a wide range of $\Fb$ which are
dominated by systematic effects.  Of course, with a different halo
model, we may find a different result, but so too is the value of
$\Fb$ likely to change.  The primary point is that $\Fb$ is
conceptually unsatisfactory given what can be observed. While future
lensing estimates may make it possible to make a statistical
determination of total mass, for the moment a more robust
observational quantity is $\Fdisk$.

\section{SUMMARY AND CONCLUSIONS}

Our entire error-budget analysis of the DMS precision and accuracy is
summarized graphically in Figure 16, illustrating the trends of random
and systematic errors on $\sddisk$, $\mlskdisk$, and $\Fdisk$ with
disk inclination. We show results adopting the nominal parameters
given in the previous section, but differentiate results for kinematic
inclinations and inclinations inferred from $K$-band iTF.

The break-down of the errors, adopting kinematic inclinations at
$i=30^\circ$, are given in Tables 1-3.  We have divided the inventory
into three groups of random (ran) measurement errors, systematic
errors for which we correct (and in this case we are interested in the
estimated residual error to this correction, including it in our
random error budget), and finally the systematic (sys) errors for
which we cannot correct. The latter are an important delineation
because they represent systematic errors beyond the control of this
and other current astronomical experiments.  These systematic errors
are comparable to our random errors for individual galaxies, but are
five times lower than the random errors when we average results over a
subset of the survey. The reason for this is simply because many of
the systematics for individual galaxies are not systematic for the
sample as a whole.

Note that the dominant errors for all quantities of interest are the
SVE axial ratio $\alpha$ and inclination (random errors), and the disk
oblateness $q$ (systematic errors); they dominate the error on
$\sddisk$, and propagate to dominate the errors on $\sds$,
$\mlsdisk$, and even $\Fdisk$.  Random errors in $\alpha$
and inclination would need to be lowered (each) by a factor of 4.5 for
other terms to become appreciable (notably $\sobs$). Random
errors in $\sddisk$ would have to decreased four-fold for
extinction uncertainties to dominate the random errors in $\mlsdisk$
in the $B$ band. Even if our extinction uncertainties are optimistic,
they are inconsequential for the error budget for our reddest bands
sampling the stellar continuum. For $\Fdisk$, the
inclination errors associated with the deprojection of the rotation
speed are about 1.7 times lower than the contributions from errors in
$\sddisk$.

Systematic errors in all cases are independent of inclination. Random
errors, however, are strongly inclination dependent, and in general
increase with inclination. The one exception is the disk fraction
$\Fdisk$, which rises steeply at smaller inclinations below
30$^\circ$ if kinematic inclinations are adopted. In general, adopting
iTF inclinations leads to smaller random errors below $i=28^\circ$,
but the gains are only significant for individual galaxies, except for
$\Fdisk$.  The reason for the reduced error sensitivity of
ensemble $\sddisk$ and $\mlskdisk$ measurements to the
choice of inclination estimator is due to the additional errors terms
which, for individual galaxies are systematic, but for the ensemble
become random, and are independent of inclination.  Because of this
modest improvement using iTF for galaxy ensemble estimates of
$\sddisk$ and $\mlskdisk$, and because $\Fdisk$
is most relevant in the context of knowing independently where a
galaxy lies on the TF relation, we prefer kinematic
inclinations.  This, however, does not preclude the interesting use of
iTF for calibrating $\sddisk$ and $\mlsdisk$ at very low
inclination.

Our findings show that typical errors for individual galaxies in our
survey will be 28\% (ran) and 26\% (sys) for $\sddisk$; 32\%
(ran) and 30\% (sys) for $\mlsdisk$, weakly dependent on band-pass;
and 19\% (ran) and 15\% (sys) for $\Fdisk$. Dividing the
sample into quartiles (e.g., 10 galaxies binned by color, luminosity,
or surface-brightness) will yield errors of 12\% (ran) and 4\% (sys)
for $\sddisk$; 14\% (ran) and 6\% (sys) for $\mlsdisk$; and
8\% (ran) and 3\% (sys) for $\Fdisk$. These numbers compare
favorably to the DMS goal of achieving 30\% errors in $\mlsdisk$.
Hence the DiskMass Survey will be able to break the disk-halo
degeneracy for individual intermediate-type spiral galaxies in the
survey, and calibrate $\mlsdisk$ to sufficient accuracy and
precision so that SPS models can be used to break the disk-halo
degeneracy in other rotation-curve samples in the nearby universe and
at high redshift. This will open the door to measuring the detailed
shape of dark-matter halos and understanding the structure and
formation of galaxies.

\acknowledgements Research was supported under NSF/AST-9618849,
NSF/AST-997078, NSF/AST-0307417, AST-0607516 (M.A.B.); Spitzer
GO-30894 (RAS and MAB); NSF/OISE-0754437 (K.B.W.).  M.A.W.V. and T.M.
acknowledge the Leids Kerkhoven-Bosscha Fonds for travel support.
M.A.B. acknowledges support from the Wisconsin Alumni Research
Foundation Vilas Fellowship and the University of Wisconsin College of
Letters \& Science Ciriacks Faculty Fellowship.

\appendix

\section{Errors on Inclination}

\subsection{Velocity-field Fitting Estimates}

Based on a preliminary analysis of roughly half the Phase-A H$\alpha$
sample, we parameterize the errors from inclined-disk model fits to
the kinematic data, described in Paper I and Andersen (2001). We find
the error distribution is roughly log-linear: $$\log \ di = a + b
\times i,$$ with the slopes and intercepts given in Table A1. These
values form a lower envelope (best, or smallest inclination errors at
a given inclination); a mid-point to the main grouping of points
(mid); and upper envelope to this main grouping (worst); an outside
envelope to the very worst cases (extreme); and are rationalized to
cross at 82 deg at a value $di = 0.2$ deg. These values for $di$ can
then be used directly with the formulae below for the logarithmic
errors in $\mtot$ and $\sddisk$.

\subsection{Inverse Tully-Fisher Estimates}

The Tully-Fisher relation is parameterized as
$$ M_j = c_{1,j} + c_{2,j} \ \log (W_R / \sin \ i)$$ where $M_j$ is
the absolute magnitude in the $j^{th}$ band, $W_R$ is the line-width,
equivalent roughly to $2 \ \vobs$, and $c_{2,j}$ is the TF
slope. With inversion and differentiation, the inclination error in
radians is given by
\begin{equation}
\frac{di}{\tan i} = \sqrt{ (\Delta \ln \vobs)^2 +
\left(\frac{\ln 10}{c_{2,j}} dM_j \right)^2 }
\end{equation}
where $dM_j$ is the scatter in the TF relationship (magnitudes). This
quantity is constant with inclination. Expression for the total mass
and disk-mass error come from inserting equation A1 into equations B1
and B11 respectively.

For the data in this survey, the dominant contributor to the iTF
inclination error comes from the scatter in the TF relation
itself. The example in Figure 16 in Paper I is typical of the
H$\alpha$ data quality, but extreme in the sense that the galaxy is
very nearly face-one, i.e., under 3.5 deg inclination assuming an
intrinsic rotation velocity of 200 km s$^{-1}$ or greater.  The
velocity error is about 2\%, a little less than the contribution from
0.1 mag scatter and a $-9$ slope for the TF relation.  In general,
galaxies in our sample are at higher inclination and the TF scatter
dominates equation B11. In this case, for large ellipsoid errors
(e.g., 50\%), differences in $\Delta \ln \sddisk$ are dominated by the
SVE shape; TF slope and scatter are almost inconsequential in the
range 0.1 and 0.5 mag and slopes between $-5$ and $-9$, typical of
what is observed in the optical through near-infrared bands. For
smaller ellipsoid errors (e.g., 10\%), TF scatter dominates the
variation in $\Delta
\ln \sddisk$ with $i$ for $\Delta \ln \sddisk > 0.1$ (or
$i>30$ deg). At smaller values, TF scatter and SVE shape lead to
comparable variations in $\Delta \ln \sddisk$ at a given $i$. TF
slope has a much smaller effect.

\section{Error Formulae}

This Appendix serves three agendas: (1) to provide analytic
expressions for random (indicated as $[ran]$) and systematic
(indicated as $[sys]$) errors of key survey quantities for individual
survey galaxies; (2) in so doing to identify which terms contribute to
these two types of errors, and how this changes when
considering results for individual survey galaxies versus the survey
as a whole; and (3) to isolate the contribution from uncertainties in
disk inclination and the SVE. The latter serves to focus on quantities
central to the rationale behind a near face-on strategy for the
DMS. Error-functions are given as logarithmic derivatives for
disk-mass surface density ($\sddisk$), the disk stellar mass-to-light
ratio ($\mlsdisk$), the disk mass-fraction ($\Fdisk$), and total mass
($\mtot$).

\subsection{Errors on $\sddisk$}

$\sddisk$ (equation 9) is estimated via line-of-sight velocity
dispersion, corrected for instrumental effects and SVE projection
(equation 6), estimated disk oblateness (equation 1), apparent disk
scale-length, and distance. Deprojection requires information on
inclination. With distinctions made in \S 5 and \S 5.1, the logarithmic
derivative of $\langle\sddisk\rangle$ for disk-averaged measurements
of an individual galaxy can be separated into a random term,
\begin{eqnarray}
\lefteqn{(\Delta \ln \langle\sddisk\rangle)^2 \ [ran] \ = \ 0.4 (\Delta \ln h_R)^2 \ + } \nonumber \\
&& \frac{4}{N_{\rm disk}} \left\{
  \left[ \left(\frac{\sobs}{\slos}\right)^2 \Delta \ln \sobs \right]^2  
+ \left[ \left(\frac{\sbs}{\slos}\right)^2 \Delta \ln \sbs \right]^2 +
\right. \nonumber \\
&&  \left. \hspace{0.265in} \left[ \left(\frac{\sinst}{\slos}\right)^2 \Delta \ln \sinst \right]^2
  + (\Delta \ln \stm)^2 \right\} + \nonumber \\
&& \frac{\tan^4 i}{ \alpha^4 \gamma}^2 \left(
\left\{ \left[2\alpha^2\gamma 
+ (1+\beta^2) \sec^2 i \right] \frac{di}{\tan i} \right\}^2 
+ [ (1+\beta^2) \Delta \ln \alpha ]^2 
+ (\beta^2 \Delta \ln \beta)^2 \right) \hspace{0.5in} 
\end{eqnarray}
and a systematic term,
\begin{equation}
(\Delta \ln \langle\sddisk\rangle)^2 \ [sys] \ = \ (\Delta \ln k)^2 \ +
  \ (\Delta \ln q)^2 \ + \ 0.4 [ (\Delta \ln D_{\rm H0})^2 + (\Delta \ln D_{\rm flow})^2],
\end{equation}
where for the simplicity of notation, the quantities $\gamma$,
$\slos$, $\sobs$, $\sbs$, $\sinst$, and $\stm$ are all averages over
azimuth {\it and} radius (e.g., $\gamma$ in equation B1 represents
$\langle\bar{\gamma}\rangle$, where $\bar{\gamma}$ is the
azimuthally-averaged value defined by equation 5). Here, $di$ is
expressed in radians, and the logarithmic derivatives of individual
terms are characteristic values for an azimuthal ring at a single
radius. The errors on $\sddisk$ at a single radius for an individual
galaxy only differ by the removal of the $N_{\rm disk}^{-1}$ factor in
the middle set of terms in equation B1; $N_{\rm disk}^{-1}$ specifies
the number of radial rings of averaged (stacked) fibers. These terms
have different dependence on $\slos$ because of the way they enter
into the calculation of $\slos$. Suitable values for each of the above
terms are itemized in Table 1. Survey systematic errors for $\sddisk$
include only uncertainty distances due to H$_0$.

Error-dependence on inclination and SVE include only the last set of
terms in equation B1, which follow from setting $\Delta \ln
\langle\sddisk\rangle = \Delta \ln(\bar{\gamma} \cos^2 i)$.  Neither
$\gamma \cos^2 i$ or $\bar{\gamma}\cos^2 i$ are rapidly varying
functions of $i$; $\bar{\gamma}\cos^2 i$ is unity for all $i$ for
$\alpha=\beta=1$, varies by less than a factor of 2 for
$\alpha=\beta=0.7$, and by less than a factor of 4 for
$\alpha=\beta=0.4$. This fact, strangely at odds with the geometric
projection of $\sigz$, arises because we are relying on the SVE
shape to derive $\sigz$ from the line-of-sight measurement. In
contrast, the geometric projection of $\sigz$ can be written
logarithmically as $\sigz \cos i / \slos = 1 /
\sqrt{\bar{\gamma}}$.  The systematic error on deriving $\sigz$
from $\slos$ (expressed as a fraction) then must be
proportional to $\slos / \sigz \cos i - 1 =
\sqrt{\bar{\gamma}} - 1.$

\subsection{Errors on $\mlsldisk$}

$\mlsldisk$ (equation 10, \S 5.2) is estimated via the stellar disk
surface-density and the extinction-corrected flux density. The former
is a function of $\sddisk$, $\sdhi$, and a parameterization
of the molecular gas component. The logarithmic derivative of
$\langle\mlsldisk\rangle$ can be written as
\begin{equation}
(\Delta \ln \langle\mlsldisk\rangle)^2 = 
(\Delta \ln \langle I_\lambda^c\rangle)^2 +
(\Delta \ln \langle\sds\rangle)^2.
\end{equation}
The expression for the first term is
\begin{equation}
(\Delta \ln \langle I_\lambda^c\rangle)^2 \ = \ \frac{0.85}{N_{\rm disk}}
\left[ \langle \Delta \mu_\lambda\rangle^2 + 0.85 \ \langle\Delta {\rm A}_\lambda\rangle^2 \right],
\end{equation}
where $\langle\Delta \mu_\lambda\rangle$ and $\langle\Delta
{\rm A}_\lambda\rangle$ are the characteristic errors in a single ring for
the surface-brightness (mag arcsec$^{-2}$) and extinction correction
(mag), respectively. This is a suitable approximation for magnitude
errors below 1. The second term can be written as:
\begin{eqnarray}
  \lefteqn{(\Delta \ln \langle\sds\rangle)^2 \ = \ \left(\frac{\xi \ \Delta \ln \xi}{1-\xi}\right)^2 + } \nonumber \\
&& (1-\xi)^2 \left[ \left(\frac{\sddisk}{\sds} \ \Delta \ln \langle\sddisk\rangle\right)^2 + \left(\frac{1.4 \ \sdhi}{\sqrt{N_{\rm disk}} \ \sds} \ \langle\Delta \ln \sdhi\rangle\right)^2 \right], 
\end{eqnarray}
and
\begin{equation}
  (\Delta \ln \xi)^2 = (\Delta \ln \eta)^2 +
  \left(\frac{\xi}{1.4 \eta} \frac{\mdisk}{\mhi} \right)^2 
  \left[(\Delta \ln \mdisk)^2 +  (\Delta \ln \mhi)^2\right].
\end{equation}
where $\mhi, \mdiskstar,$ and $\mdisk$ are the \ion{H}{1}, stellar, and
total disk-mass components. Approximating $\Delta \ln \mathcal{M}_i =
N_i^{-1/2} \langle \Delta \ln \Sigma_i \rangle = \Delta \ln
\langle\Sigma_i\rangle$ for the $i^{th}$ mass-component measured in
$N_i$ radial rings, where $\langle \Delta \ln \Sigma_i \rangle$ is
the characteristic error in $\Sigma_i$ for a single ring, and $\Delta
\ln \langle\Sigma_i\rangle$ is the error in the mean value, we then
collect random and systematic error terms to obtain:
\begin{eqnarray}
(\Delta \ln \langle\sds\rangle)^2 [ran] & = &
\left[ \frac{1}{N_{\rm disk}}\left(1.4(1-\xi) \frac{\sdhi}{\sds}\right)^2 + \frac{1}{N_{\rm atom}}
            \left(\frac{\xi^2}{1.4 \eta (1-\xi)} \frac{\mdisk}{\mhi} \right)^2 \right] 
            \langle \Delta \ln \sdhi\rangle^2 \ + \nonumber \\
&& \left[ \left((1-\xi)\frac{\sddisk}{\sds}\right)^2 + 
            \left(\frac{\xi^2}{1.4 \eta (1-\xi)} \frac{\mdisk}{\mhi} \right)^2 \right]
            (\Delta \ln \langle\sddisk\rangle)^2[ran] \hspace{0.35in}
\end{eqnarray}
\begin{eqnarray}
(\Delta \ln \langle\sds\rangle)^2 [sys] & = & \left(\frac{\xi}{1-\xi}\right)^2(\Delta \ln \eta)^2 \ + \nonumber \\ 
&& \left[ \left((1-\xi)\frac{\sddisk}{\sds}\right)^2 + 
            \left(\frac{\xi^2}{1.4 \eta (1-\xi)} \frac{\mdisk}{\mhi} \right)^2 \right]
            (\Delta \ln \langle\sddisk\rangle)^2[sys] \hspace{0.35in}
\end{eqnarray}
We relate all ratios involving mass or surface-density to $\mdiskstar/\mhi$
with the identities of $\mdisk/\mhi = \mdiskstar/\mhi + 1.4 (1+\eta)$,
$\sddisk/\sds = (1-\xi)^{-1} + 1.4(\sdhi/\sds)$,
and $\sdhi/\sds \sim (\mhi R_*^2 / \mdiskstar R_{\rm HI}^2),$ where $R_*$
and $R_{\rm HI}$ are the characteristic length-scales of the stellar and
\ion{H}{1} distribution. Suitable values for each of the above terms are
discussed in \S 5.2, and itemized in Table 2.

Random errors in $\mlsldisk$ include the sum of equations (B4) and
(B7); systematic errors are equal to equation (B8). Survey systematic
errors and inclination dependence for $\mlsldisk$, come from terms in
equation (B8) with $\sddisk[sys]$, as noted respectively in the
previous section for equation (B2).

\subsection{Errors on $\Fdisk$ and $\mtot$}

$\Fdisk$ is estimated via the mass of the stellar disk, it's
oblateness, and the projected total circular rotation speed
($\vobs-\vsys$; see \S 5.3), corrected for inclination. Using the
results in the previous section we find:
\begin{equation}
(\Delta \ln \Fdisk)^2 [ran] = \frac{1}{4} (\Delta \ln \langle\sds\rangle)^2[ran] \ +  \
(\Delta \ln \vobs)^2 \ + \ (di/\tan i)^2
\end{equation}
\begin{equation}
(\Delta \ln \Fdisk)^2 [sys] = \frac{1}{4} 
\left[  (\Delta \ln \fsd)^2 + (\Delta \ln \langle\sds\rangle)^2[sys] \right]
\end{equation}
We subsume the errors in the kinematic center and position angle of a
galaxy in the errors for $\vobs$ and inclination, and ignore the
negligible errors on $\vsys$. The last term in (B9) is roughly $2 \
di/i$ for small $i$, where $di$ is expressed in units of radians.  The
only systematic errors in this quantity for the survey ensemble stem
from distance, coupled to $\sds$ via $\sddisk$ (B2,B8).
Inclination-dependence to the random errors are from the last term in
equation B9 and terms in B1 related to $\sds$.

Errors in the total potential mass at a given radius are:
\begin{equation}
(\Delta \ln \mtot)^2 [ran] = 4 \left( \Delta \ln \vobs)^2 \ + \ (di/\tan i)^2 \right)
\end{equation}
\begin{equation}
(\Delta \ln \mtot)^2 [sys] = (\Delta \ln f_{\rm tot})^2
\end{equation}
where $f_{\rm tot}$ is a measure of the flattening of the
potential. Inclination-dependence only arises from the deprojections
of the observed tangential velocity $\vobs$.

\clearpage

\begin{deluxetable}{llcccc}
\tabletypesize{\scriptsize}
\tablewidth{0pt}
\tablenum{1} 
\tablecaption{$\sddisk$ Error Budget for Individual Galaxies at $i=30^\circ$} 
\tablehead{
\colhead{Quantity} & 
\colhead{Type} & 
\colhead{Section} & 
\colhead{Equation} &
\multicolumn{2}{c}{Log Error} \\ \cline{5-6}
\colhead{} & 
\colhead{} & 
\colhead{} & 
\colhead{} & 
\colhead{Quantity} &
\colhead{$\sddisk$}
} 
\startdata
\multicolumn{6}{c}{random} \\ \hline
$\sobs$ \dotfill\                            & measurement           & 3.3  & B1 & 0.030   & 0.027 \\ 
template mismatch \dotfill\                         & measured correction   & 3.4  & B1 & 0.040   & 0.036 \\ 
instrumental resolution \dotfill\                   & measured correction   & 3.5.1& B1 & 0.030   & 0.013 \\ 
beam smearing \dotfill\                             & measured correction   & 3.5.2& B1 & 0.500   & 0.001 \\ 
SVE deprojection, $\alpha$ \dotfill\                & measured correction   & 3.5.4& B1 & 0.250   & 0.204 \\ 
SVE deprojection, $\beta$ \dotfill\                 & measured correction   & 3.5.4& B1 & 0.050   & 0.013 \\ 
SVE deprojection, $i$ \dotfill\                     & measured correction   & 3.5.4& B1 & 0.116   & 0.184 \\ 
total SVE deprojection, $\sqrt{\bar{\gamma}}\cos i$ & measured correction   & 3.5.3& B1 & 0.138   & 0.275 \\ 
$h_R$ \dotfill\                                     & measurement           & 2.2  & B1 & 0.030   & 0.019 \\ 
$\sddisk$ \dotfill\                           & total random error    & 5.1  & B1 & $\cdots$& 0.280 \\ \hline 
\multicolumn{6}{c}{systematic} \\ \hline 
$h_R:h_z$ conversion, $q_R$ \dotfill\               & estimated correction& 2.2  & B2   & 0.250   & 0.250 \\ 
distance, $D_{\rm flow}$ \dotfill\                      & uncorrected systematic& 4    & B2 & 0.080   & 0.051 \\ 
distance, $H_0$ \dotfill\                           & uncorrected systematic& 4    & B2 & 0.070   & 0.044 \\ 
vertical distribution, $k$ \dotfill\                & uncorrected systematic& 5.1  & B2 & (0.140) &(0.140)\\ 
$\sddisk$ \dotfill\                           & total systematic error& 5.1  & B2 & $\cdots$& 0.259 \\ 
\enddata 
\tablecomments{() = excluded from final tally.}
\end{deluxetable}

\clearpage

\begin{deluxetable}{llcccc}
\tabletypesize{\scriptsize}
\tablewidth{0pt}
\tablenum{2} 
\tablecaption{$\sds$ and $\mlsdisk$ Error Budget for Individual Galaxies at $i=30^\circ$} 
\tablehead{
\colhead{Quantity} &
\colhead{Type} &
\colhead{Section} &
\colhead{Equation} &
\multicolumn{2}{c}{Log Error} \\ \cline{5-6}
\colhead{} &
\colhead{} &
\colhead{} &
\colhead{} &
\colhead{Quantity} &
\colhead{$\sds, \mlsdisk$}
} 
\startdata
\multicolumn{6}{c}{random} \\ \hline
$\sdhi$ \dotfill\                        & measurement           & 5.2 & B7 & 0.030  & 0.002 \\
$\sddisk$ \dotfill\                      & corrected measurement & 5.2 & B7 & 0.280  & 0.321 \\ 
$\sds$ \dotfill\                         & total random error    & 5.2 & B7 &$\cdots$& 0.321 \\ 
$\mu_\lambda$ \dotfill\                  & measurement           & 5.2 & B4 & 0.050  & 0.022 \\ 
internal extinction, A$_B$ \dotfill\     & estimated correction  & 5.2 & B4 & 0.180  & 0.074 \\ 
internal extinction, A$_I$ \dotfill\     & estimated correction  & 5.2 & B4 & 0.060  & 0.025 \\ 
internal extinction, A$_K$ \dotfill\     & estimated correction  & 5.2 & B4 & 0.008  & 0.003 \\ 
$\Upsilon_{*,B}^{\rm disk}$ \dotfill\    & total random error    & 5.2 & B3 &$\cdots$& 0.331 \\ 
$\Upsilon_{*,I}^{\rm disk}$ \dotfill\    & total random error    & 5.2 & B3 &$\cdots$& 0.323 \\ 
$\Upsilon_{*,K}^{\rm disk}$ \dotfill\    & total random error    & 5.2 & B3 &$\cdots$& 0.322 \\ \hline
\multicolumn{6}{c}{systematic} \\ \hline
molecular:atomic mass ratio, $\eta$ \dotfill\ & estimated correction  & 5.2 & B8 & 0.400  & 0.040 \\ 
$\sddisk$ \dotfill\                      &uncorrected measurement& 5.2 & B8 & 0.259  & 0.298 \\ 
$\sds$ and $\mlsdisk$ \dotfill\          & total systematic error& 5.2 &B8,B3&$\cdots$& 0.300 \\ 
\enddata 
\end{deluxetable}

\clearpage

\begin{deluxetable}{llcccc}
\tabletypesize{\scriptsize}
\tablewidth{0pt}
\tablenum{3} 
\tablecaption{$\Fdisk$ Error Budget for Individual Galaxies at $i=30^\circ$} 
\tablehead{
\colhead{Quantity} &
\colhead{Type} &
\colhead{Section} &
\colhead{Equation} &
\multicolumn{2}{c}{Log Error} \\ \cline{5-6}
\colhead{} &
\colhead{} &
\colhead{} &
\colhead{} &
\colhead{Quantity} &
\colhead{$\Fdisk$}
} 
\startdata
\multicolumn{6}{c}{random} \\ \hline
$\sds$ \dotfill\                     & corrected measurement & 5.2  & B9 & 0.321  & 0.161 \\
$\vobs$ \dotfill\                      & measurement           & 5.3  & B9 & 0.020  & 0.020 \\
velocity deprojection, $i$ \dotfill\     & measured correction   & 5.3  & B9 & 0.116  & 0.105 \\
$\Fdisk$ \dotfill\             & total random error    & 5.3  & B9 &$\cdots$& 0.193 \\ \hline 
\multicolumn{6}{c}{systematic} \\ \hline
$\sds$ \dotfill\                     &uncorrected measurement& 5.2  & B10& 0.300  & 0.150 \\
disk oblateness, $\fsd$ \dotfill\  & estimated correction  & 5.3  & B10& 0.019  & 0.009 \\
$\Fdisk$ \dotfill\             & total systematic error& 5.3  & B10&$\cdots$& 0.150 \\ \hline
\enddata 
\end{deluxetable}

\clearpage

\begin{deluxetable}{lcc}
\tablenum{A1}
\tablewidth{0pt}
\tablecaption{Kinematic-Inclination Error Coefficients}
\tablehead{
\colhead{} &
\colhead{a} &
\colhead{b}
}
\startdata
best    & 0.84 & -0.019 \nl
mid     & 1.26 & -0.024 \nl
worst   & 1.74 & -0.030 \nl
extreme & 2.49 & -0.039 \nl
\enddata
\end{deluxetable}

\clearpage

\begin{figure}
\figurenum{1}
\epsscale{1.0}
\plotone{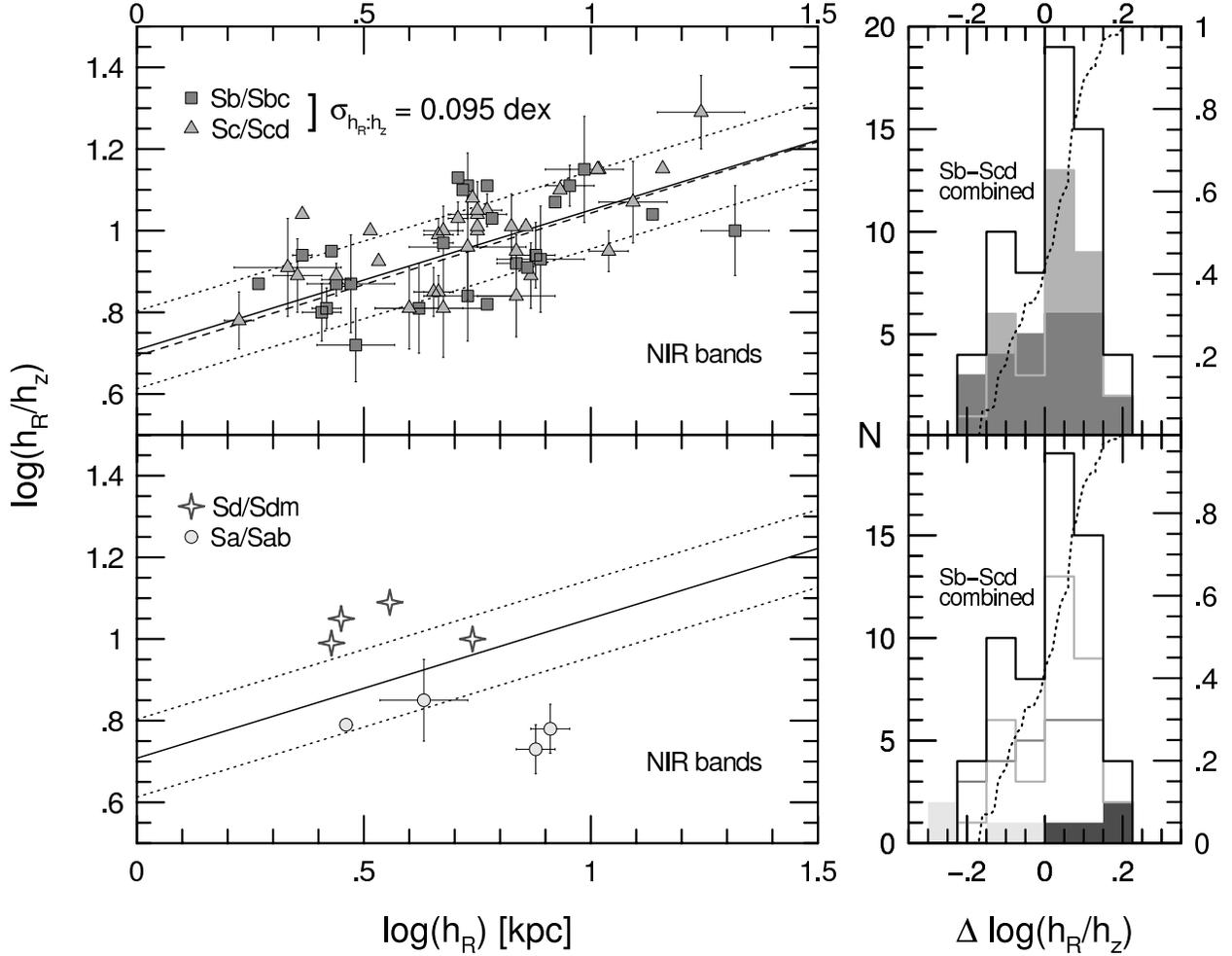}
\caption{Radial to vertical scale-length ratio ($h_R/h_z$)
  distribution in the $I$ or near-infrared bands for edge-on spiral
  galaxies.  Top panels show intermediate types while bottom panels
  show early and late types based on a compilation described in the
  text.  Left panels show the distribution versus radial scale length
  ($h_R$). The solid line represents a linear least-squares fit (no
  rejection) to the Kregel et al. sample for Sb-Scd Hubble-types,
  typical of the DMS. Dotted lines are the 1$\sigma$ dispersion about
  these fits, corresponding to 0.095 dex or 25\% random uncertainty in
  $h_z$ {\it per galaxy} based on measured radial scale-lengths. The
  dashed line is a weighted regression with intrinsic scatter
  (Akritas \& Bershady 1996) to the same subset, yielding comparable
  results in slope and dispersion. Right panels are histograms of
  deviations for all galaxies about the best-fit to the Kregel et
  al. sample, broken down by type. Black solid and dotted lines
  represent differential and normalized cumlative distributions for
  the 60 galaxies typed between Sb and Scd. Intermediate light- and
  dark-gray histograms (top) and extreme light- and dark-gray
  historgram (bottom) are the differential distributions for Sc/Scd,
  Sb/Sbc, Sd/Sdm, and Sa/Sab types respectively.}
\end{figure}

\clearpage

\begin{figure}
\figurenum{2}
\epsscale{1.0}
\plotone{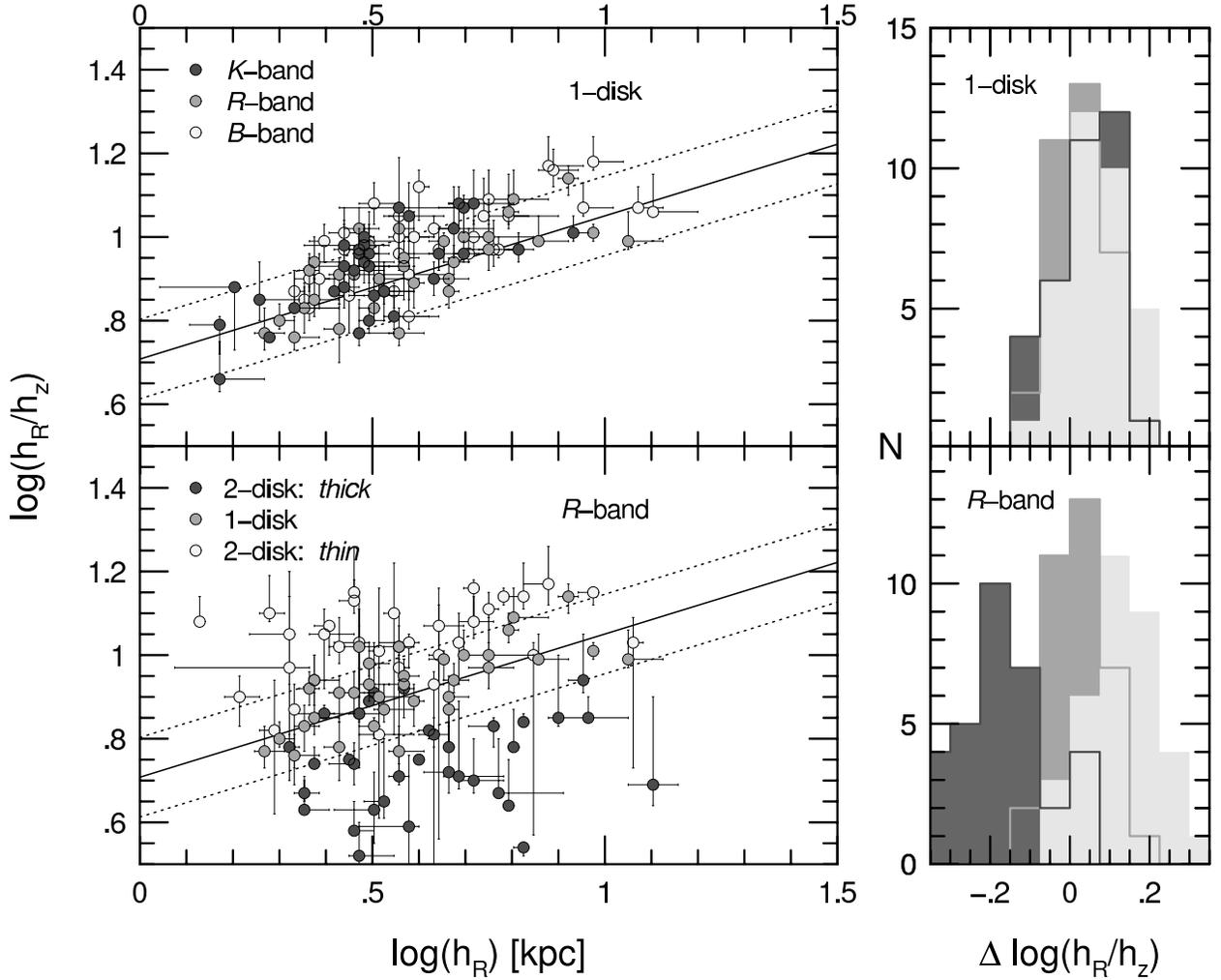}
\caption{Radial to vertical scale-length ratio ($h_R/h_z$)
  distribution for edge-on spiral galaxies from Yoachim \& Dalcanton
  (2006). Top panels show the distributions as measured for $B$, $R$,
  and $K$ bands assuming a single-component disk, as coded in the
  legend. Bottom panels show the distributions as measured in the $R$
  band for a single-component disk and a two-component disk (thick
  and thin), as coded in the legend.  Solid and dotted lines are the
  same as in Figure 1.  Right panels are histograms of deviations for
  all galaxies about the best-fit to the Kregel et al. sample, broken
  down by band (top) or disk component (bottom).}
\end{figure}

\clearpage

\begin{figure}
\figurenum{3}
\epsscale{1.0}
\plotone{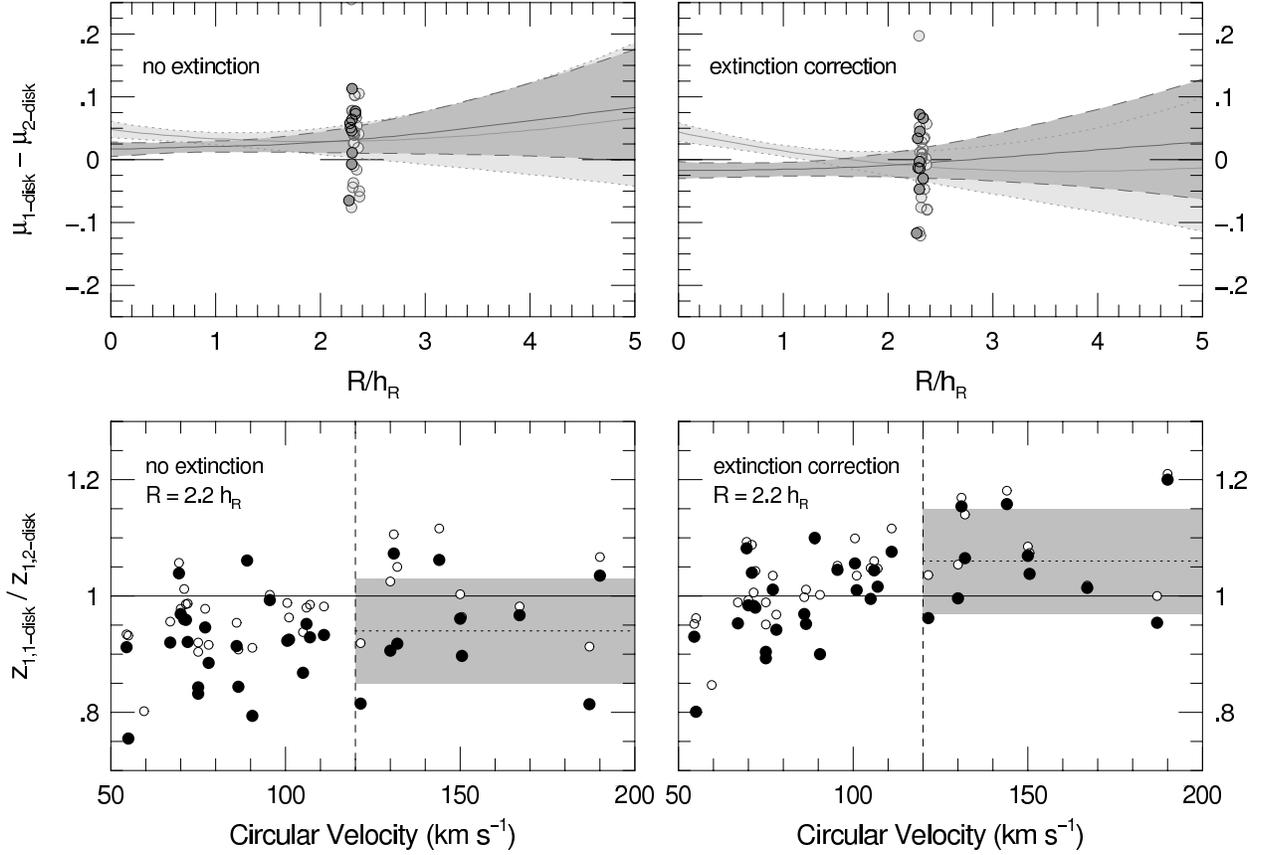}

\caption{Comparison of light-profiles for one and two-component disk
  fits. Top panels: Face-on surface-brightness difference between one
  and two-component disk fits from YD, renormalized as described in
  the text. Dark-grey region represents weighted mean and errors for
  fast-rotators ($\vrot>120$ km s$^{-1}$); light grey areas represent
  the same for slow-rotators ($\vrot<120$ km s$^{-1}$). Individual
  fits near R/h$_R$ = 2.2 are shown as shaded circles; their
  dispersion is large because of measurement error. Bottom panels:
  Ratio of the first-moment of the vertical light-profile for one and
  two-component models at R/h$_R$ = 2.2. Open circles represent moment
  integrals taken out to $z = 6h_z$ of the one-component disk; filled
  circles represent moment integrals extended to convergence. Gray
  shaded areas and dotted horizontal lines represent standard
  deviation and mean for the fast-rotators. Left-hand panels are for
  fits without correction for extinction.  Right-hand panels adopt an
  extinction model for the thin-disk component, as prescribed by YD.}
\end{figure}

\clearpage

\begin{figure}
\figurenum{4}
\epsscale{1.0}
\plotone{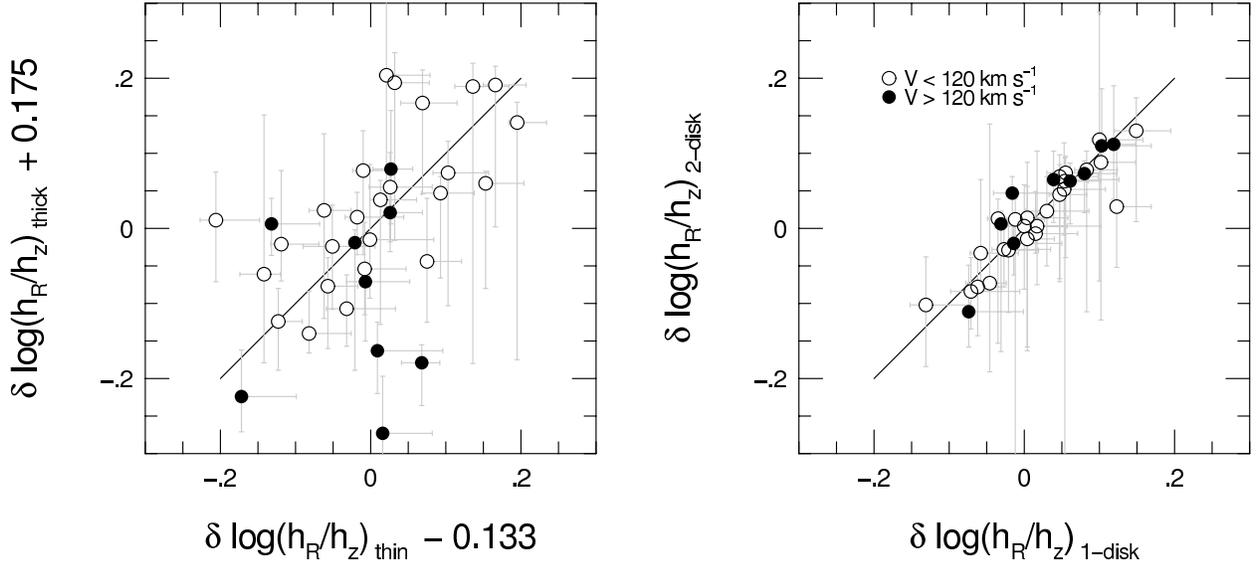}
\caption{Comparison of deviations from mean oblateness--radial
  scale-length relationship for one and two-component disk fits to the
  YD sample. The left panel compares thin and thick disk oblateness
  deviations for the two-component model, where the mean relationship
  between oblateness and disk radial scale-length is taken to be
  equation (1) with zeropoint-adjustments giving zero mean
  deviation. The right panel compares deviations for the one and
  two-component disk oblateness, where the two-component value is the
  light-weighted sum of the deviations for the two components. There
  is no zeropoint adjustment with respect to equation (1). Solid and
  open symbols are for fast and slow-rotators, as keyed in the
  figure.}
\end{figure}

\clearpage

\begin{figure}
\figurenum{5}
\epsscale{1}
\plotone{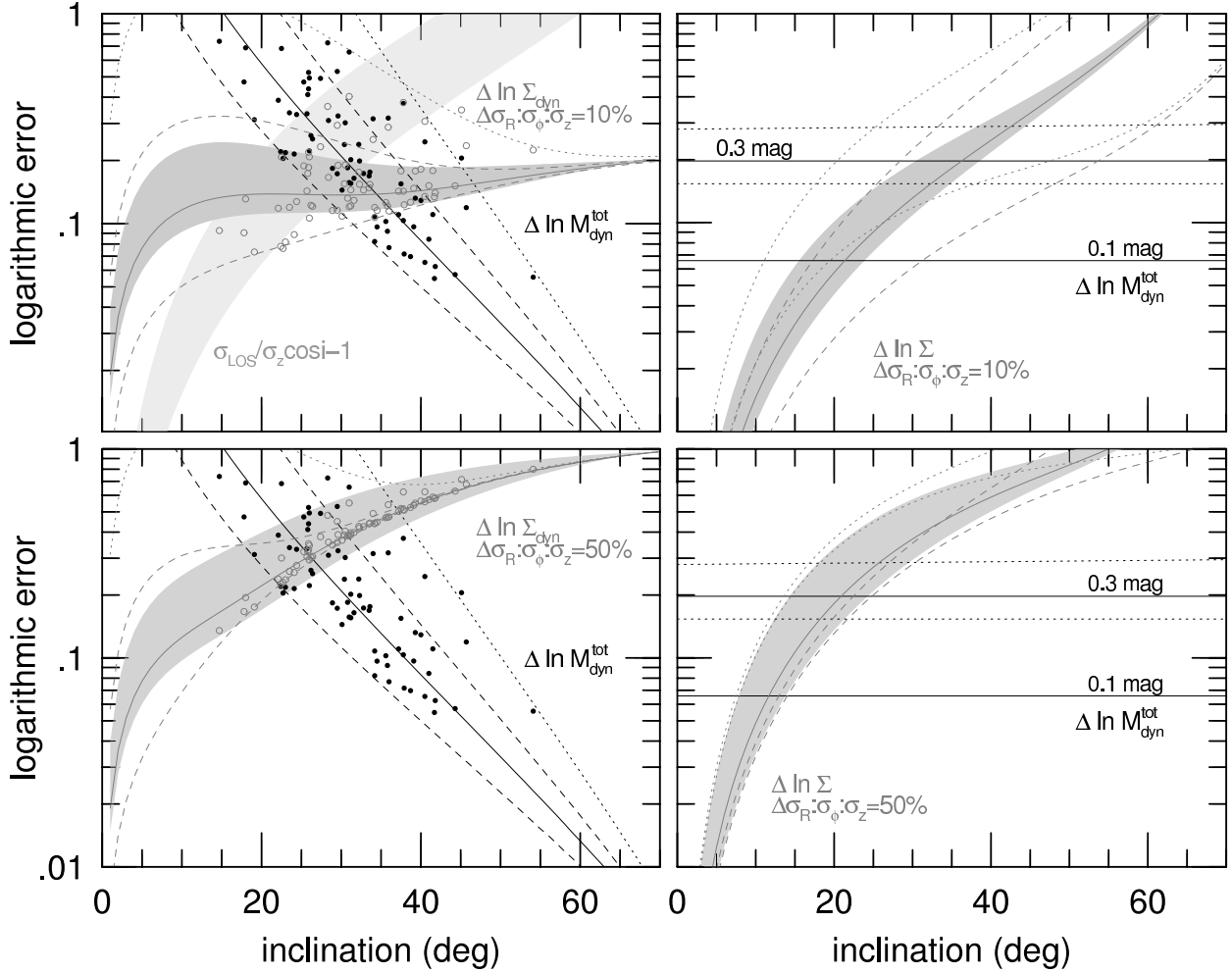}
\caption{Partial DMS error-budget, expressed as logarithmic
  derivatives, illustrating tradeoffs with inclination and SVE
  decomposition. Left-hand panels adopt kinematic inclinations and
  inclination errors; right-hand panels adopt iTF inclinations and
  errors for a range of slopes and TF-scatter (mag).  Top panels
  assume the SVE shape is known to 10\%; bottom panels
  assume a 50\% precision. In all panels, black curves, horizontal
  lines, and filled points represent total-mass errors ($\Delta \ln
  \mtot$); dark-gray open points, curves and areas represent
  disk-mass errors ($\Delta \ln \sddisk$). Light-gray area in top-left
  panel represents the fraction of the observed velocity dispersion
  ($\slos$) to the projected vertical component ($\sigz \cos
  i$) in a form proportional to the expected systematic error in
  deriving $\sigz$ from $\slos$.  See text for details.}
\end{figure}

\clearpage

\begin{figure}
\figurenum{6}
\epsscale{1}
\plotone{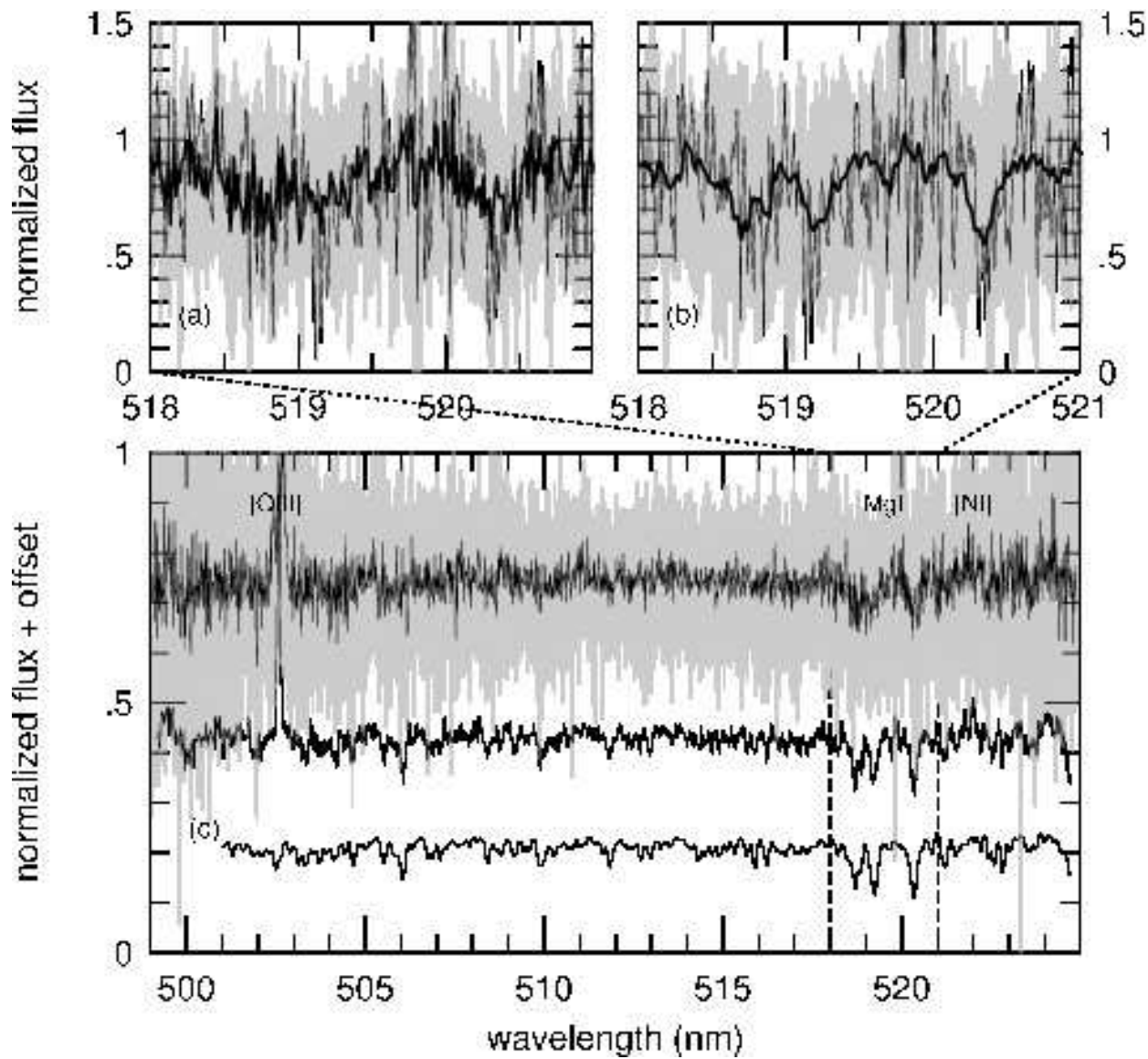}
\caption{Fiber stacking in the \ion{Mg}{1b}-region for SparsePak ``ring 4''
  observations of UGC 6918 (see text) in wavelength space for (a)
  un-registered spectra and (b) velocity-registered spectra -- both
  over the redshifted window containing only the \ion{Mg}{1b}-triplet. Spectra
  are continuum-normalized to unity. Gray lines represent individual
  fiber spectra, with one fiber highlighted by a thin black line; the
  thick, black line represents the averaged spectrum. The full spectra
  are shown in (c) from top to bottom: the un-registered spectra and
  averaged spectrum, the velocity-registered spectrum, and the
  broadened, K1~III template used to register the galaxy
  spectra. Spectra are offset in flux for clarity. The location of the
  \ion{Mg}{1b}-triplet absorption and nebular emission from
  [\ion{O}{3}]$\lambda$5007 and the [\ion{N}{1}]$\lambda\lambda$5198,5200 doublet
  are indicated. Many weak Fe and Ti absoption lines are also visible
  (see Figure 9).}
\end{figure}

\clearpage

\begin{figure}
\figurenum{7}
\epsscale{1}
\plotone{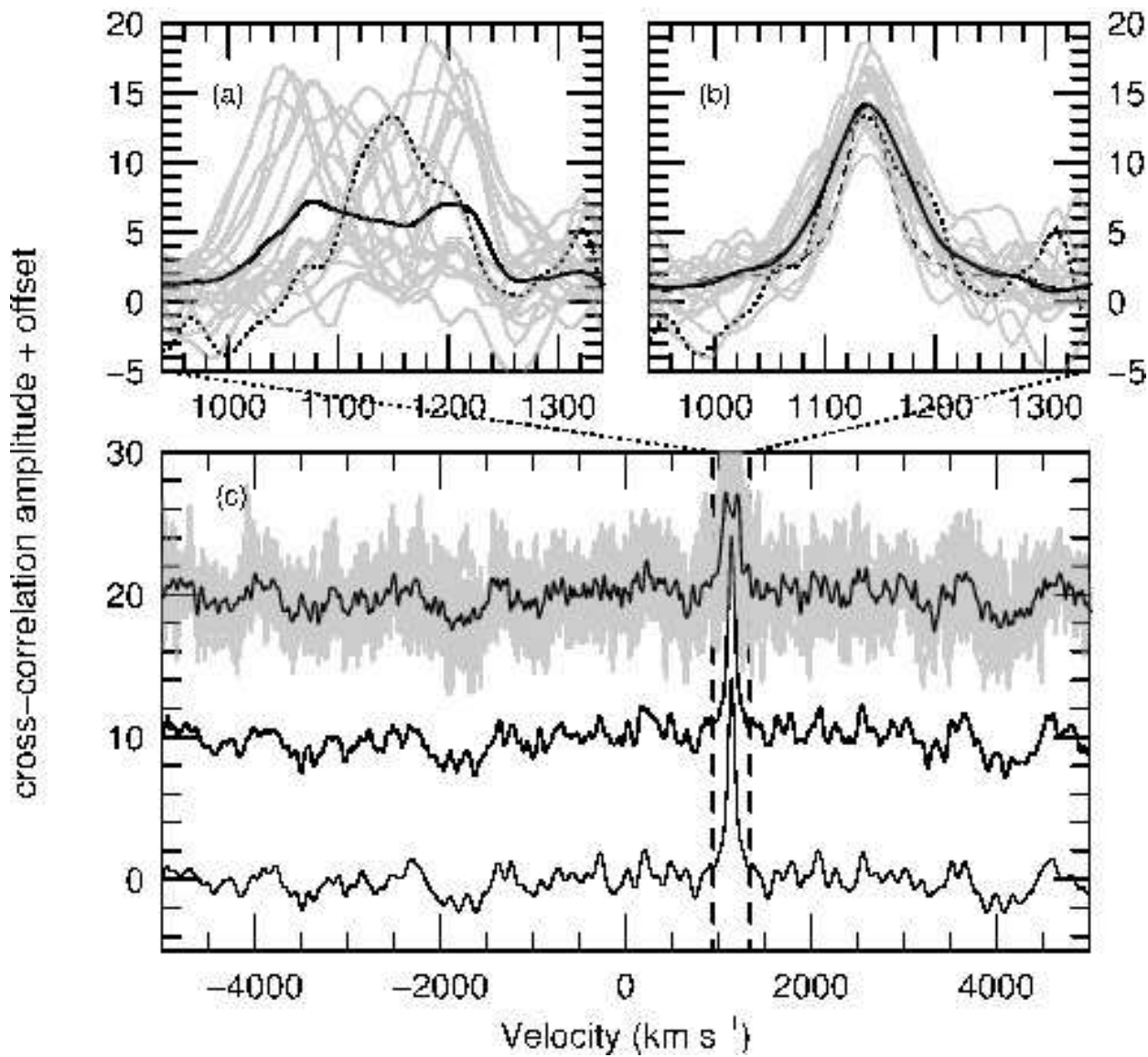}
\caption{Fiber stacking in the \ion{Mg}{1b}-region for SparsePak ``ring 4''
  observations of UGC 6918 in velocity space for (a) un-registered
  cross-correlations and (b) velocity-registered cross-correlations --
  both within $\pm200$ km s$^{-1}$ of the galaxy recession velocity.
  The cross-correlation template is the un-broadened K1~III star shown
  broadened in Figure 6. Gray lines represent individual fiber
  correlations, with one fiber highlighted by a dotted, black line, and
  the averaged spectrum correlation shown as a thick, solid line.
  Thin and dashed black lines in (b) represent the broadened-template
  correlation and the un-broadened template auto-correlation,
  respectively. Cross-correlations between $\pm 5000$ km s$^{-1}$ are
  shown in (c) from top to bottom: the un-registered spectrum, the
  velocity-registered spectrum, and the broadened, K1~III template
  used to register the spectra. Cross-correlations are offset in
  amplitude in panel (c) for clarity.}
\end{figure}

\clearpage

\begin{figure}
\figurenum{8}
\epsscale{0.8}
\plotone{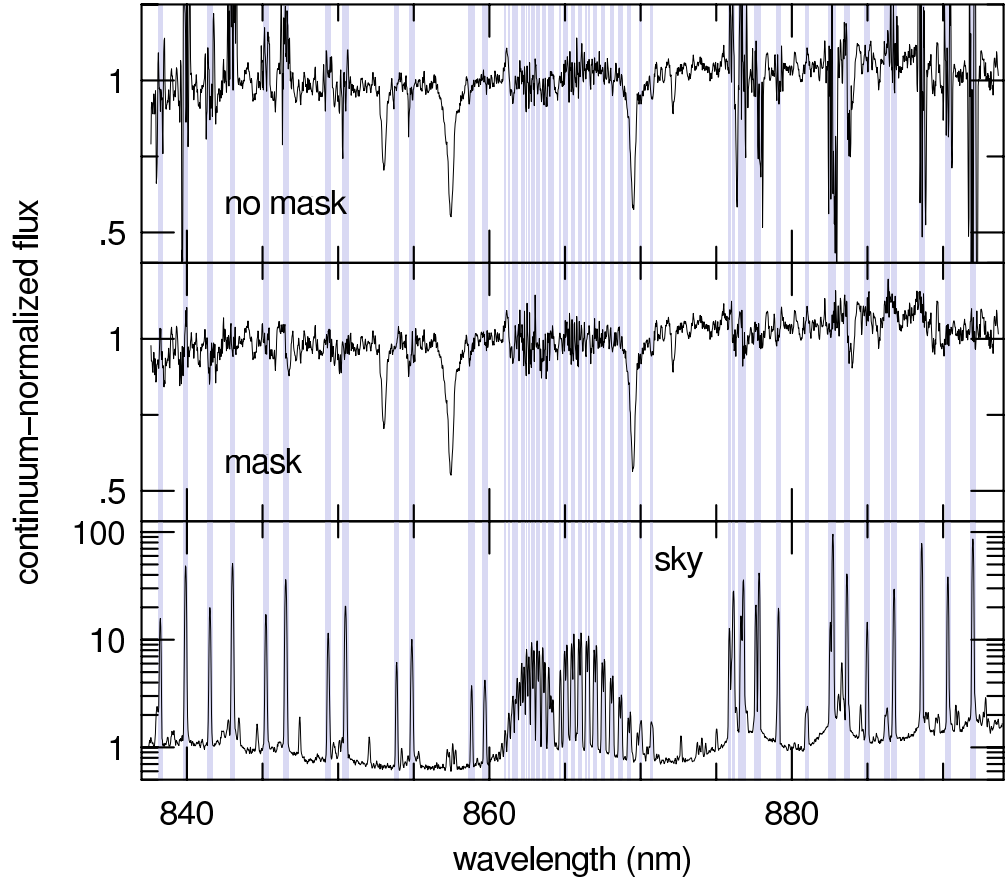}
\caption{Fiber stacking in the \ion{Ca}{2}-triplet region for SparsePak
  ``ring 3'' observations of UGC 6918, with an internal
  projected velocity spread of 153 km s$^{-1}$. The top panel shows the
  velocity-registered co-addition of 12 fiber spectra without masking
  bright sky-line regions. The middle panel has the same registration, but
  is co-added with masking. The bottom panel shows sky spectrum from the
  same data. Fluxes are normalized to the mean continuum level;
  sky-continuum and galaxy continuum at this radius are
  comparable. Gray-shaded regions indicates masks. Only the brighest
  lines have been masked (see text).}
\end{figure}

\clearpage

\begin{figure}
\figurenum{9}
\epsscale{1.0}
\plotone{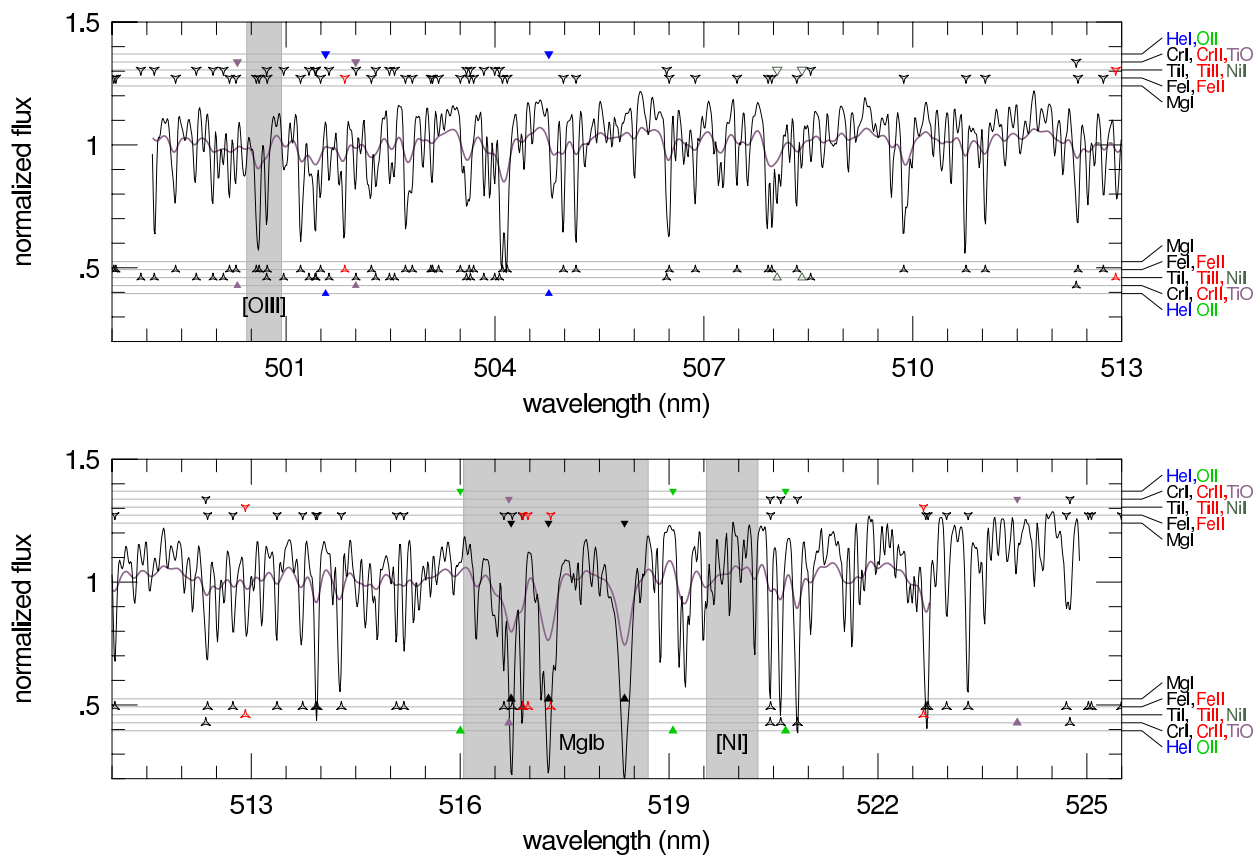}
\caption{HR 6817 (K1~III) template star observed at R = 11,750
  ($\sinst = 10.8$ km s$^{-1}$, thin black line), and smoothed to the
  measured broadening of the galaxy spectrum in Figure 6 ($\slos =
  28.9$ km s$^{-1}$; thick line). Lines contributing to absorption in
  this spectrum are marked above and below in four tiers (inside
  out): \ion{Mg}{1b} triplet (filled triangles); \ion{Fe}{1} (black
  open deltoids) and \ion{Fe}{2} (red open deltoids); \ion{Ti}{1}
  (black open deltoids), \ion{Ti}{2} (red open deltoids),
  and \ion{Ni}{1} (gray open triangles); \ion{Cr}{1} (black open
  deltoids), \ion{Cr}{2} (red open deltoids), and TiO (gray filled
  triangles). For reference a last outer tier marks HeI (blue filled
  triangles) and OII (green filled triangles) absorption features in
  hot stars. Line identifications are taken from the ILLSS Catalogue
  (Coluzzi 1993). Vertical shaded regions mark the ``\ion{Mg}{1b}''
  subregion, and the masked regions around [\ion{O}{3}]$\lambda$5007
  and [\ion{N}{1}]$\lambda\lambda$5198,5200 nebular lines. [COLOR
  FIGURE FOR ELECTRONIC EDITION ONLY.]}
\end{figure}

\clearpage

\begin{figure}
\figurenum{10a}
\epsscale{1.0}
\plotone{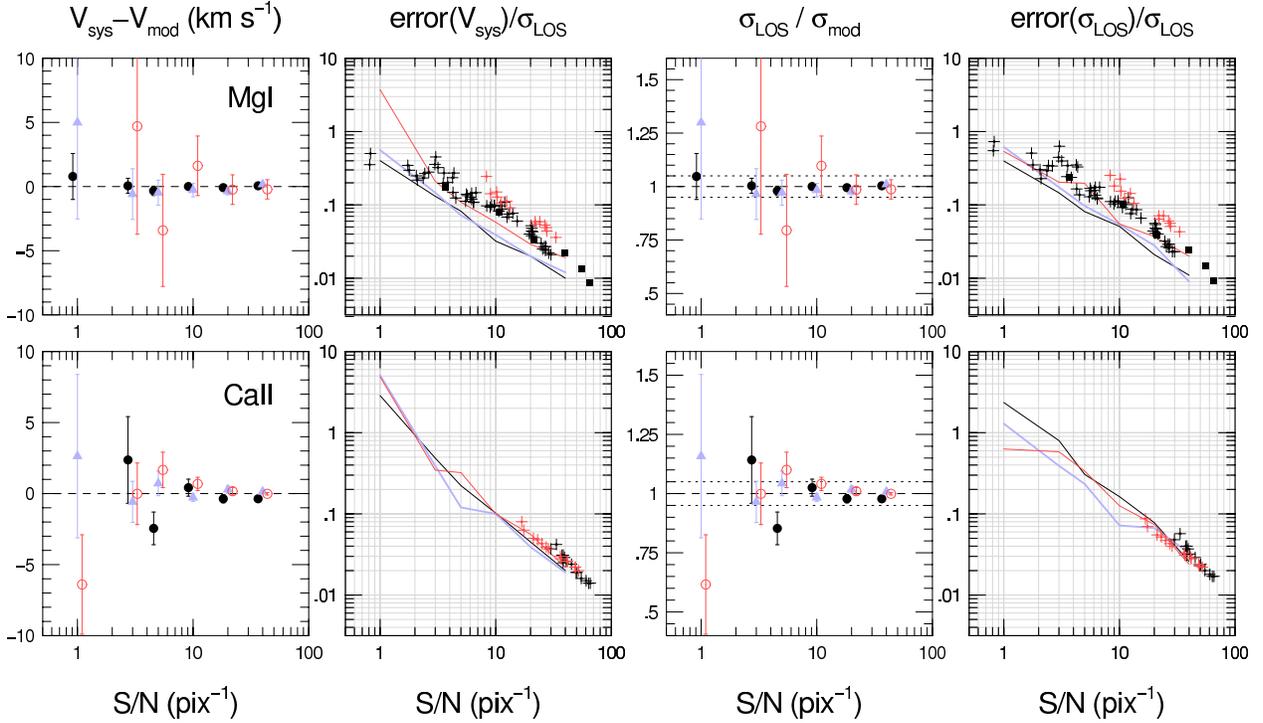}
\caption{Simulations and measurements of random errors in velocity
  ($\vsys$) and velocity dispersion ($\slos$) versus
  spectral-continuum S/N (per pixel) in the \ion{Mg}{1b} region (top
  row) and \ion{Ca}{2}-triplet region (bottom row). Measurements of
  simulated galaxy spectra (symbols in first and third column from
  left; lines in second and fourth columns from left), use the same
  methods applied to observed galaxy spectra, and are referenced to
  simulation-model values ($V_{\rm mod}$, $\sigma_{\rm mod}$).
  Simulatios use a K1~III template. Colors and symbols indicate
  ($V_{\rm mod}$, $\sigma_{\rm mod}$) in km s$^{-1}$: black and filled
  circles (1110, 20); blue or light gray and filled triangles (1110,
  60); and red or medium gray and open circles (2350, 180). The first
  two cases bracket the observed range for spiral galaxy UGC 6918; the
  latter characterizes the elliptical UGC 11356.  Note the errors in
  $\vsys$ (second column) are normalized by the {\it measured}
  velocity dispersions, $\slos$, as are the errors in $\slos$ (fourth
  column); the latter is therefore equivalent to $\Delta \ln \slos$.
  Horizontal dashed and dotted lines in first and third columns are
  for reference.  Cross-correlation measurements using a K1~III
  tempate for SparsePak fibers sampling UGC 6918 (black symbols) and
  UGC 11356 (red or medium gray symbols) are shown in the second and
  fourth columns. Individual fiber measurements are shown as pluses;
  stacks of fibers are shown as filled squares. [COLOR FIGURE FOR
  ELECTRONIC EDITION ONLY.]}
\end{figure}

\clearpage

\begin{figure}
\figurenum{10b}
\epsscale{1.0}
\plotone{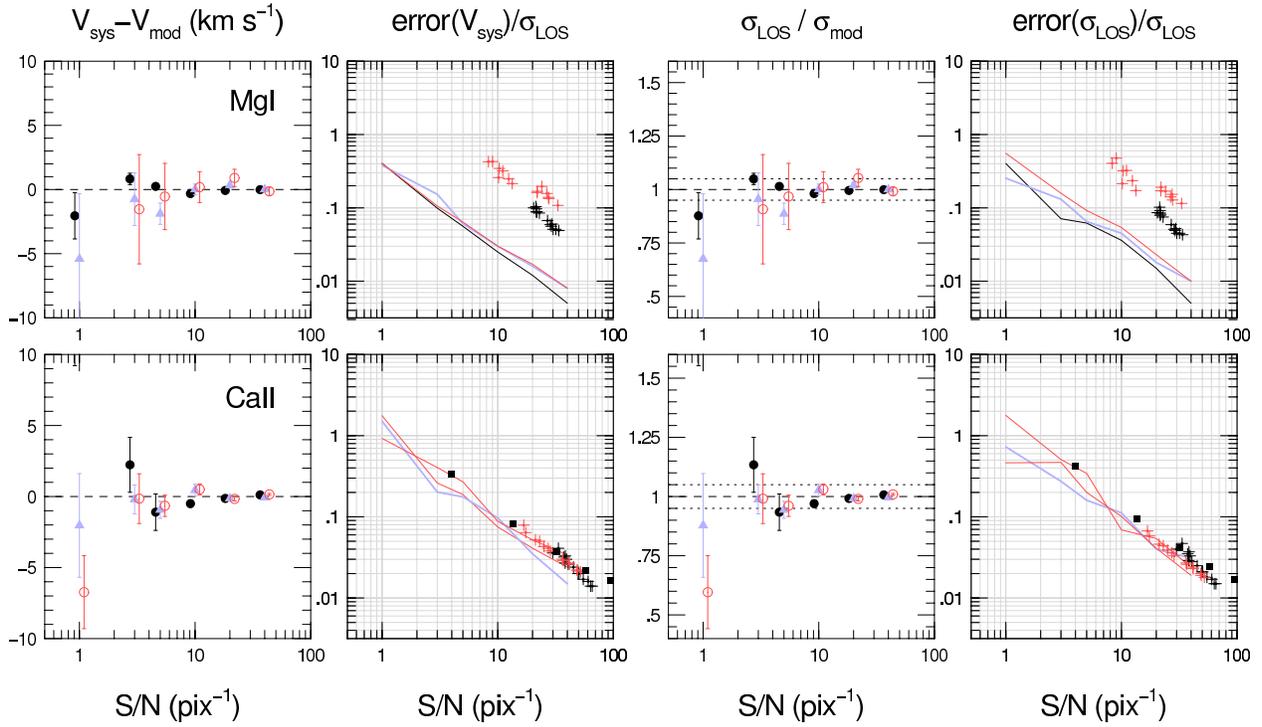}
\caption{The same as 10a, except using an M3~III template for
  simulations and cross-corretions with UGC 6918 and UGC 11356. [COLOR
  FIGURE FOR ELECTRONIC EDITION ONLY.]}
\end{figure}

\clearpage

\begin{figure}
\figurenum{11}
\epsscale{1.0}
\plotone{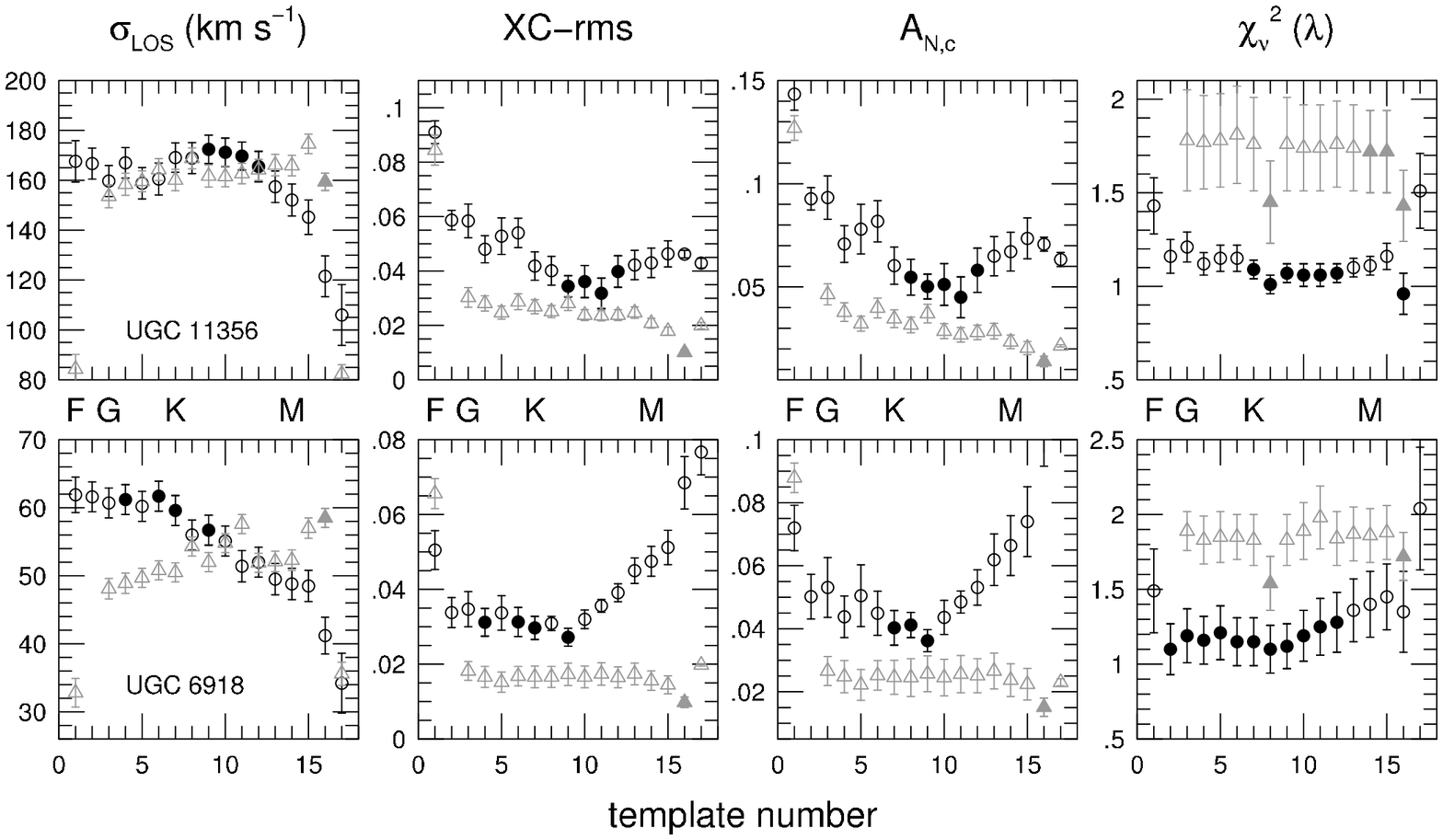}
\caption{Trends of $\slos$ (left) and mismatch indices XC-rms,
  A$_{\rm N,c}$, and $\chi^2_\nu(\lambda)$ (left to right) with
  template for the central regions of the elliptical galaxy UGC 11356
  (top row) and spiral galaxy UGC 6918 (bottom row).  Means and
  standard deviations are determined in UGC 11356 for 7 fibers between
  2 and 6.5 arcsec (effective radius between 19 and 30 arcsec; Bender
  et al. 1994, Fisher 1997, Gerhard et al. 1998), and in UGC 6918 for
  5 fibers between 5 and 10 arcsec (h$_R = 9.4$ arcsec; Verheijen
  1997). Average S/N per pixel in the spectral continuum is 25 per
  fiber in the \ion{Mg}{1b} region and 45 per fiber in
  the \ion{Ca}{2}-triplet region for both galaxies. Black circles
  represent \ion{Mg}{1b}-region measurements; gray triangles
  represent \ion{Ca}{2}-triplet region measurements for the same
  fibers in approximately the same location. Filled symbols in $\slos$
  and XC-rms columns represent templates that have XC-rms values
  statistically equivalent to the minimum XC-rms value; in other
  columns filled circles are keyed in the same way for their
  respective index. The templates are numbered and ordered by spectral
  type (giants only, luminosity class II-III), hot to cool, as
  indicated between the two rows. Templates shown here are near-solar
  metallicity with the exception of template 8, (HR 4695, K1~IIIb),
  which is substantially sub-solar with $[$Fe/H$]$=-0.48.}
\end{figure}

\clearpage

\begin{figure}
\figurenum{12}
\epsscale{1.0}
\plotone{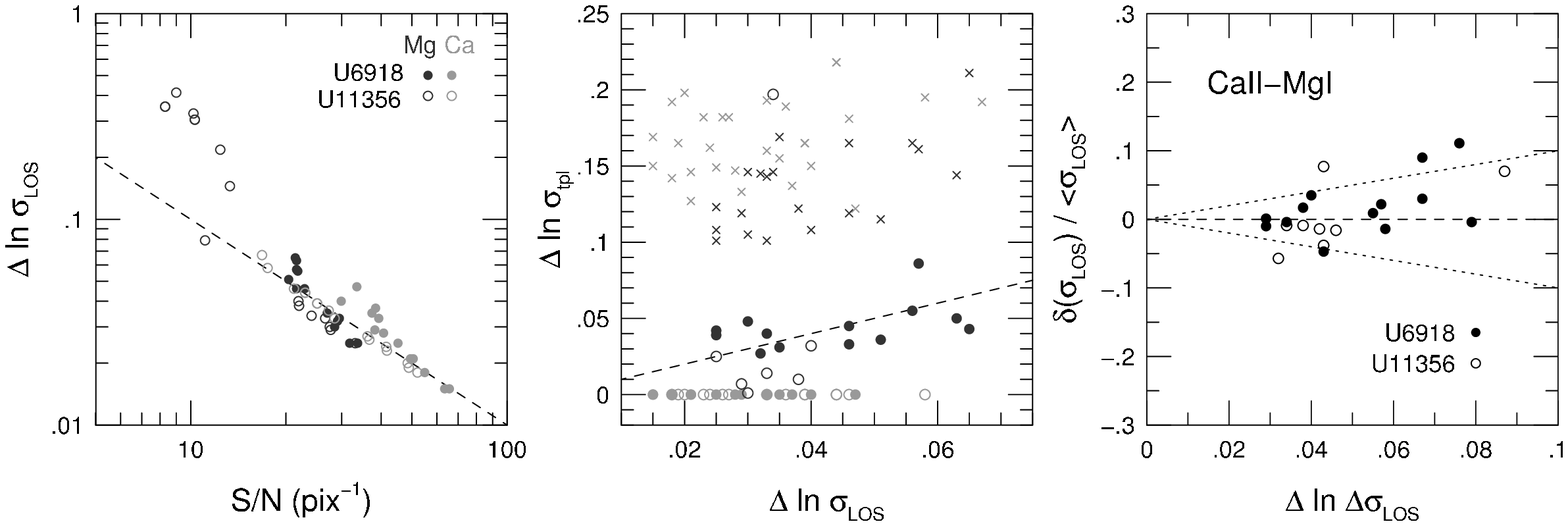}
\caption{Random errors in $\slos$ due to shot-noise in the spectra
  ($\Delta \ln \slos$) and template mismatch ($\Delta \ln \stm$) for
  14 of the brightest individual fibers in UGC 6918 and UGC 11356 in
  the \ion{Mg}{1b} and \ion{Ca}{2}-triplet spectral regions. The left
  panel illustrates the dependence of $\Delta \ln \slos$ on spectral
  continuum S/N; the dashed line depicts the relation $\Delta \ln
  \slos = (S/N)^{-1}$. Symbols are defined in the legend. The middle
  panel illustrates the amplitude of $\Delta \ln \slos$ and $\Delta
  \ln \stm$. Crosses denote systematic errors using our full range of
  templates (see text); circles denote systematic errors based on
  limiting templates using the XC-rms index; the dashed line depicts
  the 1:1 relation. The right panel shows the fractional difference
  between $\slos$ measured in the \ion{Ca}{2}-triplet and \ion{Mg}{1b}
  regions using our XC-rms index method [$\delta(\slos) /
  \langle\slos\rangle$, where $\delta(\slos)$ is the difference and
  $\langle\slos\rangle$ is the mean value for the two regions] versus
  the random error in this difference [$\Delta \ln
  \delta(\slos)$]. Dotted lines illustrate ``1$\sigma$" boundaries in
  random error.}
\end{figure}

\clearpage

\begin{figure}
\figurenum{13}
\epsscale{1.0}
\plotone{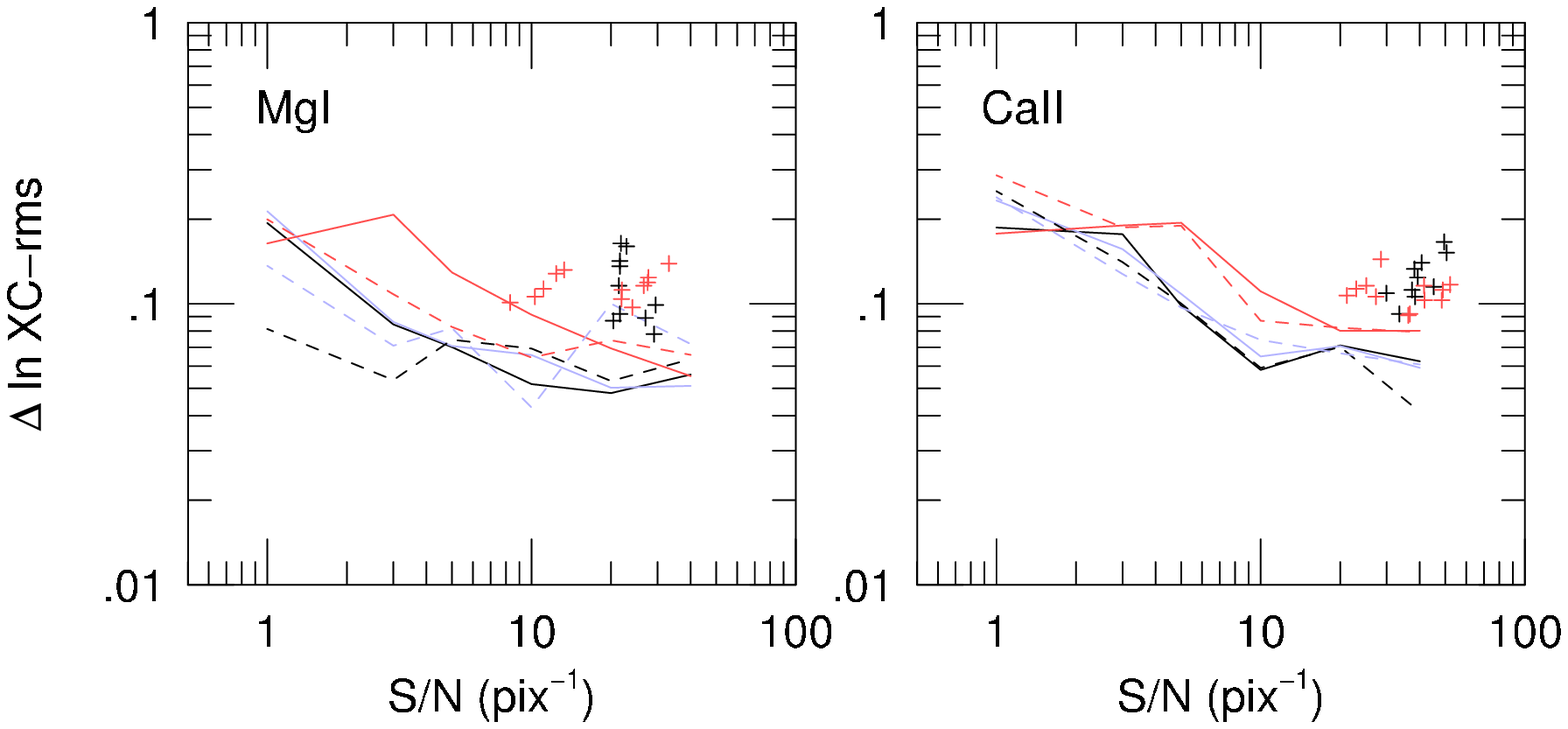}
\caption{Simulations (lines) and measurements (points) of the
  logarithmic error in the template-mismatch index XC-rms (defined in
  the text) as a function of spectral continuum S/N. Lines are
  simulations using a K1~III template (solid lines) and an M3~III
  template (dashed lines) color-coded for different broadening as
  given in Figure 10.  Measurements of $\Delta \ln$ XC-rms for
  individual fibers (plus symbols) use the K1~III template in the
  MgI-region and the M3~III template in the \ion{Ca}{2}-region. UGC
  6918 measurements are in blue; UGC 11356 measurements are in
  red. [COLOR FIGURE FOR ELECTRONIC EDITION ONLY.] }
\end{figure}

\clearpage

\begin{figure}
\figurenum{14}
\epsscale{1.0}
\plotone{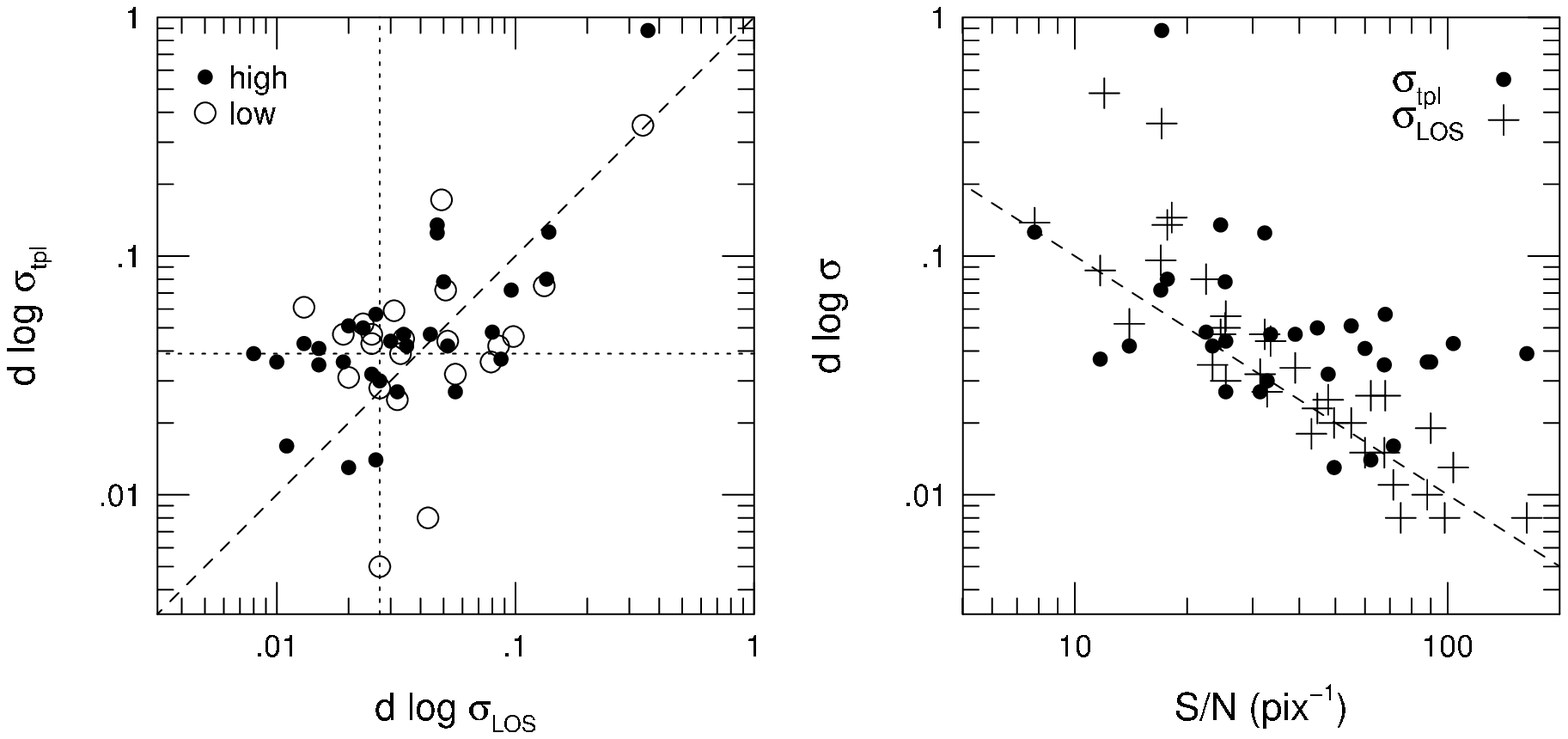}
\caption{Distribution of random errors due to shot-noise in the
  spectra ($\Delta \ln \slos$) and template mismatch
  ($\Delta \ln \stm$) for azimuthally averaged spectra for 7 spiral
  galaxies (see text) in our Phase B sample. ``High'' and ``low''
  identify template-errors assuming $\Delta \ln$ XC-rms values of 0.12
  and 0.06, respectively (Figure 13 and text). Horizontal and vetical
  dotted lines in the left-hand panel are the median values; the
  dashed line is a 1:1 relation provided for reference. The S/N range
  spanned in the right panel is representative of our survey (Paper
  I). The diagonal dashed line is the same relation between
  $\Delta \ln \slos$ and S/N as adopted in the left panel of Figure
  12.}
\end{figure}

\clearpage

\begin{figure}
\figurenum{15}
\epsscale{0.7}
\plotone{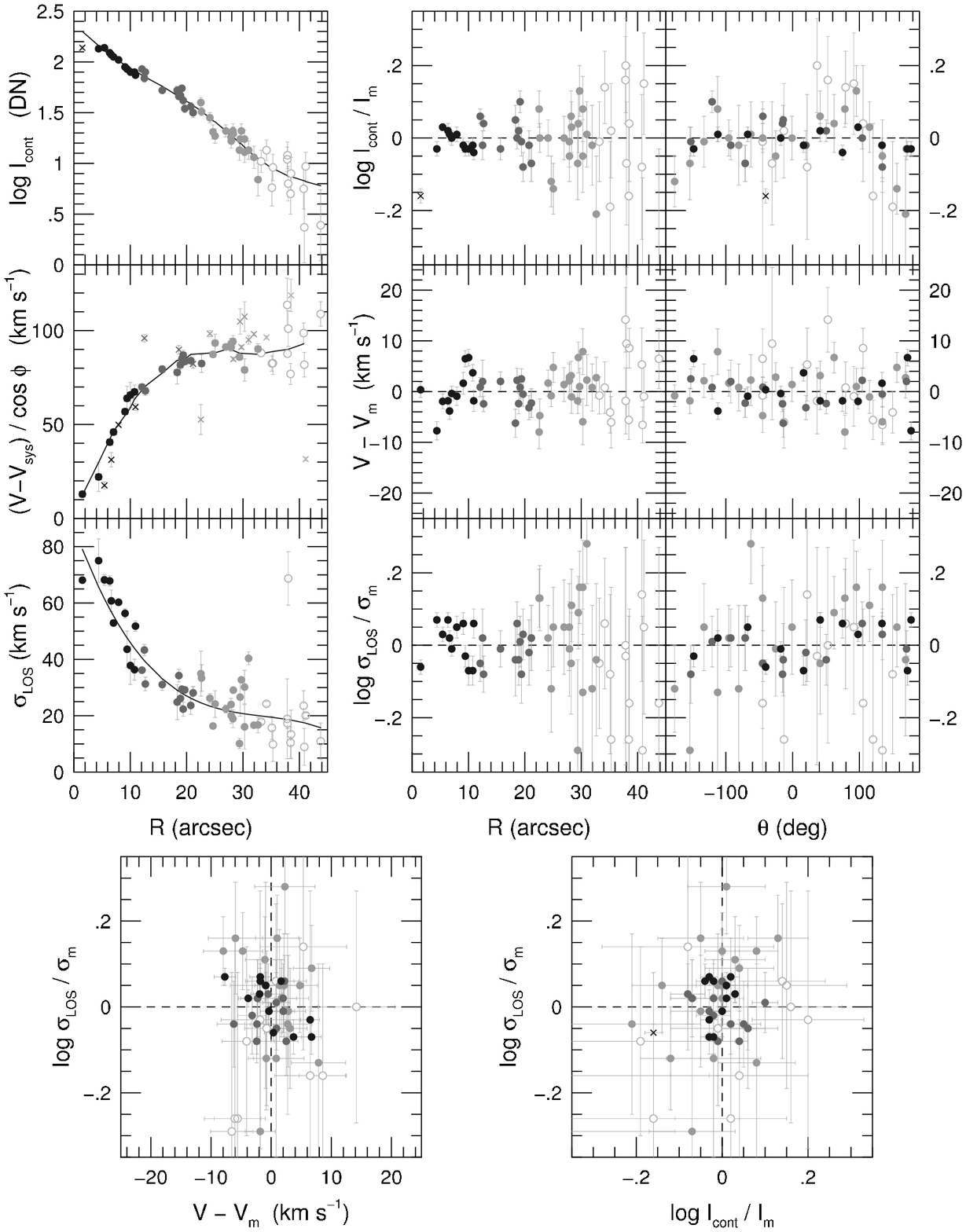}
\caption{Correlations of surface-brightness, velocity and
  velocity-dispersion residuals: checks on uncorrected systematics for
  UGC 6918.  Top three left-most panels show radial trends of spectral
  continuum intensity (I$_{\rm cont}$) in the MgI region, projected
  velocity V (deprojected for azimuth but not inclination) and
  corrected for the estimated systemic recession velocity ($\vsys$),
  and the line-of sight velocity dispersion $\sigma_{\rm
  LOS}$. Polynomial fits (curves) define smooth model values I$_{\rm
  m}$, V$_{\rm m}$, and $\sigma_{\rm m}$ respectively.  The inner-most
  surface-brightness datum is excluded from the fit and marked by x.
  Velocities are plotted as circles for azimuthal angles within
  60$^\circ$ of the major axis, and x's otherwise.  Ratios or
  differences of observed and model values for all points are shown
  versus radius ($R$) and azimuth ($\theta$) in the galaxy plane in
  the middle and right top three rows. The model $\sigma_{\rm m}$ is
  modulated in azimuth as described in the text. Bottom panels
  correlate these ratios and differences against each other. Points in
  all panels represent individual fiber measurements from SparsePak,
  shaded by radius.}
\end{figure}
\clearpage

\begin{figure}
\figurenum{16}
\epsscale{1.0}
\plotone{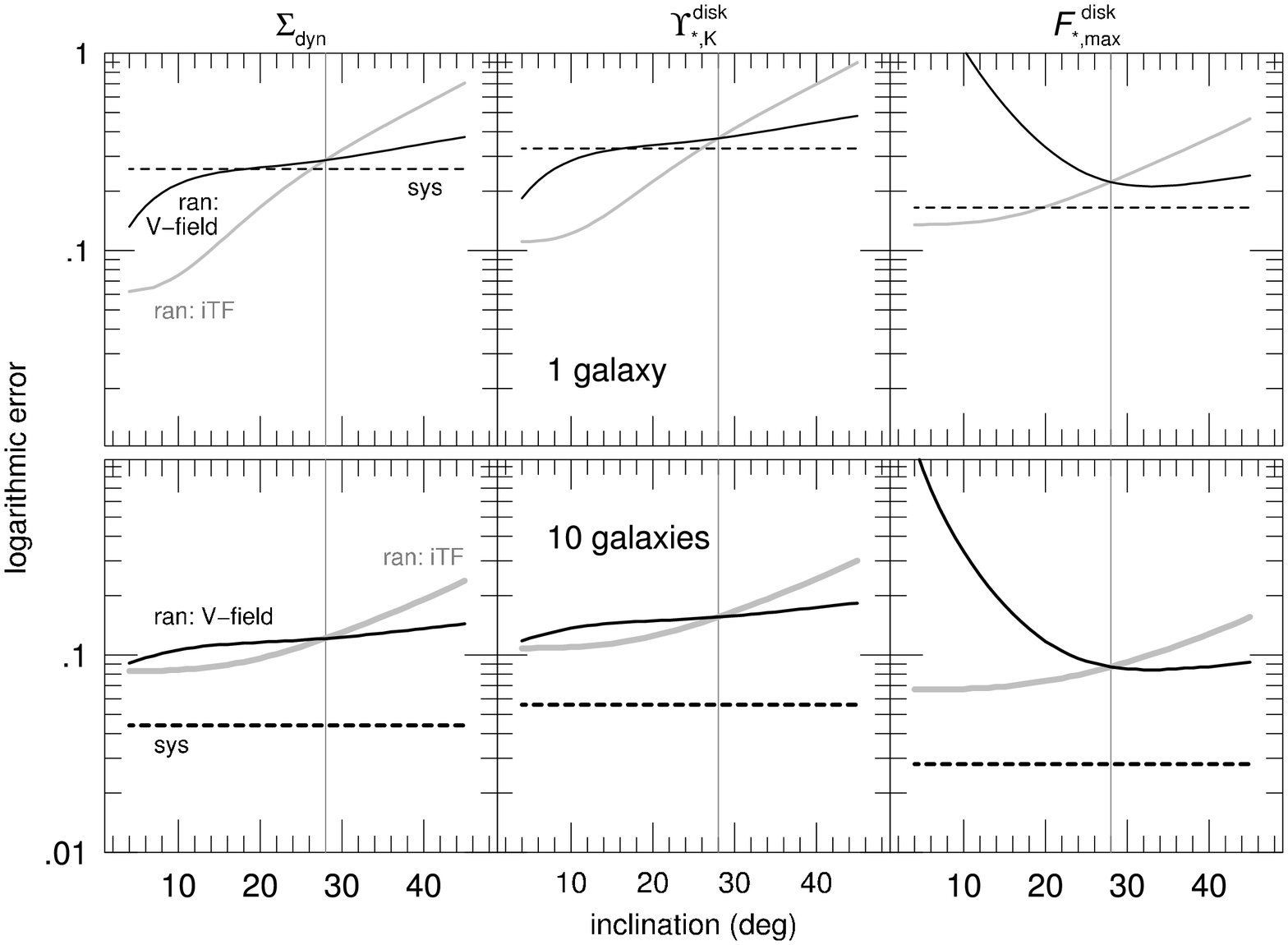}
\caption{Logarithmic errors for $\sddisk$ (left),
  $\mlskdisk$ (middle), and $\Fdisk$ (right) versus
  inclination for an single galaxy (top panels) and for averages of 10
  galaxies (bottom panels). Systematic errors (sys) are shown as
  horizontal dashed lines. Random errors (ran) are calculated using
  estimated inclination errors based on kinematic determinations
  (black curves), and inverse Tully-Fisher (iTF) determinations (gray
  curves). The vertical line at 28$^\circ$ marks the cross-over
  between these two estimates.}
\end{figure}

\clearpage

\begin{figure}
\figurenum{17}
\epsscale{0.7}
\plotone{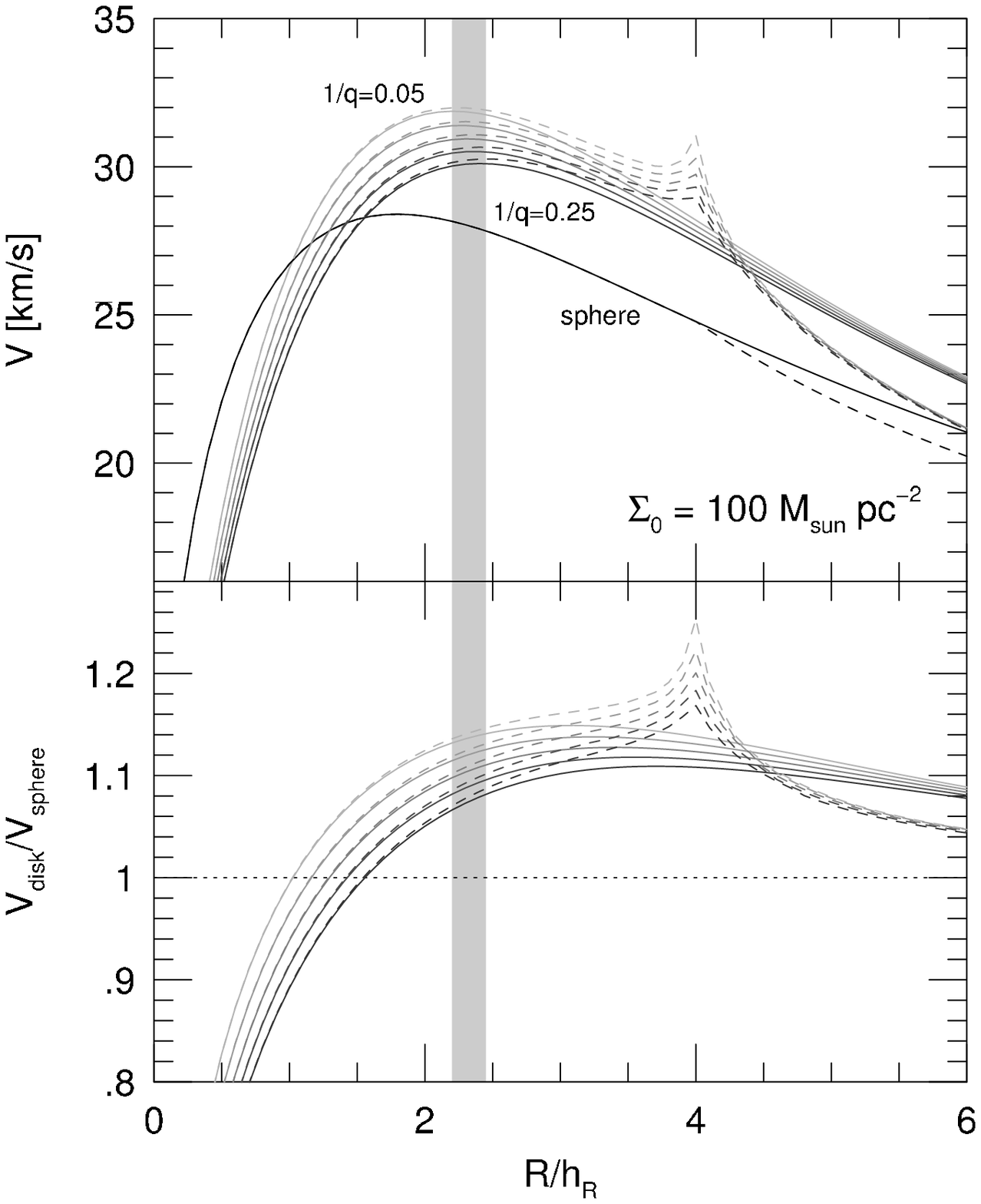}
\caption{Rotation speed of an exponential disk with central mass
  surface density of 100 $\msol$ pc$^{-2}$ and oblateness $0.05 < q <
  0.25$ versus radius normalized by scale-length, compared to a
  spherical density distribution with the same enclosed mass. Bottom
  panel shows the ratio of spherical to disk velocities. Dashed and
  solid lines show disks truncated at R/h$_R$=4 and 10, respectively.
  The radial range where these disks have peak velocities is shaded in
  gray.}
\end{figure}

\clearpage

\begin{figure}
\figurenum{18}
\epsscale{0.7}
\plotone{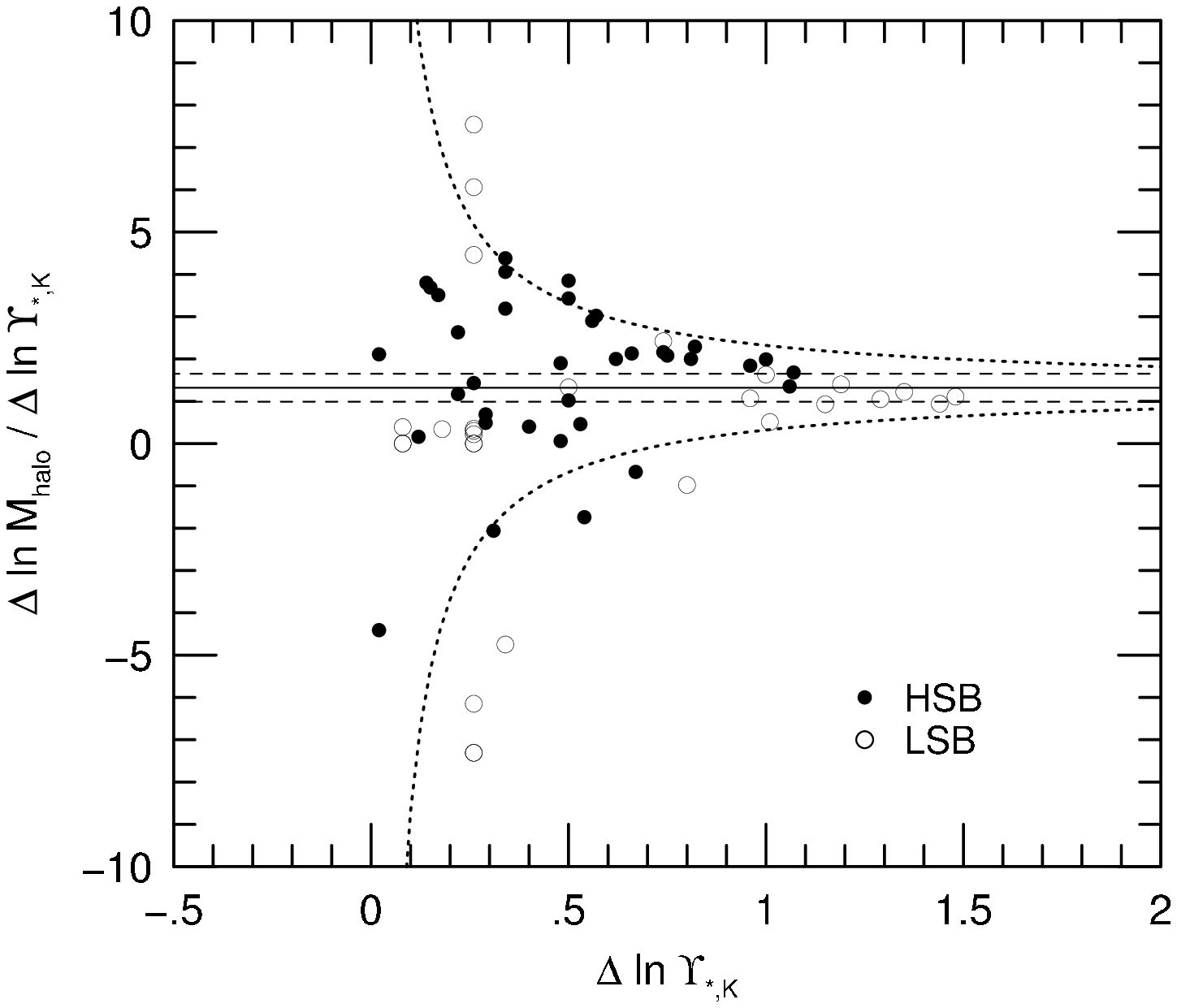}
\caption{Changes in halo mass as a function of stellar mass-to-light
  ratio for the Ursa Major galaxy sample from Verheijen (1997).
  Dotted lines illustrate a plausible error-model for this
  distribution.  Solid and dashed lines give the error-weighted mean
  and error in the mean.  Low and high surface-brightness galaxies
  (LSB and HSB) are marked.}
\end{figure}

\clearpage


\begin{references}

\reference{} Akritas, M. G., Bershady, M. A. 1996, ApJ, 470, 706

\reference{} Aoki, T. E., Hiromoto, N., Takami, H., Okamura, S. 1991, PASJ, 43, 755

\reference{} Andersen, D. R. 2001, Ph.D. Thesis, Penn State University


\reference{} Andersen, D. R., \& Bershady, M. A. 2003, ApJ, 599, L79

\reference{} Andersen, D. R., Bershady, M. A., Sparke, L. S., Gallagher, J. S., Wilcots, E. M.,
van Driel, W., Monnier-Ragaigne, D. 2006, ApJS, 166,505 

\reference{} Andersen, D. R., Walcher, C. J., Boker, T., Ho, Luis C.,
van der Marel, R. P., Rix, H.-W., Shields, J. C. 2008, ApJ, 688, 990


\reference{} Bahcall, J, \& Casertano, S. 1984, ApJ, 284, L35

\reference{} Barth, A., Ho, L. C., Sargent, W. L. W. 2002, AJ, 124, 2607

\reference{} Begeman, K. G. 1989, A\&A, 223, 47

\reference{} Bender, R., Saglia, R. P., Gerhard, O. E. 1994, MNRAS, 269, 785





\reference{} Bershady, M. A., Andersen, D. R., Verheijen, M. A. W.,
Westfall, K. M., Crawford, S. M., Swaters, R. A. 2005, ApJS, 156, 311

\reference{} Bershady, M. A., Verheijen, M. A. W., Swaters, R. A.,
Andersen, D. R., Westfall, K. M., Martinsson, T. 2009, ApJ, submitted
(Paper I)

\reference{} Binney, J., \& Merrifield, M. 1998, ``Galactic
Astronomy,'' Princeton University Press

\reference{} Binney, J., \& Tremaine, S. 1987, ``Galaxy Dynamics'',
Princeton University Press

\reference{} Bizyaev, D. \& Mitronova, S. 2002, A\&A, 389, 795

\reference{} Bizyaev, D. \& Mitronova, S. 2009, ApJ, 702, 1567

\reference{} Boissier, S., Boselli, A., Buat, V., Donas, J., Milliard,
B. 2004, A\&A, 424, 465

\reference{} Bosma, A. 1981, AJ, 86, 1825

\reference{} Bottema, R. 1993, A\&A, 275, 16

\reference{} Bottema, R. 1997, A\&A, 328, 517

\reference{} Calzetti, D., Kinney, A., L., \& Storchi-Bergmann, T. 1994, ApJ, 429, 582

\reference{} Casertano, S. 1983, MNRAS, 203, 735

\reference{} Casoli, E., et al. 1998, A\&A, 331, 451

\reference{} Coluzzi, R. 1993, Bull. Inf. Centre Donnees Stellaires, 43, 7

\reference{} Courteau, S. 1997, AJ, 114, 2402

\reference{} Dalcanton, J. J., Yoachim, P. \& Bernstein, R. A. 2004, ApJ, 608, 189

\reference{} de Grijs, R., van der Kruit, P. C. 1997, A\&AS, 327, 966

\reference{} de Grijs, R. 1998, MNRAS, 299, 595 

\reference{} de Jong, R. 1996a, A\&A, 313, 377

\reference{} de Jong, R. 1996b, A\&AS, 118, 557

\reference{} de Bruyne, V., De Rijcke, S., Dejonghe, H., Zeilinger,
W. W. 2004, MNRAS, 349, 461

\reference{} Domingue, D. L., Keel, W. C., White, R. E. 2000, ApJ, 545, 171

\reference{} Falc\'{o}n-Barroso, J. et al. 2006, MNRAS, 369, 529

\reference{} Fisher, D. 1997, AJ, 113, 950

\reference{} Fisher, D. 1997, AJ, 113, 950 

\reference{} Franx, M. \& Illingworth, G. 1988, ApJ, 327, L55

\reference{} Freeman, K. C. 1970, ApJ, 160, 811

\reference{} Fry, A. M., Morrison, H. L., Harding, P., Boroson,
T. A. 1999, AJ, 118, 1209

\reference{} Fukugita, M., Ichikawa, T., Gunn, J. E., Doi, M.,
Shimasaky, K., Schneider, D. P. 1996, AJ, 111, 1748

\reference{} Gerhard, O., Jeske, G., Saglia, R. P., Bender, R. 1998,
MNRAS, 295, 197

\reference{} Herrmann, K. A., and Ciardullo, R. 2009, ApJ, accepted
(arXiv:0910.0266v1)

\reference{} Hernquist, L. 1990, ApJ, 356, 359

\reference{} Hoekstra, H., van Albada, T. S., Sancisi, R. 2001, MNRAS, 323, 453

\reference{} Holwerda, B. W., Gonzalez, R. A., Allen, R. J., van der
Kruit, P. C. 2005, AJ, 129, 1396

\reference{} Howk, J. C., \& Savage, B. D. 1999, AJ, 117, 2077

\reference{} Kamphuis, P., Holwerda, B. W., Allen, R. J., Peletier, \&
van der Kruit, P. C. 2007, A\&A, 471, L1

reference{} Keel, W. C., \& White, R. E. 2001, AJ, 121, 1442

\reference{} Kregel, M., van der Kruit, P. C., De Grijs, R. 2002,
MNRAS, 334, 646

\reference{} Kregel, M., van der Kruit, P. C., Freeman, K. C. 2004,
MNRAS, 351, 1247

\reference{} Kregel, M., van der Kruit, P. C., Freeman, K. C. 2005,
MNRAS, 3581, 503

\reference{} Le Borgne, D., Rocca-Volmerange, B., Prugniel, P.,
Lan\c{c}on, A., Fioc, M., and Soubiran, C. 2004, A\&A, 425, 881

\reference{} Leroy, A. K., Walter, F., Brinks, E., Bigiel, F., de
Blok, W. J. G., Madore, B., Thornley, M. D. 2008, AJ, 136, 2782

\reference{} MacArthur, L, Courteau, S., Holtzman, J. 2003, ApJ, 582, 689

\reference{} Misiriotis, A., Kylafis, N. D., Papamastorakis, J.,
Xilouris, E. M.  2000, A\&A, 353, 117

\reference{} Misiriotis, A., Popescu, C. C., Tuffs, R., Kylafis,
N. D. 2001, A\&A, 372, 775

\reference{} Mould, J. et al. 2000, ApJ, 529, 786

\reference{} Morrison, H. L., Boroson, T. A., Harding, P. 1994, AJ, 108, 1191

\reference{} Nelson, C. H., Whittle, M. 1995, ApJS, 99, 67

\reference{} Obreschkow, D., \& Rawlings, S. 2009, MNRAS, 394, 1857

\reference{} Olling, R. P. 1996, AJ, 112

\reference{} Pohlen, M., Dettmar, R.-J., Lutticke, R., and
Schwarzkopf, U. 2000, A\&AS, 144, 405

\reference{} Popescu, C. C., Misiriotis, A., Kylafis, N. D., Tuffs,
R. J., Fischera, J. 2000, A\&A, 362, 138

\reference{} Regan, M. W., et al. 2001, ApJ, 561, 218

\reference{} Rix, H.-W., White, S. D. M. 1992, MNRAS, 254, 389

\reference{} Schwarzkopf, U., Dettmar, R.-J. 2000, A\&A, 361, 451

\reference{} Seth, A. C., Dalcanton, J. J., and De Jong, R. S. 2005,
AJ, 130, 1574

\reference{} Shapiro, K., Gerssen, J., van der Marel, R.P. 2003, AJ, 126, 2707

\reference{} Simkin, S. 1974, A\&A, 31, 129

\reference{} Statler, T. S., 1995, AJ, 109, 1371

\reference{} Swaters, R. A., Verheijen, M. A. W., Bershady, M. A.,
Andersen, D. R. 2003, ApJ, 587, L19

\reference{} Tamura, K., Jansen, R. A., \& Windhorst, R. A. 2009, AJ, 138, 1634

\reference{} Tully, R. B., \& Fisher, J. R. 1977, A\&A, 54, 661

\reference{} van Albada, T. S., Bahcall, J. N., Begeman, K., Sancisi,
R. 1985, ApJ, 295, 305

\reference{} van den Bergh, S., 1960, ApJ, 131, 215

\reference{} van der Kruit, P. C., \& Searle, L. 1981, A\&A, 95, 105

\reference{} van der Kruit, P. C., \& Freeman, K. C. 1984, ApJ, 278, 81

\reference{} van der Kruit, P. C., \& Freeman, K. C. 1986, ApJ, 303, 556

\reference{} van der Kruit, P. C. 1988, A\&A, 192, 117

\reference{} van der Kruit, P. C., \& de Grijs, R. 1999, A\&A, 352, 129

\reference{} van der Marel, R. P., \& Franx, M. 1993, 407, 525

\reference{} Verheijen, M. A. W. 1997, Ph.D. thesis, University of Groningen

\reference{} Verheijen, M. A. W. 2001, ApJ, 563, 694

\reference{} Verheijen, M. A. W., \& Sancisi, R. 2001, A\&A, 370, 765

\reference{} Wainscoat, R. J, Freeman, K. C., Hyland, A.R. 1989, ApJ, 337, 163

\reference{} Westfall, K. B. 2009, Ph.D. Thesis, University of Wisconsin

\reference{} Xilouris, E. M., Kylafis, N. D., Papamastorakis, J.,
Paleologou, E. G., Haerendel, G. 1997, A\&A, 325, 135

\reference{} Xilouris, E. M., Alton, P. B., Davies, J. I., Kylafis,
N. D., Papamastorakis, J., Trewhella, M. 1998, A\&A, 331, 894

\reference{} Xilouris, E. M., Byun, Y. I., Kylafis, N. D., Papamastorakis
1999, A\&A, 344, 868

\reference{} Yoachim, P. \& Dalcanton, J. J. 2006, AJ, 131, 226

\reference{} Young, J. D., Knezek, P. M. 1989, ApJ, 347, L55

\end{references}
\end{document}